\documentclass[twocolumn,showpacs,preprintnumbers,amsmath,amssymb,superscriptaddress]{revtex4-1}

\usepackage{graphicx,footnote}
\usepackage{enumerate}
\usepackage[format=plain,singlelinecheck=off]{caption}
\usepackage{subcaption}
\usepackage{ngerman}
\usepackage[applemac]{inputenc}
\usepackage{amsmath}
\usepackage[stable]{footmisc}
\usepackage{dsfont}
\usepackage{textcomp}
\usepackage{listliketab}
\usepackage{tabularx}
\usepackage{multirow}
\usepackage{color}
\usepackage{tabto}
\usepackage{stmaryrd}
\usepackage{amssymb}
\usepackage{here}
\usepackage{footnote}
\usepackage{pstricks-add}
\usepackage{eepic}
\usepackage{tikz}
\usepackage{atbegshi}
\usepackage[titletoc,title]{appendix}

\newcommand*\teo{  \begin{minipage}[htp]{5cm}
\begin{tikzpicture}[scale=1] \draw (0,0) -- (5,0); \draw [ultra thick] (0,-0.25) -- (0,0.25); \draw (1.6,-0.25) -- (1.6,0.25); \draw (1.66,-0.25) -- (1.66,0.25); \draw (2.6,-0.25) -- (2.6,0.25); \draw (2.66,-0.25) -- (2.66,0.25); \draw [fill] (4.9,0.08) -- (4.9,-0.08) -- (5,0) -- (4.9,0.08);  \draw [fill]  (1.2,0)circle (2pt); \draw [ultra thick] (1.2,0) -- (4.9,0); \end{tikzpicture}
 \end{minipage}}
 \newcommand*\mbo{ \begin{minipage}[htp]{5cm}
\begin{tikzpicture}[scale=1] \draw (0,0) -- (5,0); \draw [ultra thick] (0,-0.25) -- (0,0.25); \draw (1.6,-0.25) -- (1.6,0.25); \draw (1.66,-0.25) -- (1.66,0.25); \draw (2.6,-0.25) -- (2.6,0.25); \draw (2.66,-0.25) -- (2.66,0.25); \draw [fill] (4.9,0.08) -- (4.9,-0.08) -- (5,0) -- (4.9,0.08); \draw [fill]  (1.2,0)circle (2pt); \draw [fill]  (3.06,0)circle (2pt); \draw [ultra thick] (1.2,0) -- (3.06,0); \end{tikzpicture}
 \end{minipage}}

 \newcommand*\tos{ \begin{minipage}[htp]{5cm}
\begin{tikzpicture}[scale=1]
\draw (0,0) -- (5,0); \draw [ultra thick] (0,-0.25) -- (0,0.25); \draw (1.6,-0.25) -- (1.6,0.25); \draw (1.66,-0.25) -- (1.66,0.25); \draw (2.6,-0.25) -- (2.6,0.25); \draw (2.66,-0.25) -- (2.66,0.25);\draw [fill] (4.9,0.08) -- (4.9,-0.08) -- (5,0) -- (4.9,0.08); \draw[fill] (0,-2pt) arc (-90:90:2pt);  \draw [ultra thick] (0,0) -- (4.9,0); \end{tikzpicture}
 \end{minipage}}
\newcommand*\teomin{  \begin{minipage}[htp]{5cm}
\begin{tikzpicture}[scale=1]
\draw (0,0) -- (5,0); \draw [ultra thick] (0,-0.25) -- (0,0.25); \draw (1.6,-0.25) -- (1.6,0.25); \draw (1.66,-0.25) -- (1.66,0.25); \draw (2.6,-0.25) -- (2.6,0.25); \draw (2.66,-0.25) -- (2.66,0.25); \draw [fill] (4.9,0.08) -- (4.9,-0.08) -- (5,0) -- (4.9,0.08); 
 \draw [fill]  (1.63,0)circle (2pt); \draw [ultra thick] (1.63,0) -- (4.9,0); 
\end{tikzpicture}
 \end{minipage}}
\newcommand*\mboeo{ \begin{minipage}[htp]{5cm}
\begin{tikzpicture}[scale=1]
\draw (0,0) -- (5,0); \draw [ultra thick] (0,-0.25) -- (0,0.25); \draw (1.6,-0.25) -- (1.6,0.25); \draw (1.66,-0.25) -- (1.66,0.25); \draw (2.6,-0.25) -- (2.6,0.25); \draw (2.66,-0.25) -- (2.66,0.25); \draw [fill] (4.9,0.08) -- (4.9,-0.08) -- (5,0) -- (4.9,0.08); 
 \draw [fill]  (1.2,0)circle (2pt); \draw [fill]  (3.8,0)circle (2pt); \draw [fill]  (3.06,0)circle (2pt); \draw [ultra thick] (1.2,0) -- (3.06,0);  \draw [ultra thick] (3.8,0) -- (4.9,0); 
\end{tikzpicture}
 \end{minipage}}
\newcommand*\mboeomax{ \begin{minipage}[htp]{5cm}
\begin{tikzpicture}[scale=1]
\draw (0,0) -- (5,0); \draw [ultra thick] (0,-0.25) -- (0,0.25); \draw (1.6,-0.25) -- (1.6,0.25); \draw (1.66,-0.25) -- (1.66,0.25); \draw (2.6,-0.25) -- (2.6,0.25); \draw (2.66,-0.25) -- (2.66,0.25); \draw [fill] (4.9,0.08) -- (4.9,-0.08) -- (5,0) -- (4.9,0.08); 
 \draw [fill]  (1.2,0)circle (2pt); \draw [fill]  (3.8,0)circle (2pt); \draw [fill]  (2.63,0)circle (2pt); \draw [ultra thick] (1.2,0) -- (2.63,0);  \draw [ultra thick] (3.8,0) -- (4.9,0); 
\end{tikzpicture}
 \end{minipage}}
\newcommand*\mboeomin{ \begin{minipage}[htp]{5cm}
\begin{tikzpicture}[scale=1]
\draw (0,0) -- (5,0); \draw [ultra thick] (0,-0.25) -- (0,0.25); \draw (1.6,-0.25) -- (1.6,0.25); \draw (1.66,-0.25) -- (1.66,0.25); \draw (2.6,-0.25) -- (2.6,0.25); \draw (2.66,-0.25) -- (2.66,0.25); \draw [fill] (4.9,0.08) -- (4.9,-0.08) -- (5,0) -- (4.9,0.08); 
 \draw [fill]  (1.63,0)circle (2pt); \draw [fill]  (3.8,0)circle (2pt); \draw [fill]  (3.06,0)circle (2pt); \draw [ultra thick] (1.63,0) -- (3.06,0);  \draw [ultra thick] (3.8,0) -- (4.9,0); 
\end{tikzpicture}
 \end{minipage}}
\newcommand*\mboeomaxmin{ \begin{minipage}[htp]{5cm}
\begin{tikzpicture}[scale=1]
\draw (0,0) -- (5,0); \draw [ultra thick] (0,-0.25) -- (0,0.25); \draw (1.6,-0.25) -- (1.6,0.25); \draw (1.66,-0.25) -- (1.66,0.25); \draw (2.6,-0.25) -- (2.6,0.25); \draw (2.66,-0.25) -- (2.66,0.25); \draw [fill] (4.9,0.08) -- (4.9,-0.08) -- (5,0) -- (4.9,0.08); 
 \draw [fill]  (1.63,0)circle (2pt); \draw [fill]  (3.8,0)circle (2pt); \draw [fill]  (2.63,0)circle (2pt); \draw [ultra thick] (1.63,0) -- (2.63,0);  \draw [ultra thick] (3.8,0) -- (4.9,0); 
\end{tikzpicture}
 \end{minipage}}
\newcommand*\mbomax{ \begin{minipage}[htp]{5cm}
\begin{tikzpicture}[scale=1]
\draw (0,0) -- (5,0); \draw [ultra thick] (0,-0.25) -- (0,0.25); \draw (1.6,-0.25) -- (1.6,0.25); \draw (1.66,-0.25) -- (1.66,0.25); \draw (2.6,-0.25) -- (2.6,0.25); \draw (2.66,-0.25) -- (2.66,0.25); \draw [fill] (4.9,0.08) -- (4.9,-0.08) -- (5,0) -- (4.9,0.08); 
 \draw [fill]  (1.2,0)circle (2pt); \draw [fill]  (2.63,0)circle (2pt); \draw [ultra thick] (1.2,0) -- (2.63,0); 
\end{tikzpicture}
 \end{minipage}}
\newcommand*\mbomin{ \begin{minipage}[htp]{5cm}
\begin{tikzpicture}[scale=1]
\draw (0,0) -- (5,0); \draw [ultra thick] (0,-0.25) -- (0,0.25); \draw (1.6,-0.25) -- (1.6,0.25); \draw (1.66,-0.25) -- (1.66,0.25); \draw (2.6,-0.25) -- (2.6,0.25); \draw (2.66,-0.25) -- (2.66,0.25); \draw [fill] (4.9,0.08) -- (4.9,-0.08) -- (5,0) -- (4.9,0.08); 
 \draw [fill]  (1.63,0)circle (2pt); \draw [fill]  (3.06,0)circle (2pt); \draw [ultra thick] (1.63,0) -- (3.06,0); 
\end{tikzpicture}
 \end{minipage}}
\newcommand*\mbomaxmin{ \begin{minipage}[htp]{5cm}
\begin{tikzpicture}[scale=1]
\draw (0,0) -- (5,0); \draw [ultra thick] (0,-0.25) -- (0,0.25); \draw (1.6,-0.25) -- (1.6,0.25); \draw (1.66,-0.25) -- (1.66,0.25); \draw (2.6,-0.25) -- (2.6,0.25); \draw (2.66,-0.25) -- (2.66,0.25); \draw [fill] (4.9,0.08) -- (4.9,-0.08) -- (5,0) -- (4.9,0.08); 
 \draw [fill]  (1.63,0)circle (2pt); \draw [fill]  (2.63,0)circle (2pt); \draw [ultra thick] (1.63,0) -- (2.63,0); 
\end{tikzpicture}
 \end{minipage}}
\newcommand*\boteo{ \begin{minipage}[htp]{5cm}
\begin{tikzpicture}[scale=1]
\draw (0,0) -- (5,0); \draw [ultra thick] (0,-0.25) -- (0,0.25); \draw (1.6,-0.25) -- (1.6,0.25); \draw (1.66,-0.25) -- (1.66,0.25); \draw (2.6,-0.25) -- (2.6,0.25); \draw (2.66,-0.25) -- (2.66,0.25); \draw [fill] (4.9,0.08) -- (4.9,-0.08) -- (5,0) -- (4.9,0.08); 
\draw [fill]  (0.3,0)circle (2pt); \draw [fill]  (0.8,0)circle (2pt); \draw [fill]  (1.2,0)circle (2pt); \draw [ultra thick] (0.3,0) -- (0.8,0);  \draw [ultra thick] (1.2,0) -- (4.9,0); 
\end{tikzpicture}
 \end{minipage}}
\newcommand*\boteomin{ \begin{minipage}[htp]{5cm}
\begin{tikzpicture}[scale=1]
\draw (0,0) -- (5,0); \draw [ultra thick] (0,-0.25) -- (0,0.25); \draw (1.6,-0.25) -- (1.6,0.25); \draw (1.66,-0.25) -- (1.66,0.25); \draw (2.6,-0.25) -- (2.6,0.25); \draw (2.66,-0.25) -- (2.66,0.25); \draw [fill] (4.9,0.08) -- (4.9,-0.08) -- (5,0) -- (4.9,0.08); 
\draw [fill]  (0.3,0)circle (2pt); \draw [fill]  (0.8,0)circle (2pt); \draw [fill]  (1.63,0)circle (2pt); \draw [ultra thick] (0.3,0) -- (0.8,0);  \draw [ultra thick] (1.63,0) -- (4.9,0); 
\end{tikzpicture}
 \end{minipage}}

\newcommand*\toeo{ \begin{minipage}[htp]{5cm}
\begin{tikzpicture}[scale=1]
\draw (0,0) -- (5,0); \draw [ultra thick] (0,-0.25) -- (0,0.25); \draw (1.6,-0.25) -- (1.6,0.25); \draw (1.66,-0.25) -- (1.66,0.25); \draw (2.6,-0.25) -- (2.6,0.25); \draw (2.66,-0.25) -- (2.66,0.25); \draw [fill] (4.9,0.08) -- (4.9,-0.08) -- (5,0) -- (4.9,0.08); 
\draw[fill] (0,-2pt) arc (-90:90:2pt); \draw [fill]  (1.2,0)circle (2pt); \draw [fill]  (3.2,0)circle (2pt);  \draw [ultra thick] (0,0) -- (1.2,0);  \draw [ultra thick] (3.2,0) -- (4.9,0); 
\end{tikzpicture}
 \end{minipage}}
\newcommand*\toteo{ \begin{minipage}[htp]{5cm}
\begin{tikzpicture}[scale=1]
\draw (0,0) -- (5,0); \draw [ultra thick] (0,-0.25) -- (0,0.25); \draw (1.6,-0.25) -- (1.6,0.25); \draw (1.66,-0.25) -- (1.66,0.25); \draw (2.6,-0.25) -- (2.6,0.25); \draw (2.66,-0.25) -- (2.66,0.25); \draw [fill] (4.9,0.08) -- (4.9,-0.08) -- (5,0) -- (4.9,0.08); 
\draw[fill] (0,-2pt) arc (-90:90:2pt);\draw [fill]  (0.8,0)circle (2pt); \draw [fill]  (1.2,0)circle (2pt); \draw [ultra thick] (0,0) -- (0.8,0);  \draw [ultra thick] (1.2,0) -- (4.9,0); 
\end{tikzpicture}
 \end{minipage}}

\begin{document}

\title{Dynamics of test particles in the general five-dimensional Myers-Perry spacetime}

\author{Valeria Diemer (n\'{e}e Kagramanova)}
\email{valeriya.diemer@uni-oldenburg.de}
\affiliation{Institut f\"{u}r Physik, Universit\"{a}t Oldenburg, 26111 Oldenburg, Germany}
\author{Jutta Kunz}
\email{jutta.kunz@uni-oldenburg.de}
\affiliation{Institut f\"{u}r Physik, Universit\"{a}t Oldenburg, 26111 Oldenburg, Germany}
\author{Claus L\"{a}mmerzahl}
\email{claus.laemmerzahl@zarm.uni-bremen.de}
\affiliation{ZARM, Universit\"{a}t Bremen, Am Fallturm, 28359 Bremen, Germany}
\affiliation{Institut f\"{u}r Physik, Universit\"{a}t Oldenburg, 26111 Oldenburg, Germany}
\author{Stephan Reimers}
\email{stephan.reimers@uni-oldenburg.de}
\affiliation{Institut f\"{u}r Physik, Universit\"{a}t Oldenburg, 26111 Oldenburg, Germany}

\date{\today}

\begin{abstract}

\noindent
We present the complete set of analytical solutions of the geodesic equation 
in the general five-dimensional Myers-Perry spacetime in terms of the 
Weierstrass $\wp$-, $\zeta$- and $\sigma$-functions. 
We analyze the underlying polynomials in the polar and radial equations, 
which depend on the parameters of the metric and the conserved quantities 
of a test particle, and characterize the motion by their zeros. 
We exemplify the efficiency of the analytical method on the basis of the
explicit construction of test particle orbits and by addressing observables 
in this spacetime.

\end{abstract}

\maketitle

\section{Introduction}

A surge of interest in higher-dimensional gravity has
resulted from the search for a theory of quantum gravity. 
With String- and M-theory as candidates, 
it finds important applications in the AdS/CFT correspondence \cite{Maldacena:1997re}.
Also the first successful statistical counting of black hole entropy 
in string theory could be performed for a
black hole in five spacetime dimensions \cite{Strominger:1996sh}.\\
Discovered in 1986, the Myers-Perry solutions \cite{Myers:1986un} 
represent the higher-dimensional generalizations of the Kerr solution
\cite{Kerr:1963ud}. Like the Kerr black holes,
the Myers-Perry black holes possess an event horizon with spherical topology.
However, depending on the number of dimensions $D$,
they may possess $N=[(D-1)/2]$ independent angular momenta,
associated with rotation in $N$ orthogonal planes.\\
Further generalizations of the Myers-Perry solutions
include the general Kerr-de Sitter and Kerr-NUT-AdS metrics
in all higher dimensions \cite{Gibbons:2004uw,Chen:2006xh}.
Reviews of higher-dimensional black hole solutions 
in vacuum or in supergravity theory are, for instance, found
in \cite{Emparan:2008eg,Emparan:2008zz}. Here also black objects
with non-spherical horizon topology are discussed
together with the associated nonuniqueness
of higher-dimensional black holes.\\
Another motivation for the study of higher-dimensional black holes 
was proposed in \cite{Kodama:2009bf}. Here, 
the search for stable rotating black holes -
among the various black objects that higher dimensions offer 
(black holes, black rings, black strings, etc.) -
was advocated, in order to see which of them might be observable in nature.
(In this connection see also, e.g.,~\cite{Kanti:2004nr}.)\\
The only way to explore a gravitational field and to study its properties is through the observation of orbits of test particles and light rays. Therefore, the related geodesic equations need to be solved, either numerically or analytically. If possible, analytical solutions are preferable since they provide a basis for a systematical study of the spacetime properties and allow accurate calculations for e.g. spacetime observables. The method of separation of variables has proved to be a very useful tool in order to obtain the geodesic equations of a gravitational field, yielding, if separable, a set of $D$ ordinary differential equations.\\
The Hamilton-Jacobi equations of motion in the Kerr spacetime
are separable \cite{Chandrasekhar:1983,Oneill:1995}.
As discussed in
\cite{Kraniotis:2003ig,Kraniotis:2004cz,Kraniotis:2005zm,Kraniotis:2007zz,Fujita:2009bp,Hackmann:2009nh,Hackmann:2010zz},
the solutions of the geodesic equations in the Kerr spacetime
can be expressed in terms of elliptic functions.
In contrast, in the Kerr-de Sitter spacetime
in four dimensions hyperelliptic functions are required.\\
Separability of the Hamilton-Jacobi equations of motion 
in the Myers-Perry  and the Kerr-de Sitter spacetimes 
in all higher dimensions 
was shown in \cite{Vasudevan:2004ca,Vasudevan:2004mr} and in \cite{Krtous:2008tb,Frolov:2008jr,Frolov:2010cr,Frolov:2006pe,Kubiznak:2006kt} (see also \cite{Kubiznak:2007ca}). 
A method for obtaining the general analytical solution 
of the geodesic equations in terms of the hyperelliptic 
theta- and sigma-functions in the Myers-Perry spacetimes 
with one rotation parameter was presented in \cite{Enolski:2010if}.
Geodesic stability of circular orbits
in the singly-spinning $D$-dimensional Myers-Perry spacetime
was studied in \cite{Cardoso:2008bp}.
The explicit construction of the general solutions is, however, still missing.\\
Here we focus on the construction of geodesics 
in the five-dimensional Myers-Perry spacetime,
where the extra dimension leads to the presence 
of two independent angular momenta. 
Some special cases for motion in the five-dimensional Myers-Perry spacetime
were studied in \cite{Frolov:2003en,Frolov:2004pr},
where it was shown that there are no stable circular orbits
in the equatorial plane.
In \cite{Gooding:2008tf} scattering and capture of particles
by five-dimensional rotating black holes were investigated.\\
In the recent work \cite{Kagramanova:2012hw}, two of us studied the motion 
of both massive and massless particles in the special
five-dimensional Myers-Perry spacetime with equal rotation parameters. In this case the analytical solutions of the geodesic equations simplify. 
The present work shall be understood as the generalization of this study,
elaborating the fully general case of two independent angular momenta.\\
After discussing some general features of the five-dimensional
Myers-Perry spacetime in Sec.~II,
we derive the equations of motion for test particles
in this spacetime.
By analyzing the polar and radial equations
for the test particle motion,
we classify the possible types of orbits in Sec.~III
and discuss the effective potential in Sec.~IV.
The complete set of analytical solutions of the geodesic equations is given 
in terms of the Weierstrass $\wp$-, $\zeta$- and $\sigma$-functions in Sec.~V.
In Sec.~VI we exhibit orbits of massive and massless test particles 
for selected sets of parameter values. 
We present some observables for the five-dimensional Myers-Perry spacetime 
in Sec.~VII, and apply the results to the found orbits.

\section{Geodesic equations}

We here briefly recall the general Myers-Perry black hole solution 
in five dimensions and the corresponding set of equations
of motion for test particles and light in this spacetime.

\subsection{Myers-Perry spacetime}

\begin{figure}[htbp]
\centering
   \includegraphics[width=0.44\textwidth]{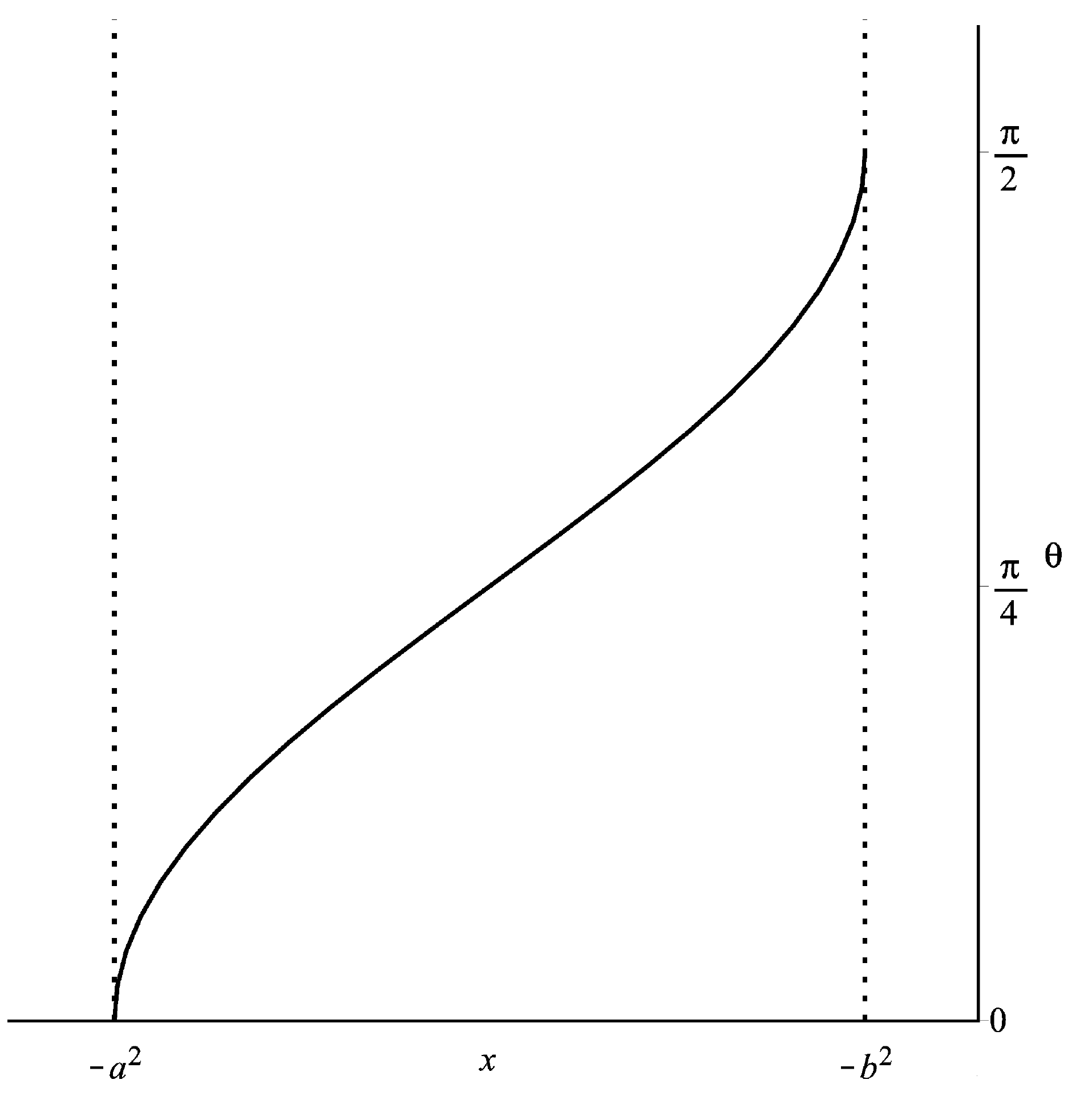}
   \caption{Values of $x = x_{\rm s} \in [-a^2, -b^2]$ 
and $\theta = \theta_{\rm s} \in [0, \frac{\pi}{2}]$ 
for which $\rho^2$ vanishes.}
\label{fig:thetasing}
\end{figure}

The five-dimensional Myers-Perry metric depends on three parameters
\cite{Myers:1986un},
where the mass parameter $\mu$ 
is proportional to the mass $M$ of the black hole 
\begin{align}
M = \frac{3 \pi \mu}{8 G_5},
\end{align}
with the five-dimensional gravitational coupling constant $G_5$,
and the two possible rotation parameters $a$ and $b$
are proportional to the angular momenta $J_a$ and $J_b$, respectively,
\begin{align}
J_a= \frac{2}{3} M a , \qquad J_b= \frac{2}{3} M b .
\end{align}
Without loss of generality, we will choose $a \ge b$ in the following.\\ 
The line element in Boyer-Lindquist coordinates \cite{Boyer:1966qh} is given by \cite{Frolov:2003en}
\begin{align}
\label{eq:linelementx}
\begin{aligned}
\mathrm ds^2 = &-\mathrm dt^2 + \rho^2 \left(\frac{\mathrm dx^2}{4 \Delta} + \mathrm d\theta^2 \right)\\ &+ \alpha \sin^2\theta \,\mathrm d\phi^2+ \beta \cos^2\theta \, \mathrm d\psi^2\\ &+ \frac{\mu}{\rho^2} \left(\mathrm dt + a \sin^2 \theta \, \mathrm d\phi + b \cos^2 \theta \, \mathrm d\psi \right)^2,
\end{aligned}
\end{align}
with
\begin{align}
\label{eq:3}
\begin{aligned}
\alpha &:= x + a^2,\\
\beta &:= x + b^2,\\
\Delta &:= \alpha \beta - \mu x,\\
\rho^2 &:= x + a^2 \cos^2\theta + b^2 \sin^2\theta.
\end{aligned}
\end{align}
The radial coordinate $x=r^2$ was introduced by Myers and Perry 
in order to cover the whole spacetime \cite{Myers:1986un, Gibbons:2009um}.
Note that the radial coordinate $x$ is then defined for values 
$x \in [-a^2, \infty)$. 
Although the transformation becomes complex for $x<0$, 
the line element \eqref{eq:linelementx} remains real \cite{Anabalon:2010ns}.
The angular coordinates cover the ranges $0 \le \theta \le \pi/2$,
$0 \le \phi \le 2 \pi$ and $0 \le \psi \le 2 \pi$.\\
The Kretschmann scalar diverges at $\rho^2 = 0$ \cite{Gibbons:2009um}. 
Therefore, this condition defines the curvature singularity $(\theta_{\rm s}, x_{\rm s})$. 
But this function does not vanish at a single point. 
In fact, for each $x \in [-a^2, -b^2]$ 
there is a $\theta \in [0, \frac{\pi}{2}]$ 
so that $\rho^2$ vanishes (see Fig.~\ref{fig:thetasing}).
Thus $\rho^2 = 0$ does not describe a ring-shaped singularity 
as in the Kerr spacetime but a closed surface which is not traversable. Note that this is true even in the case of a vanishing rotation parameter $(a\neq0, b=0)$.
In the special case of equal rotation parameters,
the singularity becomes point-like. Of course, the singularity is point-like for the Schwarzschild-case of vanishing rotation parameters, as well.

\begin{figure}[htbp]
\centering
   \includegraphics[width=0.47\textwidth]{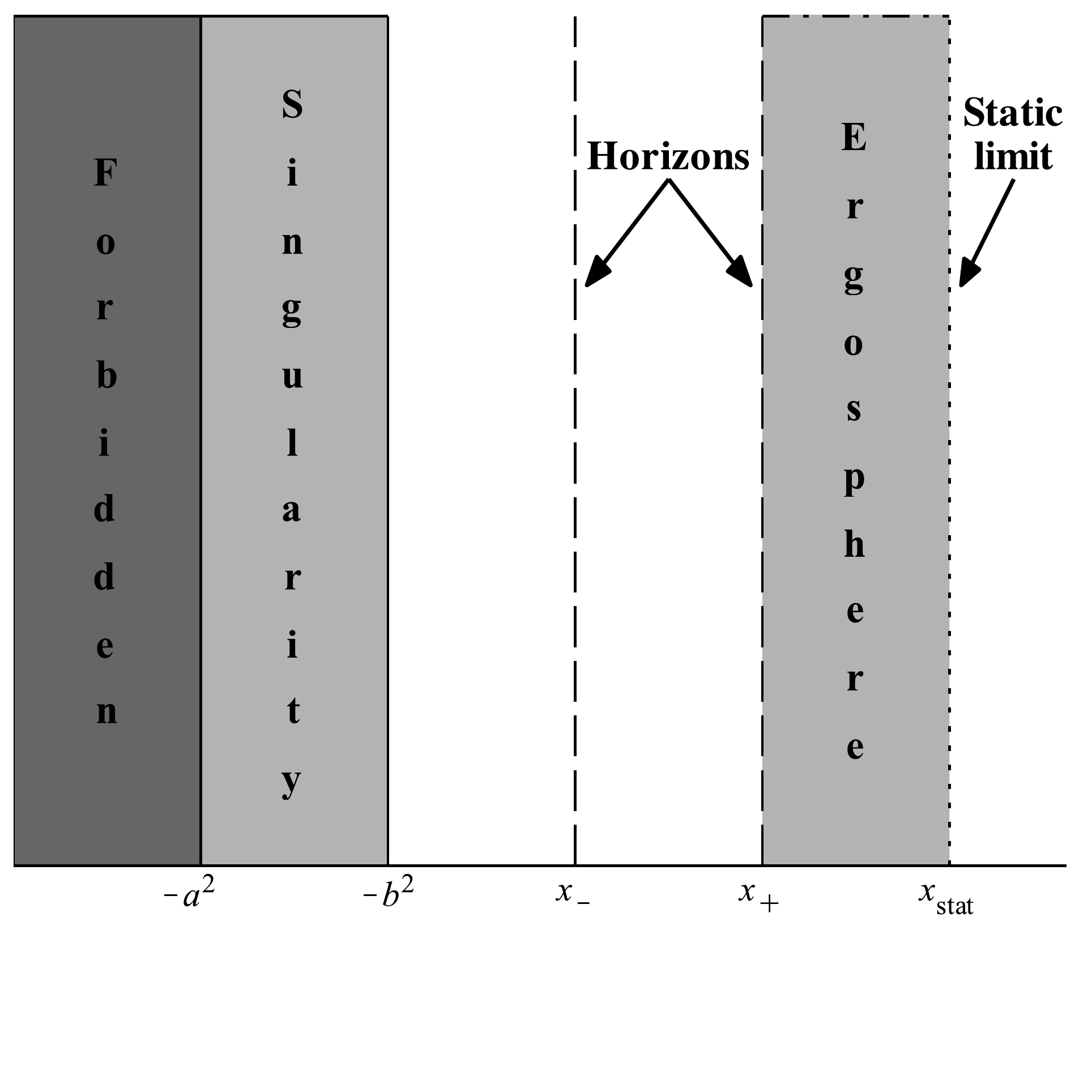}\vspace{-1.2cm}
   \caption{Singularity, horizons and static limit of the five-dimensional Myers-Perry spacetime.}
\label{fig:spacetime}
\end{figure}

The horizons $x_\pm$ are defined by the zeros of $\Delta$ \cite{Myers:1986un}
\begin{align}
x_{\pm} = \frac{1}{2} \left(\mu - a^2 - b^2 \pm \sqrt{\left(\mu - a^2 - b^2 \right)^2 - 4 a^2 b^2} \right).
\end{align}
If $\mu > \left(a + b \right)^2$, 
then both roots are real and positive, 
yielding two regular horizons $x_\pm$, 
where $x_-$ is the Cauchy horizon and $x_+$ is the event horizon. 
In the special case $\mu = \left(a + b \right)^2$, 
a real double root occurs and thus both horizons merge,
leading to an extremal five-dimensional Myers-Perry black hole. 
For $\left(a - b \right)^2 < \mu < \left(a + b \right)^2$, 
there are no real roots any longer 
and therefore a naked singularity appears, 
which is considered unphysical by the cosmic censorship hypothesis. 
If $0 < \mu < \left(a + b \right)^2$, then both roots are real and negative. 
For $\mu = \left(a-b \right)^2$, there is a negative double root with $x_\pm\in(-a^2,-b^2]$.\\
For $\mu < 0$, i.e.~for a negative black hole mass, 
both zeros are real and negative, but one is always larger than $-b^2$. 
This seems to indicate that, even for negative mass, 
there is a regular horizon. 
As pointed out in \cite{Gibbons:2009um}, however, this is not true.  
In fact, there is a causality violating region outside this surface, 
analogous to the case of the repulson discussed
by Gibbons and Herdeiro \cite{Gibbons:1999uv}.\\
The static limit $x_{\rm stat}$ is defined by $g_{tt} = 0$, 
which leads to a single solution
\begin{align}
x_{\rm stat} = \mu - a^2 \cos^2 \theta - b^2 \sin^2 \theta.
\end{align}
This is in contrast to the Kerr spacetime, where two solutions 
of the equation $g_{tt}=0$ are found.\\
In the Kerr spacetime, the ergosphere has an oblate spheroidal shape 
that touches the event horizon at the poles. 
In the five-dimensional Myers-Perry spacetime, 
this is only the case if 
one of the angular momenta vanishes.
In general ($a\neq b\neq 0$), 
the event horizon is located entirely inside the ergosphere. 
In the special case $|a|=|b|$, the ergosphere is independent of $\theta$.\\
The radial positions of the singularity, both horizons 
and the static limit in the five-dimensional Myers-Perry spacetime 
are represented in Fig.~\ref{fig:spacetime}.

\subsection{Equations of motion}

Being stationary and axisymmetric, 
the general five-dimensional Myers-Perry metric 
admits three Killing vector fields 
$\partial_t$, $\partial_\phi$ and $\partial_\psi$. 
The related conserved quantities for the geodesic motion 
are the conjugate momenta \cite{Frolov:2003en}
\begin{align} 
\begin{aligned} 
p_t  &= g_{tt} \dot t + g_{t\phi} \dot \phi + g_{t\psi} \dot \psi =: -E,\\
p_\phi &= g_{t\phi} \dot t + g_{\phi \phi} \dot \phi + g_{\phi \psi} \dot \psi =: \Phi,\\
p_\psi &=  g_{t \psi} \dot t + g_{\phi \psi} \dot \phi + g_{\psi \psi} \dot \psi=: \Psi,
\end{aligned}
\end{align}
where the dot denotes the derivative with respect to 
an affine parameter $\lambda$. Note, that in the case of a massive test particle, we assumed a normalized rest mass. Furthermore, we obtain
\begin{align} 
\begin{aligned}
p_x  &= g_{xx} \dot x,\\
p_\theta &= g_{\theta \theta} \dot \theta.
\end{aligned}
\end{align}
The Hamiltonian is then given by \cite{Misner:1974qy}
\begin{align} 
\mathcal{H} = \frac{1}{2} g^{\mu \nu} p_\mu p_\nu.
\end{align}
As addressed in \cite{Krtous:2006qy, Frolov:2003en, Frolov:2007cb, Frolov:2007nt}, 
there is also a Killing tensor $K^{\mu \nu}$, 
which is related to a hidden symmetry. 
The associated constant of motion $K$ 
can be derived by separating the Hamilton-Jacobi equation
\begin{align}
\label{eq:hamjaceq}
\mathcal H = \frac{1}{2} g^{\mu \nu} \frac{\partial S}{\partial x^\mu} \frac{\partial S}{\partial x^\nu} = - \frac{\partial S}{\partial \lambda},
\end{align}
as first discovered by Carter for the Kerr spacetime 
\cite{Carter:1968rr}. The action $S$ can be separated using the ansatz
\begin{align}
S = \frac{1}{2} \delta \lambda - E t + S_x (x) + S_\theta (\theta) + \Phi \phi + \Psi \psi.
\end{align}
Here we introduced a parameter $\delta$, 
which describes the norm of the four-velocity 
$g_{\mu \nu} \dot x^\mu \dot x^\nu = -\delta$, 
in order to investigate massive $(\delta=1)$ 
and massless $(\delta=0)$ test particle motion simultaneously.\\
It is now straightforward to obtain the geodesic equations
\begin{align}
\label{eq:geo1}
\dot x^2 &= \left(\frac{\mathrm dx}{\mathrm d\tau}\right)^2 &&\!\!\!\!\!\! = 16 \Delta^2 \mathcal{X},\\
\label{eq:geo2}
\dot \theta^2 &= \left(\frac{\mathrm d\theta}{\mathrm d\tau}\right)^2 &&\!\!\!\!\!\! = \Theta,\\
\label{eq:geo3}
\dot \phi &= \; \; \, \frac{\mathrm d\phi}{\mathrm d\tau} &&\!\!\!\!\!\! = \frac{\Phi}{\sin^2\theta} - \frac{a \beta \mu }{\Delta} \mathcal E - \frac{a^2-b^2}{\alpha} \Phi,\\
\label{eq:geo4}
\dot \psi &=  \; \; \,  \frac{\mathrm d\psi}{\mathrm d\tau}  &&\!\!\!\!\!\! = \frac{\Psi}{\cos^2\theta} - \frac{\alpha b \mu}{\Delta} \mathcal E - \frac{b^2-a^2}{\beta} \Psi,\\
\label{eq:geo5}
\dot t &= \; \; \,  \frac{\mathrm dt}{\mathrm d\tau} &&\!\!\!\!\!\! = E \rho^2 + \frac{ \alpha \beta \mu}{\Delta} \mathcal E,
\end{align}
where the Mino time $\tau$ \cite{Mino:2003yg} was introduced by
\begin{align}
\rho^2 \mathrm d\tau = \mathrm d\lambda,
\end{align}
in order to separate the $\theta$- and $x$-equations, 
together with the abbreviations
\begin{align}
\begin{aligned}
\label{eq:abbrev}
\mathcal E := &\,E + \frac{a\Phi}{\alpha} + \frac{b \Psi}{\beta},\\
\mathcal Q := &\,\left(a^2 -b^2 \right)\left(\frac{\Phi^2}{\alpha} - \frac{\Psi^2}{\beta} \right),\\
\mathcal X := &\,\frac{1}{4 \Delta} \left[\left(E^2- \delta \right) x - K  + \mathcal Q  + \frac{\alpha \beta \mu}{ \Delta} \mathcal E^2 \right],\\
\Theta := &\,\left(E^2 - \delta \right) \left(a^2 \cos^2\theta + b^2 \sin^2\theta \right) - \frac{\Phi^2}{\sin^2\theta} \\ &- \frac{\Psi^2}{\cos^2\theta}+ K.
\end{aligned}
\end{align}

\section{Complete classification}

Before solving the geodesic equations, we will study several properties of the test particle motion by investigating the right-hand sides of Eq.~\eqref{eq:geo1} - \eqref{eq:geo5}.

\subsection{$\theta$-motion}

\begin{figure*}[htbp]
\centering
\begin{minipage}[htbp]{0.5\textwidth}
   \includegraphics[width=0.73\textwidth]{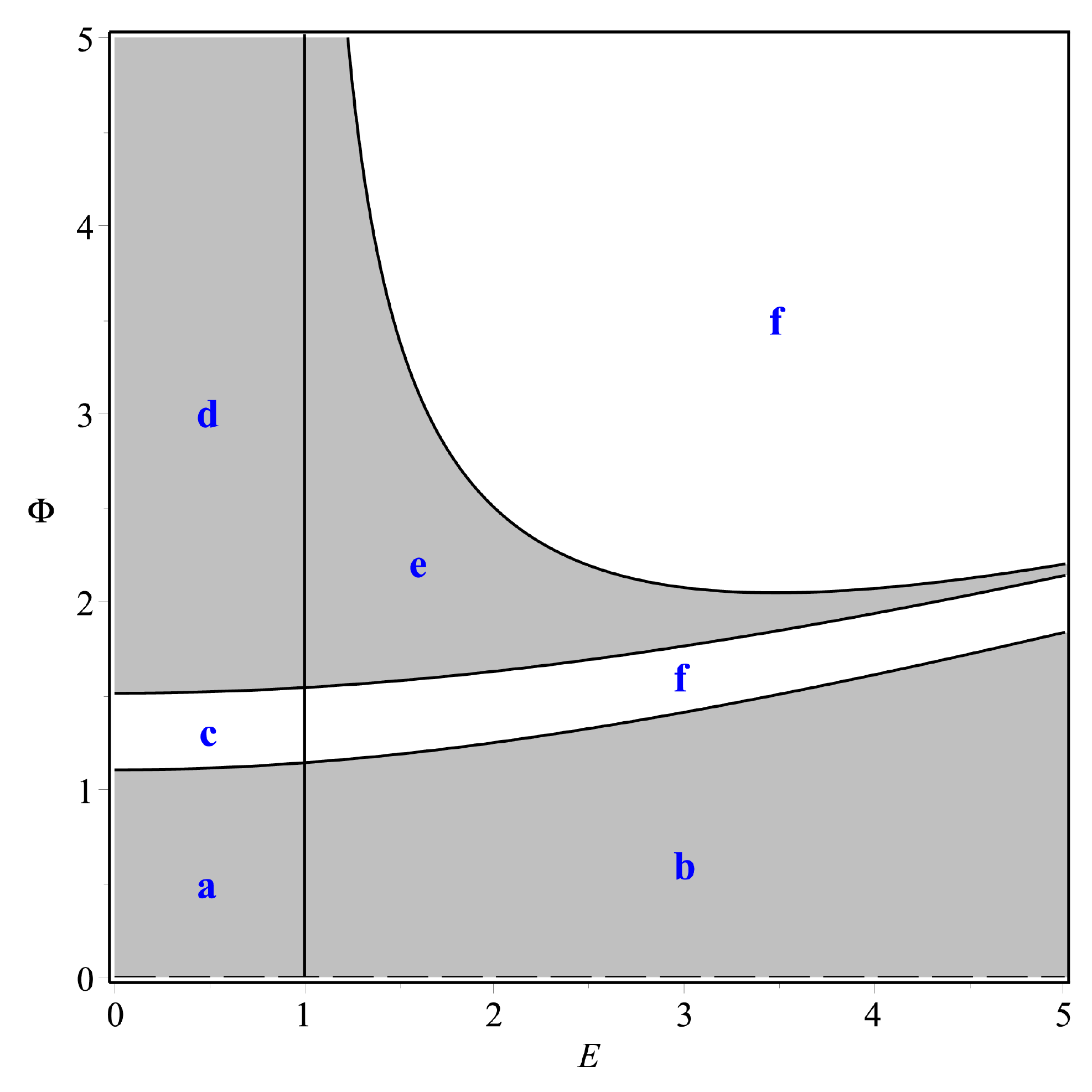}
\subcaption{$K = 1.8, \Psi = -0.2,  a = 0.4, b=0.3, \mu=1, \delta = 1$.}
   \end{minipage}\hfill
   \begin{minipage}[htbp]{0.5\textwidth}
   \includegraphics[width=0.73\textwidth]{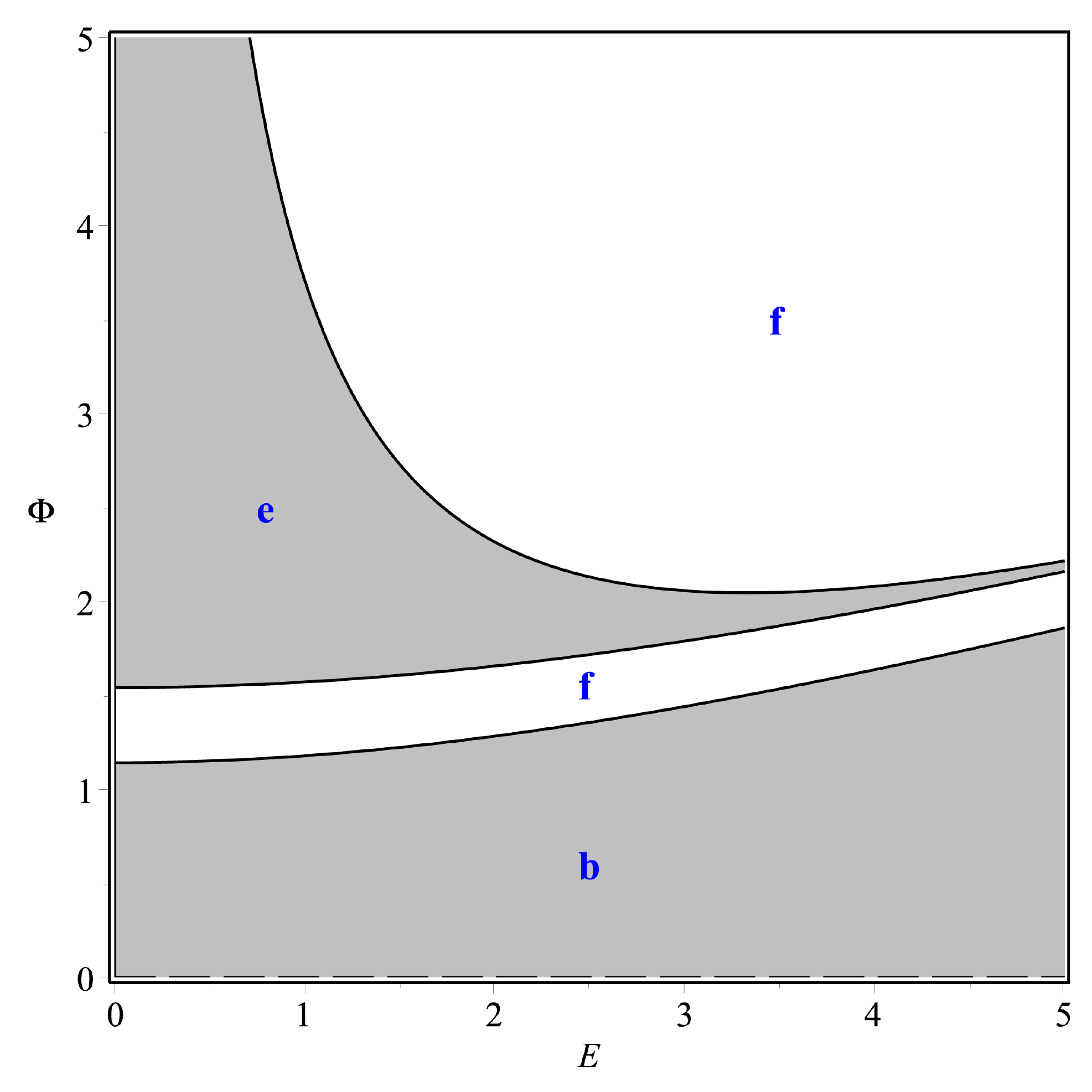}
\subcaption{$K = 1.8, \Psi = -0.2,  a = 0.4, b=0.3, \mu=1, \delta = 0$.}
   \end{minipage}\\[10pt]
   
   \begin{minipage}[htbp]{0.5\textwidth}
   \includegraphics[width=0.73\textwidth]{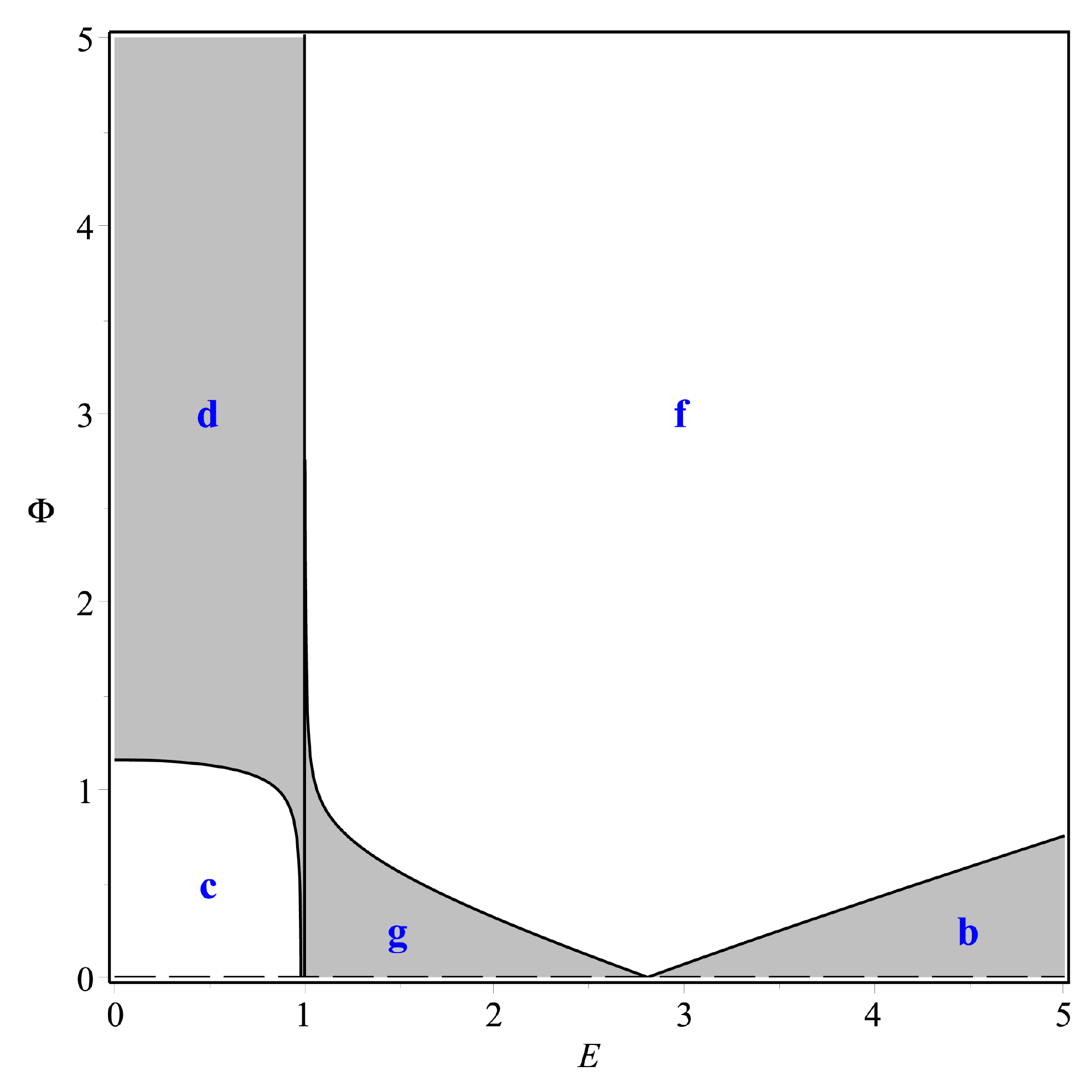} 
\subcaption{$K = -0.1, \Psi = 1,  a = 0.4, b=0.3, \mu=1, \delta = 1$.}
\label{fig:2c}
\end{minipage}\hfill
\begin{minipage}[htbp]{0.5\textwidth}
   \includegraphics[width=0.73\textwidth]{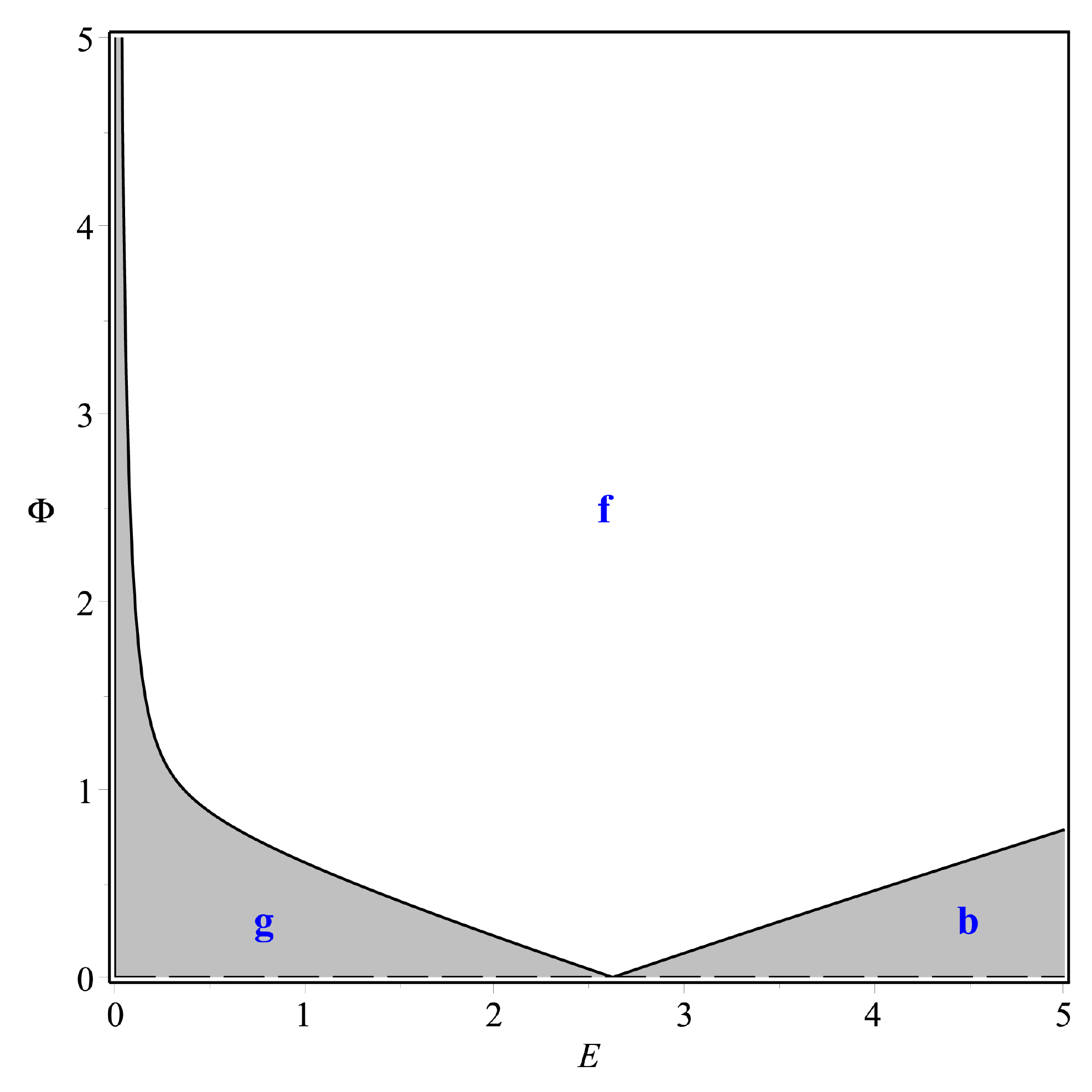}
\subcaption{$K = -0.1, \Psi = 1,  a = 0.4, b=0.3, \mu=1, \delta = 0$.}
\label{fig:2d}
   \end{minipage}
\caption{Definition of regions with different numbers of zeros for massive (a, c) and massless (b, d)  test particle motion using $E$-$\Phi$-plots. The special case $\xi_0 = 1$ $(\Phi = 0)$ is represented by a dashed line. Grey regions denote three real zeros, white regions one real and two complex conjugate zeros. The discriminant of $\Xi$ vanishes for values at the boundary lines so that only a constant $\theta$-motion is possible.}
\label{fig:2}
\end{figure*}

The $\theta$-motion is described by Eq.~\eqref{eq:geo2}. The subspace $\theta = 0$ can only be reached if $\Phi = 0$ and the subspace $\theta = \frac{\pi}{2}$ can only be reached if $\Psi = 0$ \cite{Frolov:2003en}. For $\Phi = 0$ and $\theta = 0$, Eq.~\eqref{eq:geo2} can be rewritten as
\begin{align}
K = \Psi^2 - \left(E^2 - \delta \right) a^2
\label{eq:Cartera}
\end{align}
and for $\Psi = 0$ and $\theta = \frac{\pi}{2}$ Eq.~\eqref{eq:geo2} yields
\begin{align}
K = \Phi^2 - \left(E^2 - \delta \right) b^2.
\label{eq:Carterb}
\end{align}
Other constant $\theta$-motions with $\theta_0 \in \left(0, \frac{\pi}{2}\right)$ are given by
\begin{align}
\Theta(\theta_0) = 0 \qquad \text{and} \qquad \left. \frac{\mathrm d \Theta}{\mathrm d\theta} \right|_{\theta_0} = 0.
\end{align}
For further investigation of the $\theta$-motion, we perform the substitution $\xi = \cos^2 \theta$, which is bijective for $\theta \in [0, \frac{\pi}{2}]$. Thus \eqref{eq:geo2} is related to a polynomial $\Xi$ of order three
\begin{align}
\dot \xi^2 = a_3 \xi^3 + a_2 \xi^2 + a_1 \xi + a_0 =: \Xi
\label{eq:thetasubs}
\end{align}
with the coefficients
\begin{align}
\begin{aligned}
a_3 &= -4 \left(E^2 - \delta \right) \left(a^2 - b^2 \right),\\
a_2 &=  4 \left(E^2 - \delta \right) \left(a^2 - 2 b^2 \right) - 4K,\\
a_1 &=  4 \left(E^2 - \delta \right) b^2 - 4 \Phi^2 + 4 \Psi^2 + 4K,\\
a_0 &= -4 \Psi^2.
\end{aligned}
\label{eq:thetacoeffs}
\end{align}
For $E^2 = \delta$ or $a = b$, this will reduce to a second order polynomial. The case $a = b$ has already been investigated in \cite{Kagramanova:2012hw}.\\
The zeros of $\Xi$ are the turning points of the $\theta$-motion, which need to be real to be physically relevant. In order to obtain real values for $\xi$ from $\Xi$, we have to require $\Xi \ge 0$. The regions for which $\Xi \ge 0$ are bounded by the zeros of $\Xi$. The number of zeros depends both on the parameters of the black hole $(a, b)$ and on the parameters of the test particle $(E, K, \delta, \Phi, \Psi)$. The conditions $\Xi = 0$ and $\frac{\mathrm d\Xi}{\mathrm d\xi} = 0$ define the double zeros of $\Xi$ and thus the boundaries between regions where $\Xi$ has a different number of real zeros. Since $\Xi$ is a polynomial of order three with real coefficients, there are either three real or one real and two complex conjugate zeros.\\
For various parameters $(K, \delta, \Psi, a, b)$, we can plot the remaining parameters $E$ and $\Phi$ in a parametric diagram. Since $\Xi$ is symmetric in $E$ and $\Phi$ it is sufficient to restrict the plots to the first quadrant. Additionally, only the absolute value of the rotation parameters $a$ and $b$ is relevant. Because of the substitution $\xi = \cos^2 \theta$, we need to confine the valid zeros by $\xi_0 \in [0, 1]$. For $\Phi = 0$ one zero is always $\xi_0 = 1$ and, likewise, for $\Psi = 0$ one zero is always $\xi_0 = 0$. The interesting regions of the parametric diagram are shown in Fig.~\ref{fig:2}.\\
There are seven different regions, representing different variations of possible zeros. The grey regions belong to parameters where $\Xi$ has three real zeros and the white regions represent parameters where $\Xi$ has one real and two complex conjugate zeros.\\
Unlike in the Kerr spacetime, where $K$ must be non-negative, we can also have a $\theta$-motion for a negative Carter constant (see Fig.~\ref{fig:2c} \& \ref{fig:2d}).\\
For massless test particle motion, the regions (a), (c) and (d) vanish, but the general structure of the $E$-$\Phi$-plots is only slightly different.\\
All possible combinations are listed in Tab.~\ref{tab:thetazeros}. Eventually, the physically valid number of polar turning points is given, concerning the restriction $\xi \in [0, 1]$:

\begin{table}[H]
   \centering
   \begin{tabular}{|c|c|c|}
\hline
Region & Number of zeros & Polar turning points\\
\hline
\hline
      (a)      & $3 \in \mathbb{R}^+$, \ $0 \in \mathbb{R}^-$  & 2 \\
 \hline
       (b)     &  $2 \in \mathbb{R}^+$, \ $1 \in \mathbb{R}^-$   &  2 \\
        \hline
      (c)       &  $1 \in \mathbb{R}^+$, \ $0 \in \mathbb{R}^-$ &  0\\
 \hline
      (d)      & $1 \in \mathbb{R}^+$, \ $2 \in \mathbb{R}^-$ &  0\\
 \hline
       (e)     &  $0 \in \mathbb{R}^+$, \ $3 \in \mathbb{R}^-$ &  0\\
 \hline
      (f)      & $0 \in \mathbb{R}^+$, \ $1 \in \mathbb{R}^-$ & 0\\
 \hline
       (g)     & $2 \in \mathbb{R}^+$, \ $1 \in \mathbb{R}^-$   & 0\\
 \hline
      \end{tabular}
   \caption{Zeros of $\Theta$ for different regions of the $E$-$\Phi$-plots.}
   \label{tab:thetazeros}
\end{table}

Tab.~\ref{tab:thetazeros} points out that only region (a) and (b) will yield a physical $\theta$-motion.

\subsection{$x$-motion}

The $x$-motion is described by the equation
\begin{align}
\dot x^2 = 16 \Delta^2 \mathcal{X},
\end{align}
which can be written as a polynomial of the form
\begin{align}
\dot x^2 = b_3 x^3 + b_2 x^2 + b_1 x + b_0 =: X
\label{eq:xsubs}
\end{align}
with coefficients
\begin{align}
\begin{aligned}
b_3 =&\,4 \left(E^2 - \delta \right),\\
b_2 =& \,4 \left(E^2 - \delta \right) \left(a^2 + b^2 \right) - 4K + 4 \delta \mu,\\
b_1 =& \,4 \left(E^2 - \delta \right) a^2 b^2 + 4 E^2 \mu \left(a^2 + b^2 \right) + 8\mu E \left(a \Phi + b \Psi \right)\\ &+ 4K \left(\mu - a^2 - b^2 \right)+ 4\left(\Phi^2 - \Psi^2 \right) \left(a^2 - b^2 \right),\\
b_0 =&\,4\mu \left(abE + a \Psi + b \Phi \right)^2 - 4 a^2 b^2 \left(K - \Phi^2 - \Psi^2 \right)\\ &- 4 a^4\Psi^2 -4b^4 \Phi^2.
\end{aligned}
\label{eq:xcoeffs}
\end{align}
The real zeros of $X$ will determine the radial turning points of the physical motion and therefore correspond to different orbit types. Basically, there are five different types of orbits in the five-dimensional Myers-Perry spacetime :\\

{\bf Escape orbit (EO):} $x$ starts from infinity and approaches a periapsis (closest radial turning point) and goes back to infinity without crossing any horizon.\\

{\bf Two-world escape orbit (TEO):} A special case of an escape orbit, where the radial turning point lies behind both horizons. Due to general causal restrictions, it cannot repass both horizons to the former universe but to a different, second universe.\\

{\bf Bound orbit (BO):} $x$ oscillates between two radial turning points $x_1, x_2$, where $x_1, x_2 \le x_-$.\\

 {\bf Many-world bound orbit (MBO):} A special case of a bound orbit, where $x_1 \le x_-$ and $x_2 \ge x_+$. For the same reasons mentioned considering the two-world escape orbit, each time both horizons are passed through, the former universe can't be reentered. So after every oscillation, the test particle enters a different universe.\\
 
{\bf Terminating orbit (TO):} $x$ ends in the curvature singularity at $x = x_{\rm s}$.\\

\begin{figure*}[htbp]
\begin{minipage}[htbp]{0.5\textwidth}
   \includegraphics[width=0.73\textwidth]{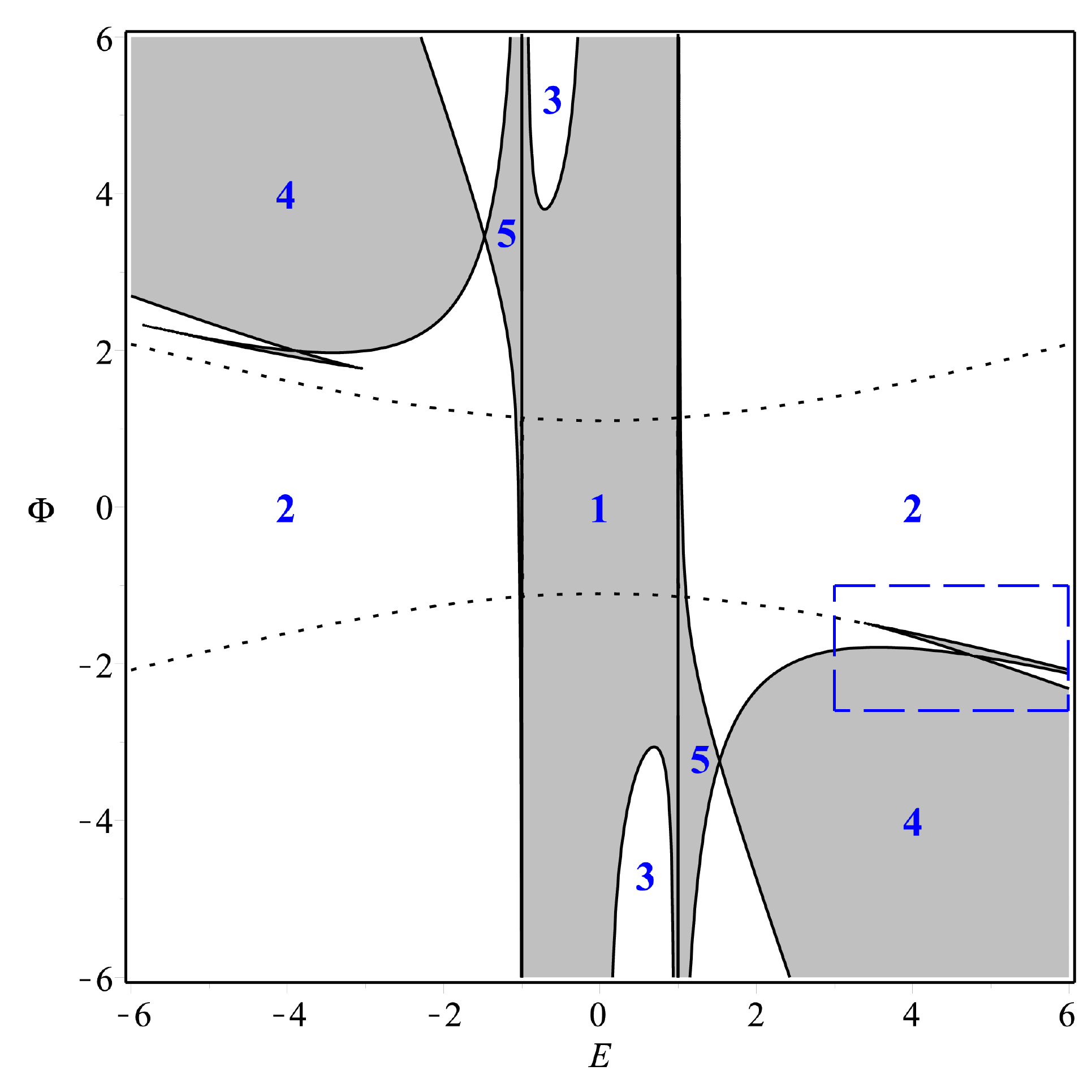}
 \subcaption{$K=1.8, \Psi=-0.2, a=0.4, b=0.3, \mu=1, \delta=1$.}
    \label{fig:xzerosa}
   \end{minipage}\hfill
   \begin{minipage}[htbp]{0.5\textwidth}\vspace{-0.2cm}
\begin{tikzpicture}
\node[inner sep=0pt] (russell) at (0,0)
    {\includegraphics[width=0.7\textwidth]{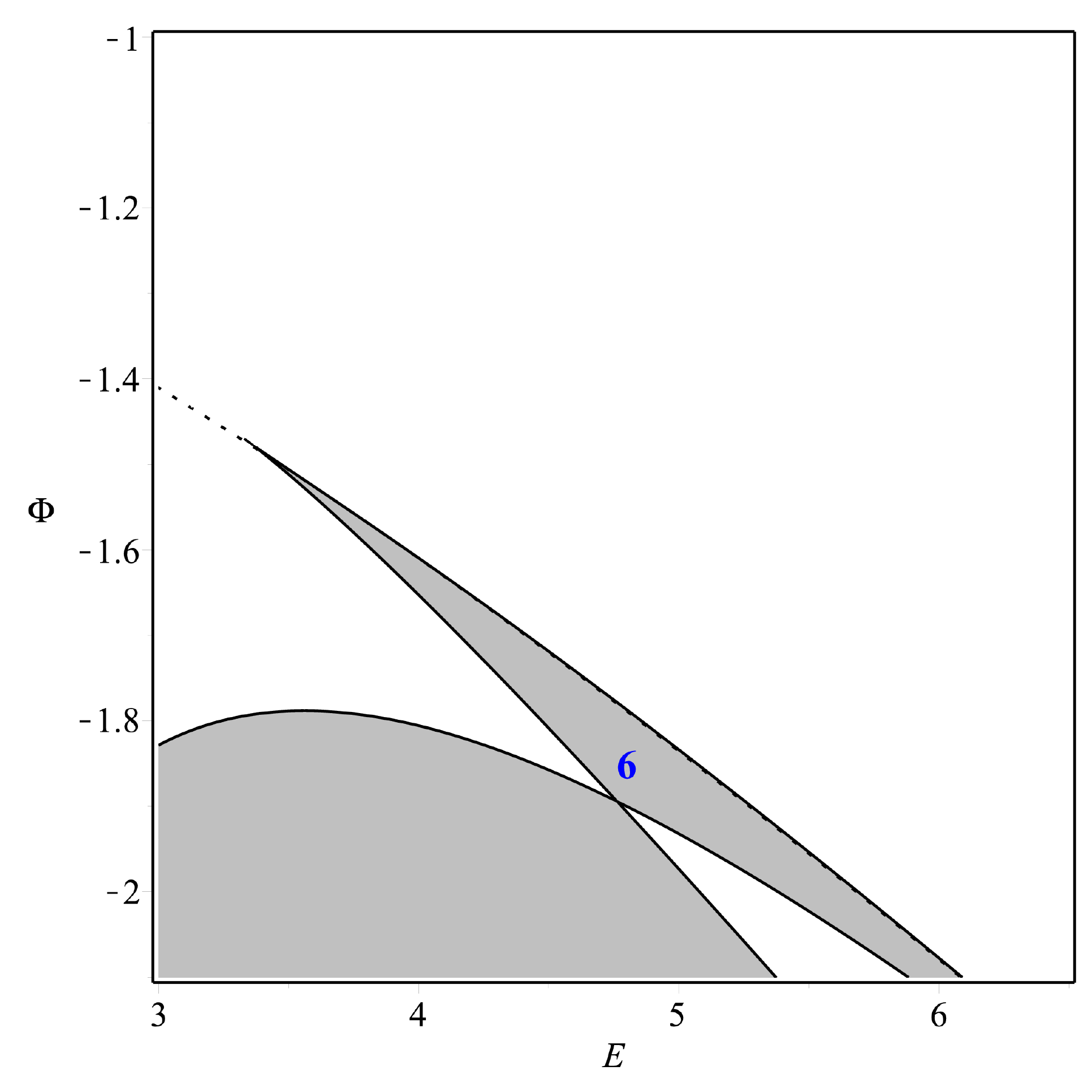}};
\draw[->, thick] (-1.3,0.6) -- (-0.21,1.5);
\node[inner sep=0pt] (whitehead) at (1.8,1.57)
    {\includegraphics[width=.45\textwidth]{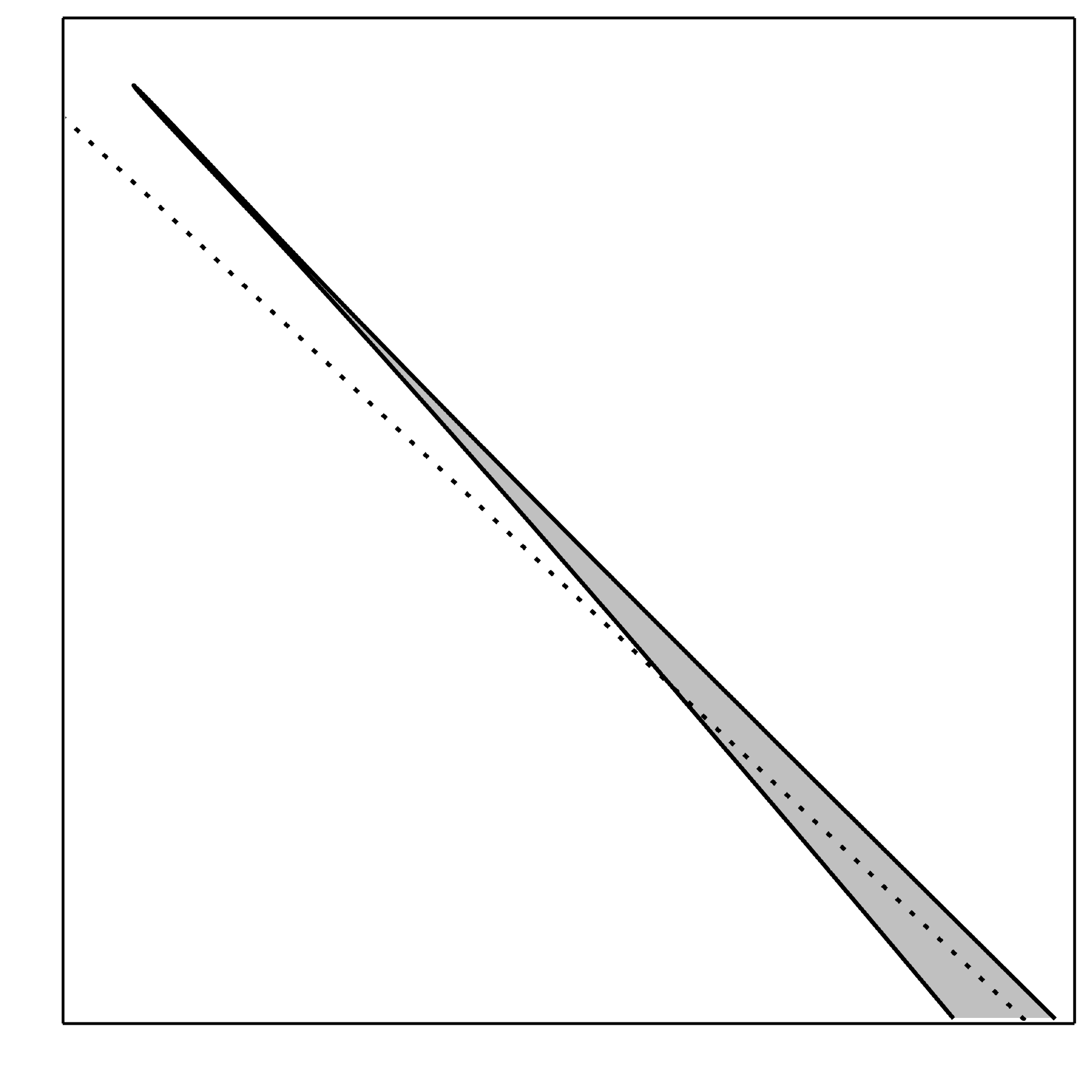}};
\end{tikzpicture}
 \subcaption{Close-up of blue dashed box in (a).}
     \label{fig:xzerosb}
\end{minipage}\\[20pt]

\begin{minipage}[htbp]{0.5\textwidth}
\includegraphics[width=0.73\textwidth]{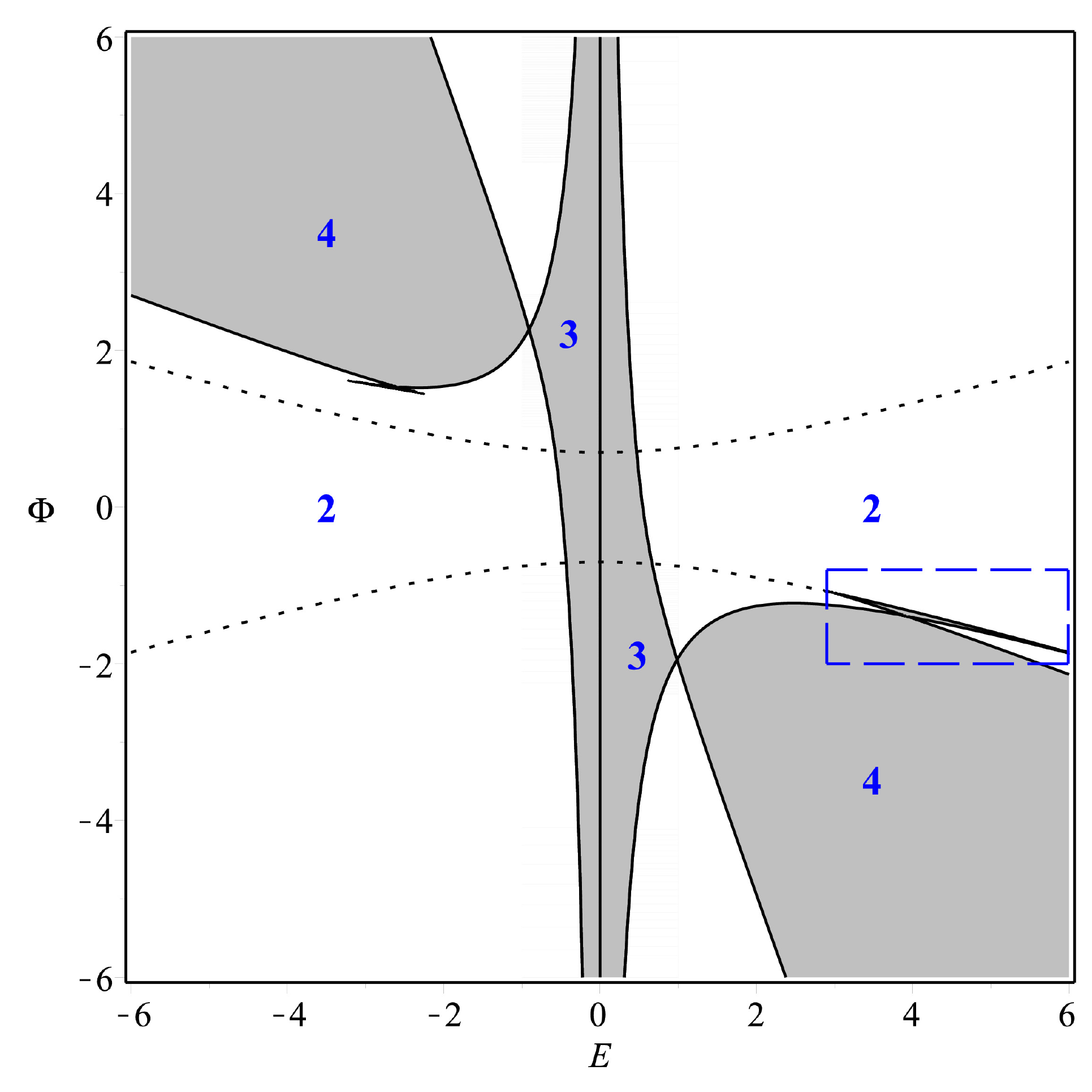}
 \subcaption{$K=1, \Psi=-0.3,a=0.4, b=0.3, \mu=1, \delta=0$.}
     \label{fig:xzerosc}
   \end{minipage}\hspace{-0.12cm}
   \begin{minipage}[htbp]{0.5\textwidth}\vspace{-0.2cm}
\begin{tikzpicture}
\node[inner sep=0pt] (russell) at (0,0)
    {\includegraphics[width=0.7\textwidth]{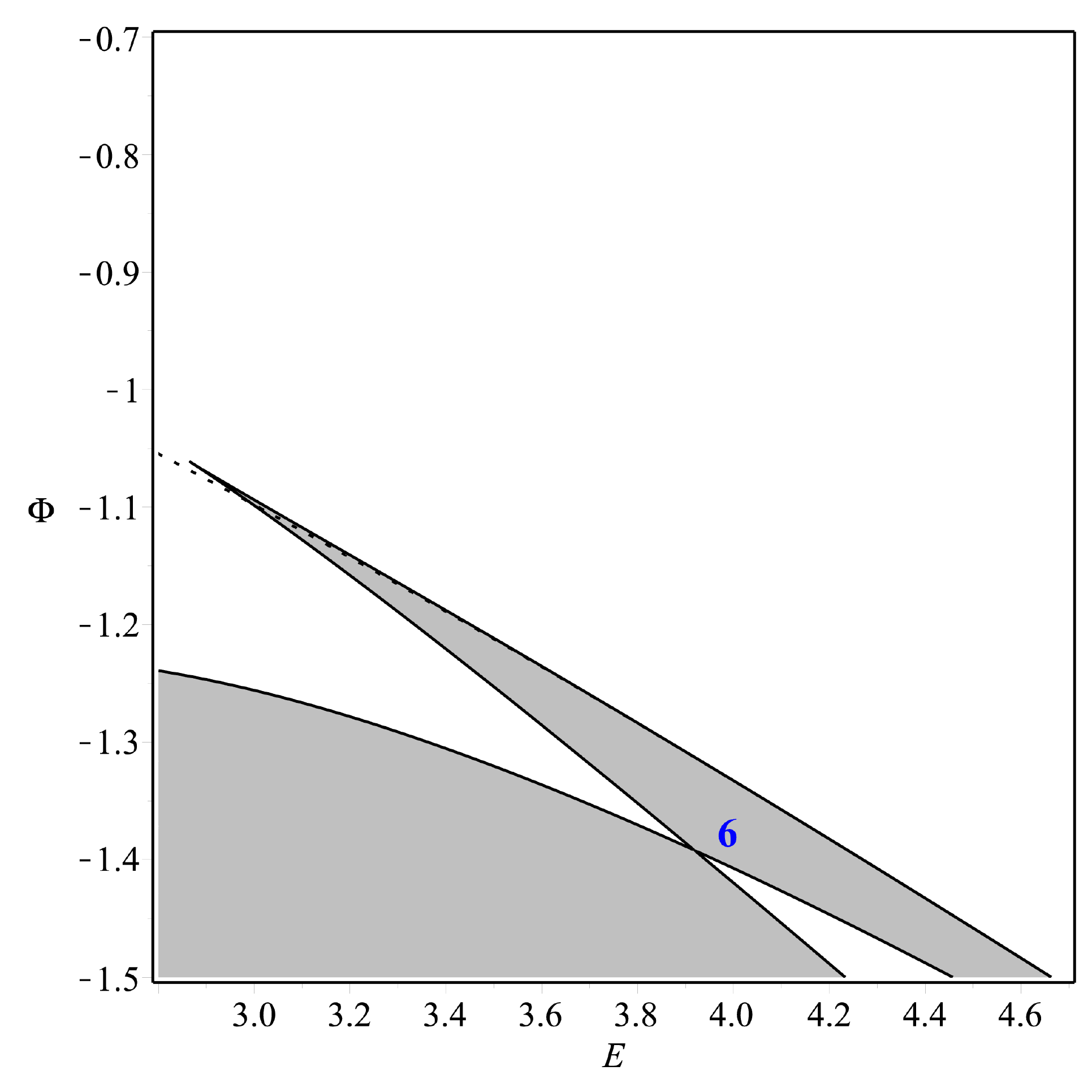}};
\draw[->, thick] (-1.65,0.5) -- (-0.21,1.5);
\node[inner sep=0pt] (whitehead) at (1.8,1.57)
    {\includegraphics[width=.45\textwidth]{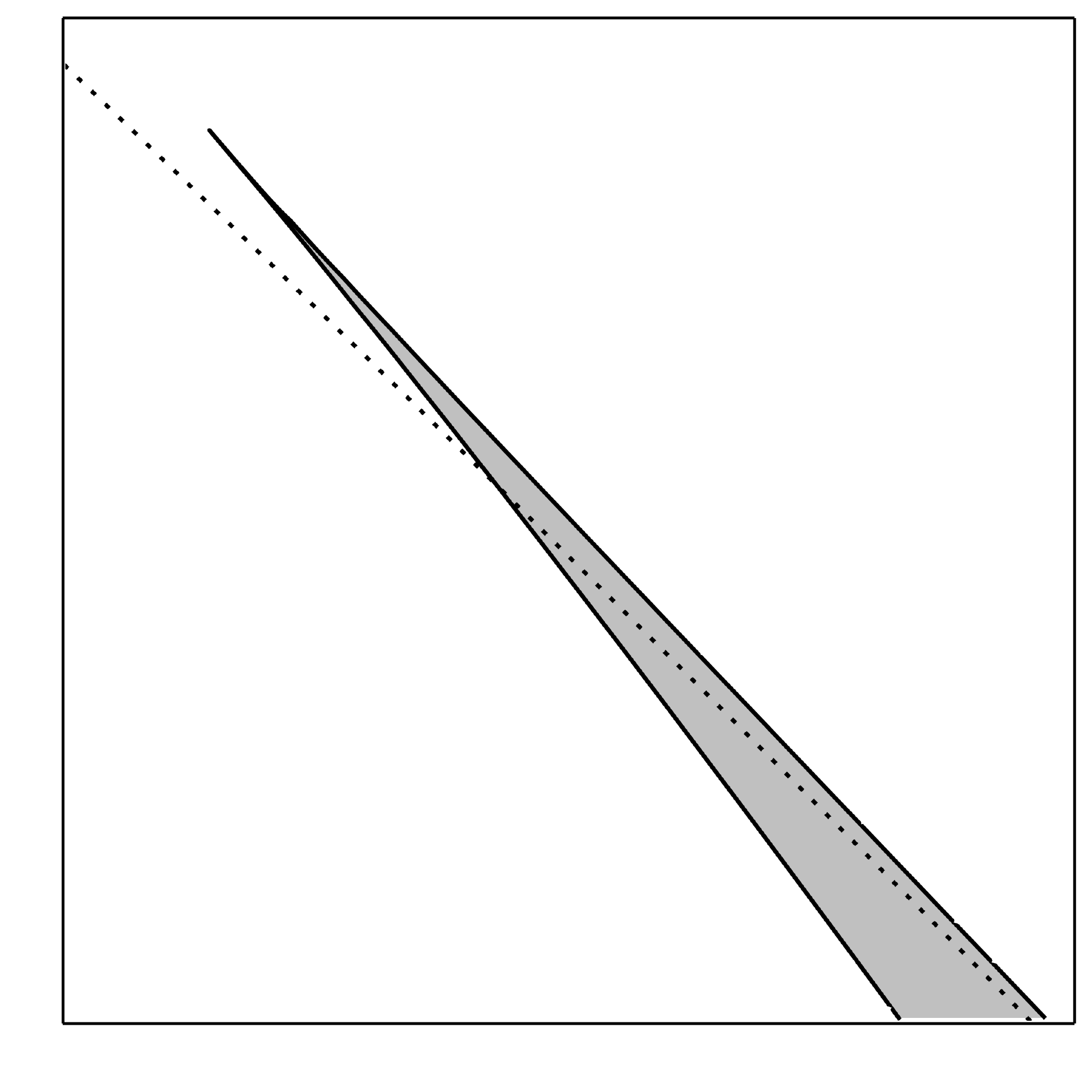}};
\end{tikzpicture}
 \subcaption{Close-up of blue dashed box in(c).}
     \label{fig:xzerosd}
   \end{minipage}
   \caption{$E$-$\Phi$-plots for massive (a, b) and massless (c, d) test particle motion. Grey regions denote three real zeros, white regions one real and two complex conjugate zeros. The dotted lines represent the boundaries of region (a) and (b) of the $\theta$-related $E$-$\Phi$-plots, so that only the parameter values between these dotted lines are physically valid.}
   \label{fig:xzeros}
\end{figure*}

We will visualize these types of orbits in the next section by introducing an effective potential. However, we first want to investigate possible combinations of parameters for the radial motion in the same way as we did for the $\theta$-equation because, eventually, a set of parameters is only viable if it holds true for both the $\theta$- and the $x$-equation.\\
Since $X$ is also a polynomial of order three, we can again differentiate between regions where $X$ has three real zeros or one real and two complex conjugate zeros (see Fig.~\ref{fig:xzeros}). Additionally, the restrictions of the $\theta$-equation are visualized by the dotted line, representing the boundary of region (a) and (b). There are further restrictions on the zeros of $X$ as well. At least, a physical turning point is only relevant if $x_0 \in (x_{\rm s}, \infty)$. Tab.~\ref{tab:xzeros} provides the physical relevant zeros of all regions:

\begin{table}[H]
   \centering
   \begin{tabular}{|c|c|c|}
\hline
Region & Number of zeros & Radial turning points\\
\hline
\hline
    (1)      & $3 \in \mathbb{R}$, \ $2 \in (-a^2, \infty)$ & 2\\
 \hline
       (2)     &  $1 \in \mathbb{R}$, \ $1 \in (-a^2, \infty)$ &  1\\
        \hline
      (3)       &  $1 \in \mathbb{R}$, \ $1 \in (-a^2, \infty)$  & 1\\
 \hline
      (4)      & $3 \in \mathbb{R}$, \ $3 \in (-a^2, \infty)$& 3\\
 \hline
       (5)      & $3 \in \mathbb{R}$, \ $3 \in (-a^2, \infty)$& 3\\
 \hline
       (6)      & $3 \in \mathbb{R}$, \ $3 \in (-a^2, \infty)$& 3\\
 \hline
      \end{tabular}
   \caption{Zeros of $X$ for different regions of the $E$-$\Phi$-plots. The indicated number of zeros refers to the maximum number of zeros in the related region.}
   \label{tab:xzeros}
\end{table}

Region (1) and (2) obviously overlap with regions (a) and (b) of allowed $\theta$-motion and we will see that these regions are related to many-world bound and two-world escape orbits. Furthermore, there might be terminating orbits for parameter values of these regions. Region (3) and (4) will not coincide with regions of allowed $\theta$-motion. Fig.~\ref{fig:xzerosa} indicates that region (5) may extend into regions of possible $\theta$-motion and therefore it will be of physical interest. The same is true for region (6) in Fig.~\ref{fig:xzerosb}. The parameter values of region (5) will belong to escape orbits and many-world bound orbits. Region (6) is related to two-world escape orbits and bound orbits which are hidden behind the Cauchy horizon. Both regions may also contain terminating orbits as well. We obtain similar diagrams for massless test particle motion (see Fig.~\ref{fig:xzerosc} \& \ref{fig:xzerosd}).\\
Note that for a test particle motion restricted to the equatorial $\theta=\frac{\pi}{2}$-plane \eqref{eq:Carterb} one zero of $X$ is always given by $x_0 = -b^2$ and for a motion restricted to the $\theta=0$-plane \eqref{eq:Cartera} one zero is always given by $x_0 = -a^2$, which represents the singularity $x_{\rm s}$ in each case. Therefore, the two remaining roots are either complex conjugate or completely real. In the first case, the only possible orbit type is given by terminating orbits, if $X>0$ for $x>x_{\rm s}$. At least in the massless case this is true due to

\begin{align}
\begin{aligned}
\theta&=\frac{\pi}{2}: \quad &&\left.\frac{\mathrm dX}{\mathrm dx}\right \vert_{x=-b^2} = 4 \mu \left(Ea + \Phi \right)^2,\\
\theta&=0: \quad &&\left.\frac{\mathrm dX}{\mathrm dx}\right \vert_{x=-a^2} = 4 \mu \left(Eb + \Psi \right)^2,
\label{eq:boundlightmotion}
\end{aligned}
\end{align}

which is non-negative (for a positive black hole mass $\mu$), so that $X>0$ for $x>x_{\rm s}$ in both planes. If the derivatives vanish, the corresponding zeros are given by a real double root at $x_0^{1,2}=x_{\rm s}$ and $x_0^3 = 0$. Furthermore, this implies the absence of bound light motion in the equatorial planes, since a region with $X>0$ between two radial turning points is always bounded by the singularity due to the conditions \eqref{eq:boundlightmotion}. As shown in \cite{Frolov:2003en}, there are also no bound orbits in the massive case.

\subsection{$\phi$- and $\psi$-motion}

The azimuthal motion is given by the  $\phi$- and $\psi$-equation
\begin{align}
\dot \phi &= \frac{\Phi}{\sin^2\theta} - \frac{a \beta \mu }{\Delta} \mathcal E - \frac{a^2-b^2}{\alpha} \Phi,\\
\dot \psi &= \frac{\Psi}{\cos^2\theta} - \frac{\alpha b \mu}{\Delta} \mathcal E - \frac{b^2-a^2}{\beta} \Psi.
\end{align}
It is obvious that both equations diverge at the horizons. This is not a physical effect but purely an artifact of the Boyer-Lindquist coordinates. In the Kerr spacetime, one could avoid this by using Kerr-Schild coordinates. However, in the five-dimensional case, the Kerr-Schild coordinates cover a smaller portion of the five-dimensional Myers-Perry spacetime than the Boyer-Lindquist ones (if $b \neq 0$) \cite{Anabalon:2010ns}.\\
In the case of a regular spacetime, $\Delta$ changes its sign when crossing a horizon and therefore also $\dot \phi$ and $\dot \psi$ will change their sign \cite{Chandrasekhar:1983,Oneill:1995} because the other terms are negligible near the horizons.

\section{Effective potential}

Another way of analyzing the possible geodesics of a spacetime is to investigate the effective potential. As usual, we will define the effective potential by rewriting the radial $x$-equation. Furthermore, we need to consider the $\theta$-equation as well, which may provide additional restrictions on the test particle's energy. Finally, we will investigate the possible test particle orbit types in these effective potentials and relate them to the $E$-$\Phi$-regions of the radial motion investigated above.

\subsection{Definition}

We can rewrite the $x$-equation in the following form
\begin{align}
\dot x^2 = \gamma_2 E^2 + \gamma_1 E + \gamma_0,
\end{align}
where
\begin{align}
\begin{aligned}
\gamma_2 &= 4 \left( \Delta x+ \mu \alpha \beta \right),\\[9pt]
\gamma_1 &= 8\mu \left(a \beta \Phi + \alpha b \Psi \right),\\
\gamma_0&= -4 \Delta \left(\delta x + K - \mathcal{Q} \right) + 4\mu \alpha \beta \left(\frac{a \Phi}{\alpha} + \frac{b \Psi}{\beta} \right)^2.
\end{aligned}
\label{eq:gammacoeffs}
\end{align}
The zeros of this quadratic polynomial define an effective potential $V_{\rm eff}$
\begin{align}
\dot x^2 = \gamma_2 \left(E-V_{\rm eff}^+\right) \left(E-V_{\rm eff}^- \right),
\label{eq:effpot}
\end{align}
where
\begin{align}
V_{\rm eff}^\pm := \frac{-\gamma_1 \pm \sqrt{\gamma_1^2 - 4\gamma_2 \gamma_0}}{2 \gamma_2}.
\label{eq:Veff}
\end{align}
This means that, for a given set of parameters, there might be restrictions on $x$ depending on the test particle's energy $E$. The right-hand side of Eq.~\eqref{eq:effpot} needs to be equal to or greater than zero. The condition $\dot x^2 = 0$ defines the radial turning points for some value of $E$.

\subsection{Properties}

First, we will investigate the general properties of the effective potential. Both parts of the effective potential will merge if the radicand
\begin{align}
64 \Delta \left[(\delta x + K - \mathcal{Q}) (\Delta x + \alpha \beta \mu) - x \mu \alpha \beta \left(\frac{a \Phi}{\alpha} + \frac{b \Psi}{\beta}\right)^2 \right]
\label{eq:Veffre}
\end{align}
of Eq.~\eqref{eq:Veff} vanishes, which is obviously true for $\Delta=0$. Thus, the potential parts $V_{\rm eff}^\pm$ merge at the horizons attaining the values
\begin{align}
\begin{aligned}
V_{\rm eff}^\pm (x_+) &= - \frac{a}{x_+ + a^2} \Phi - \frac{b}{x_+ + b^2} \Psi = -\Omega_a \Phi - \Omega_b \Psi,\\
V_{\rm eff}^\pm (x_-) &= - \frac{a}{x_- + a^2} \Phi - \frac{b}{x_- + b^2} \Psi.
\end{aligned}
\end{align}
At infinity, the effective potential parts $V_{\rm eff}^\pm$ take the values
\begin{align}
\lim \limits_{x \rightarrow \infty} V_{\rm eff}^\pm = \pm \sqrt{\delta}
\label{eq:Vefflarge}
\end{align}
so that we have a different behavior for massive and massless test particle motion at infinity.\\
In the case of vanishing rotation parameters ($a=b=0$), one regains the five-dimensional Schwarzschild-Tangherlini effective potential \cite{Hackmann:2008tu}. 

\subsection{Plots}

\begin{figure*}[htbp]
\begin{minipage}[htbp]{0.5\textwidth}
\centering
   \includegraphics[width=0.73\textwidth]{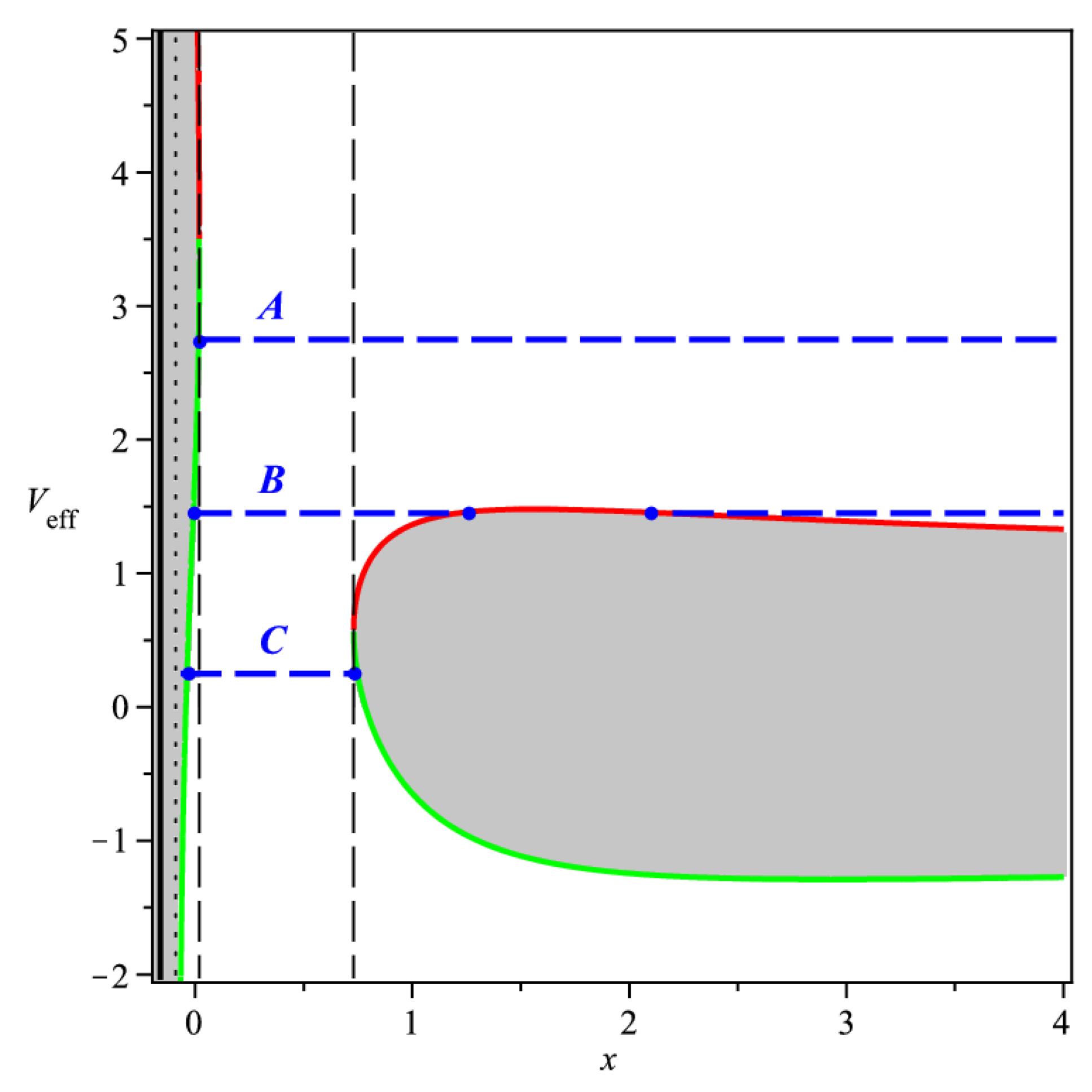}
 \subcaption{$K = 5, \Phi = -0.7, \Psi = -0.7, a = 0.4, b = 0.3$,\\ $\mu = 1, \delta=1$.}
    \label{fig:V1a}
   \end{minipage}\hfill
   \begin{minipage}[htbp]{0.5\textwidth}
   \centering
   \includegraphics[width=0.73\textwidth]{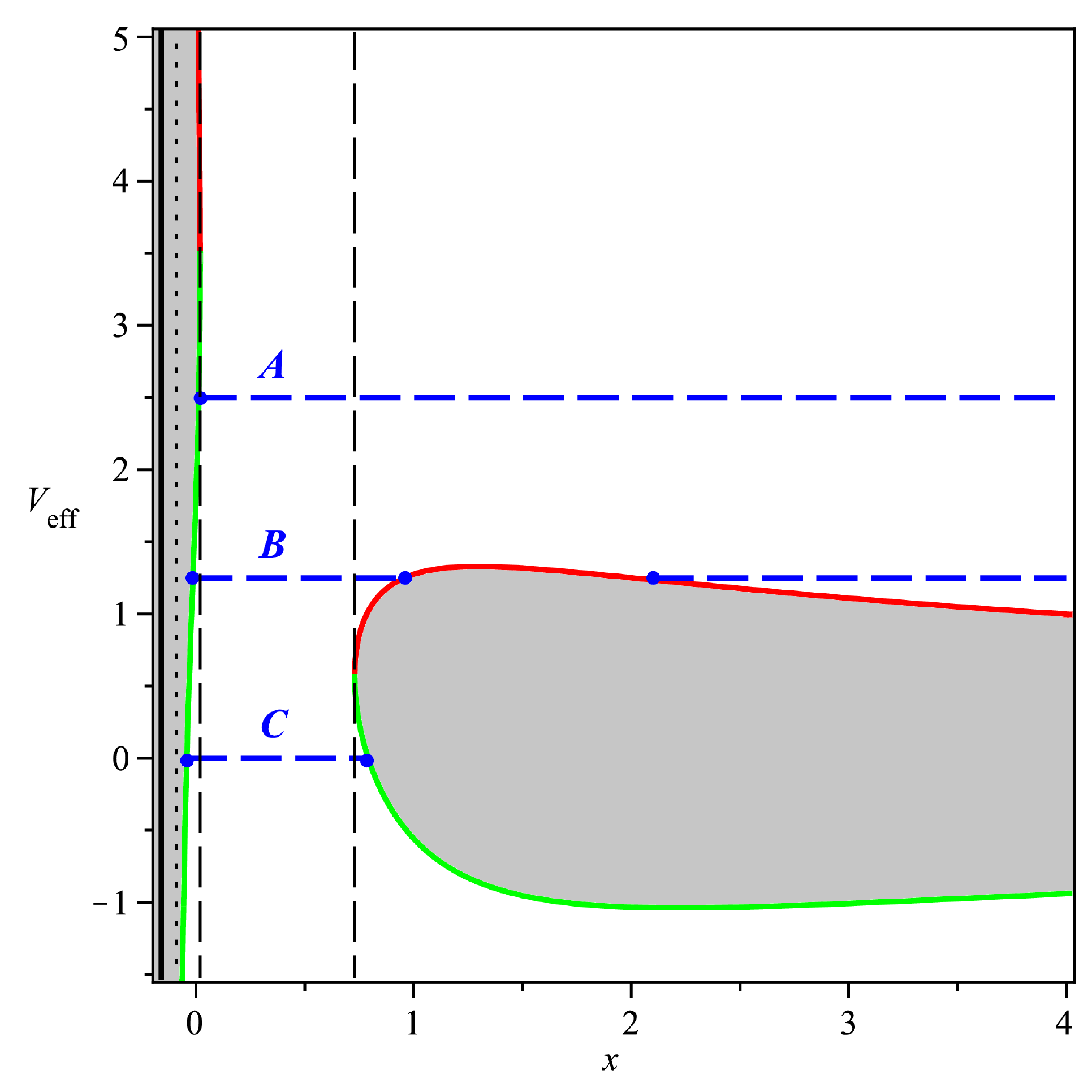} 
 \subcaption{$K = 5, \Phi = -0.7, \Psi = -0.7, a = 0.4, b = 0.3$,\\ $\mu = 1, \delta=0$.}
    \label{fig:V1b}
\end{minipage}\\[10pt]

\begin{minipage}[htbp]{0.5\textwidth}
\centering
\includegraphics[width=0.73\textwidth]{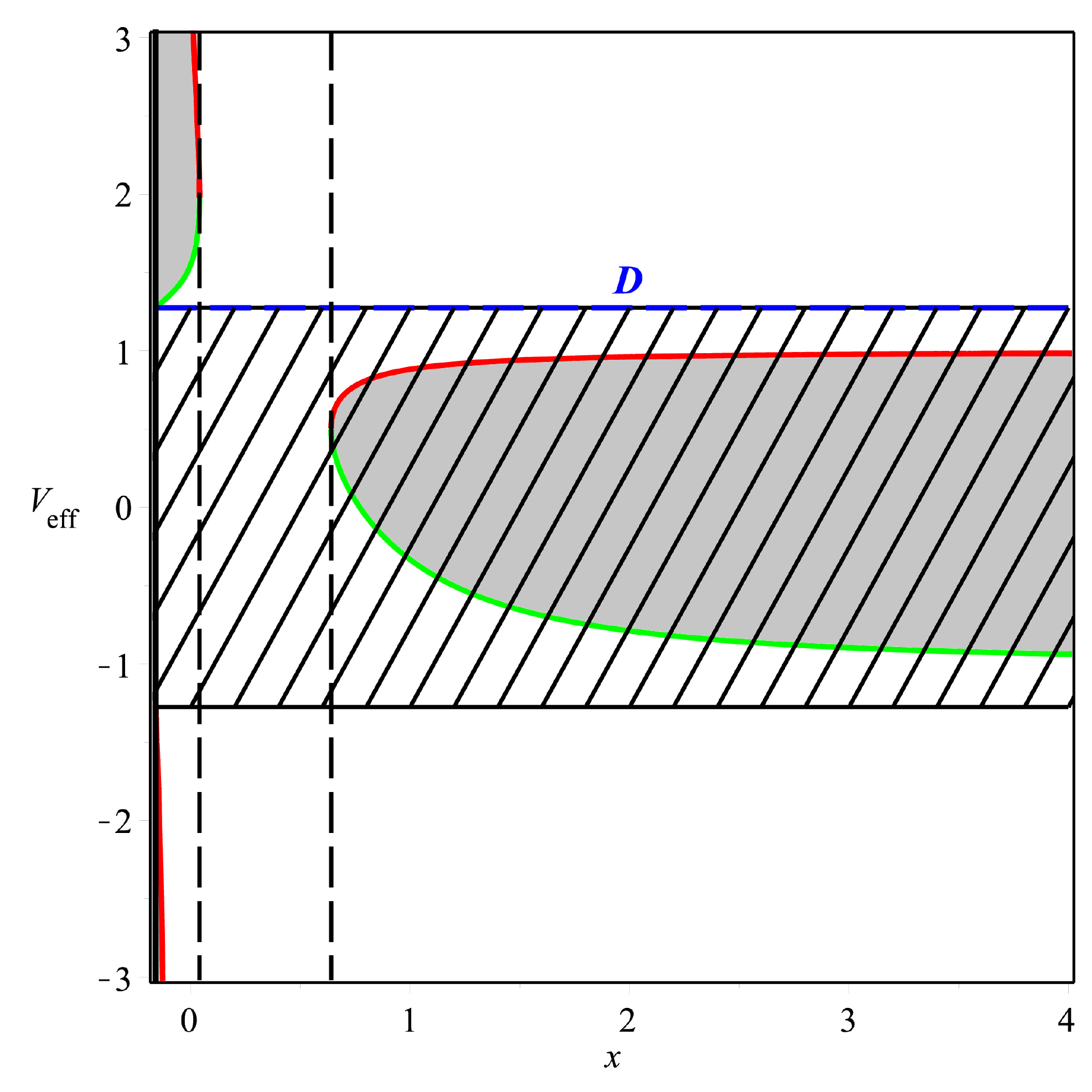}
 \subcaption{$K = 0.9, \Phi = -0.5, \Psi = -0.5, a = b = 0.4$,\\ $\mu = 1, \delta=1$.}
    \label{fig:V1c}
   \end{minipage}\hspace{-0.12cm}
\begin{minipage}[htbp]{0.5\textwidth}
\centering
 \includegraphics[width=0.73\textwidth]{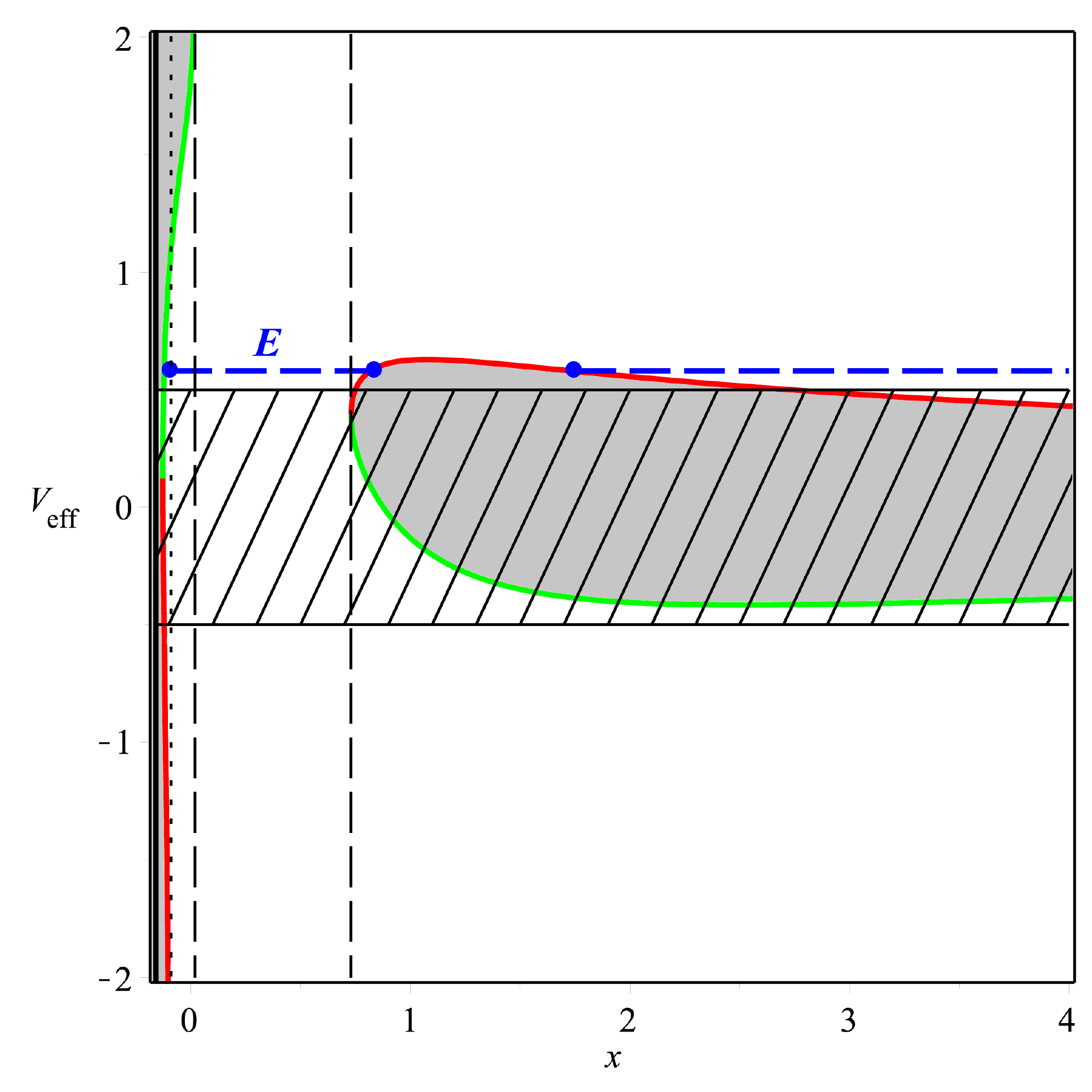}
 \subcaption{$K = 0.9, \Phi = -0.485, \Psi = -0.48, a = 0.4, b = 0.3$,\\ $\mu = 1, \delta=0$.}
    \label{fig:V1d}
   \end{minipage}
  \caption{Typical effective potentials in the five-dimensional Myers-Perry spacetime for both massive (a, c) and massless (b, d) test particle motion. Both parts of the effective potential $V_{\rm eff}^+$ (red) and $V_{\rm eff}^-$ (green) enclose a forbidden area (grey) within which $X$ is negative. The horizons are represented by dashed lines. The boundary of the singularity at $x=-b^2$ is depicted by a dotted line and the one at $x=-a^2$ by a thick line. Five types of possible orbits are represented by thick blue dashed lines of constant energy, which are called $A$, $B$, $C$, $D$ and $E$.}
   \label{fig:V1}
\end{figure*}

\begin{figure*}[htbp]
\begin{minipage}[htbp]{0.5\textwidth}
\centering
   \includegraphics[width=0.73\textwidth]{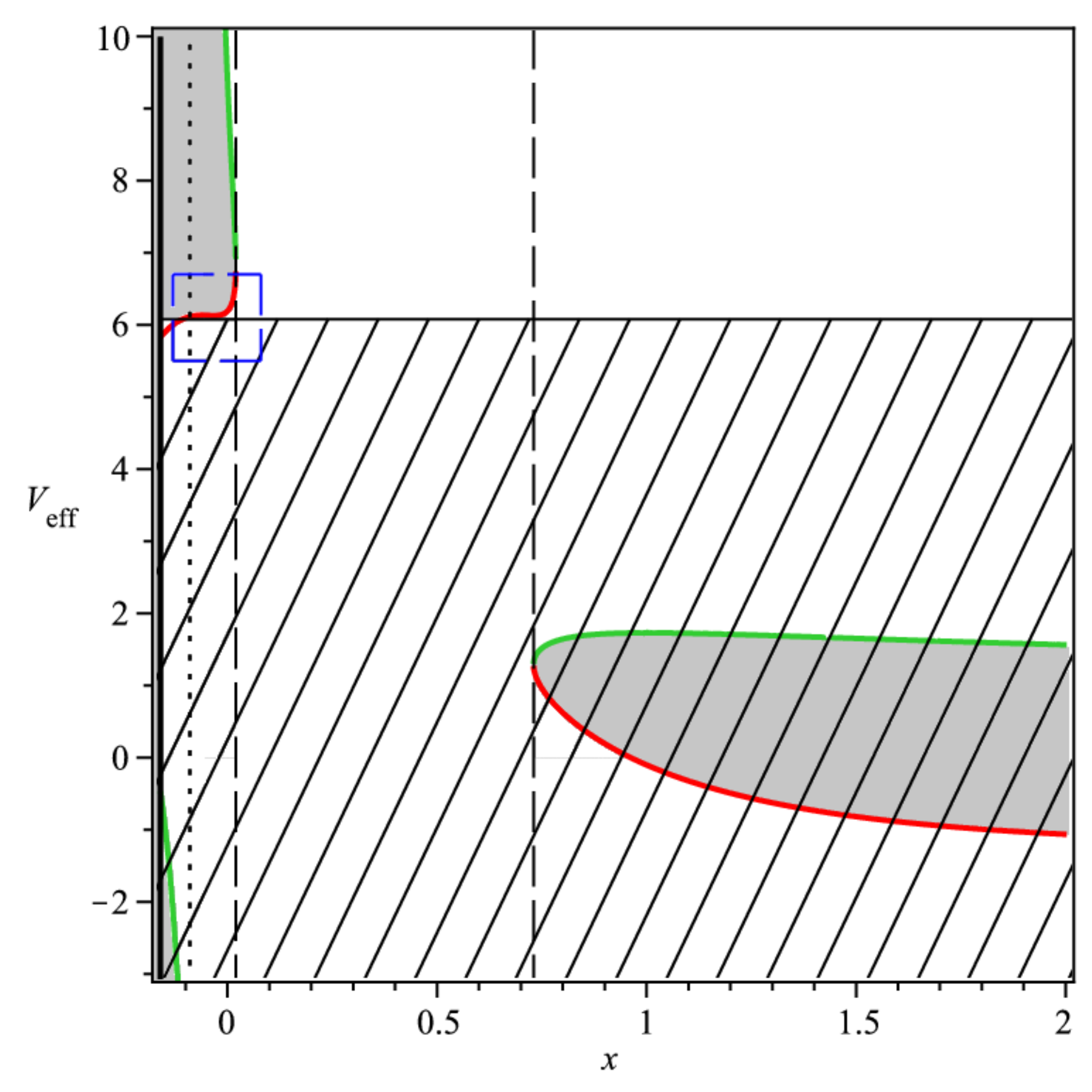}
 \subcaption{$K = 5, \Phi = -2.45, \Psi = -0.5, a = 0.4, b = 0.3$,\\ $\mu = 1, \delta=1$.}
    \label{fig:V2a}
   \end{minipage}\hfill
   \begin{minipage}[htbp]{0.5\textwidth}
   \centering
   \includegraphics[width=0.73\textwidth]{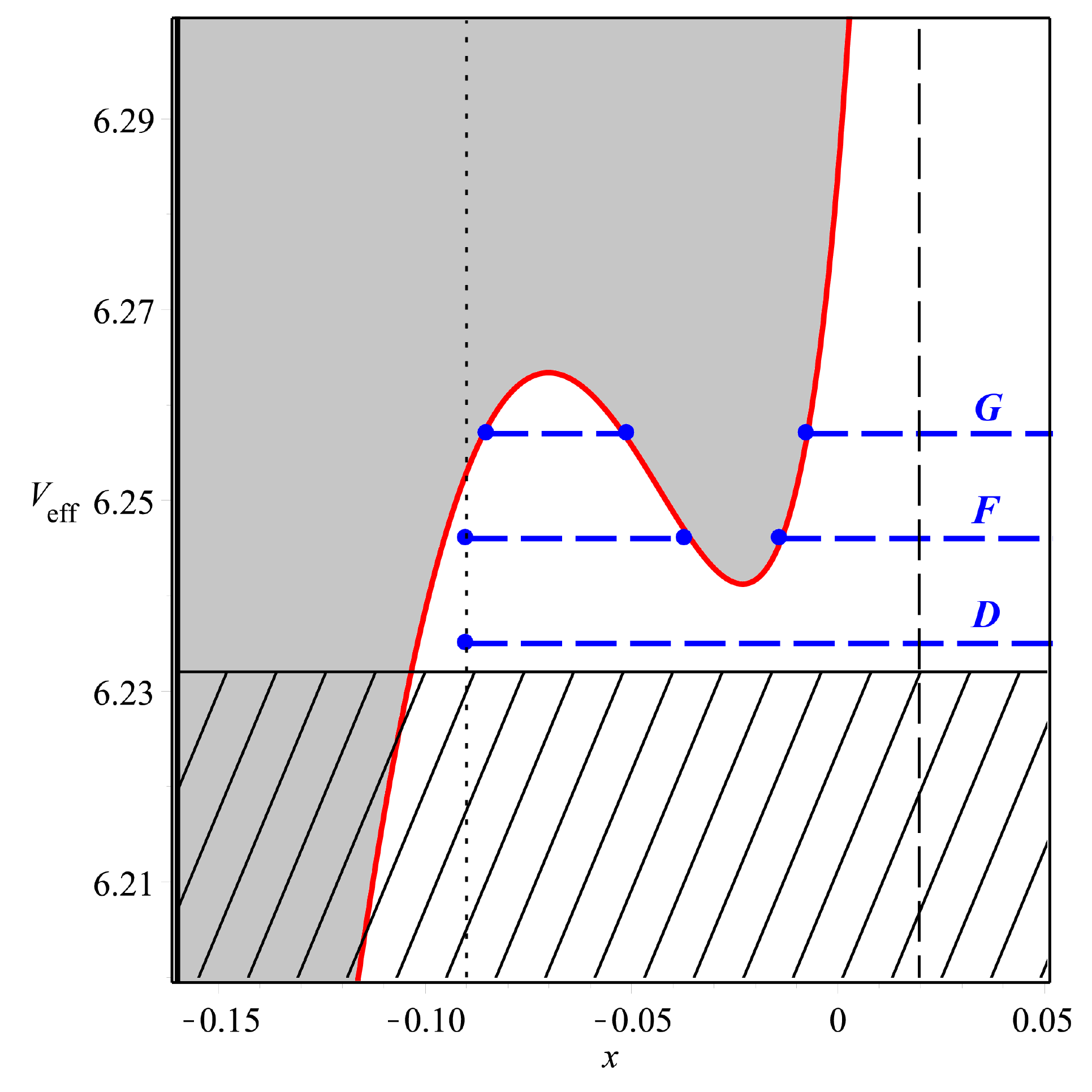} 
 \subcaption{Close-up of blue dashed box in (a).}
    \label{fig:V2b}
\end{minipage}\\[10pt]

\begin{minipage}[htbp]{0.5\textwidth}
\centering
\includegraphics[width=0.73\textwidth]{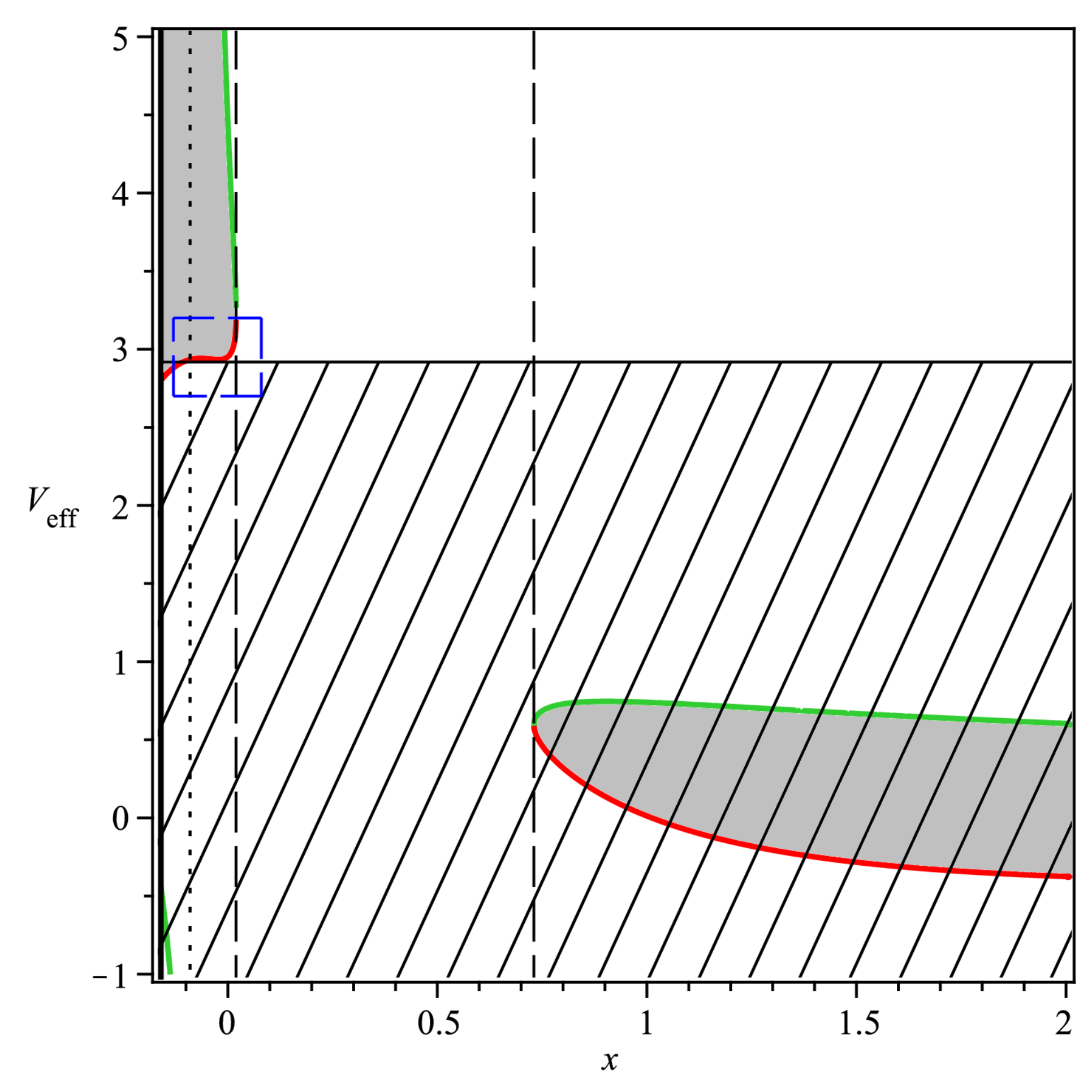}
 \subcaption{$K = 1, \Phi = -1.08, \Psi = -0.3, a = 0.4, b = 0.3$,\\ $\mu = 1, \delta=0$.}
    \label{fig:V2c}
   \end{minipage}\hspace{-0.12cm}
\begin{minipage}[htbp]{0.5\textwidth}
\centering
 \includegraphics[width=0.73\textwidth]{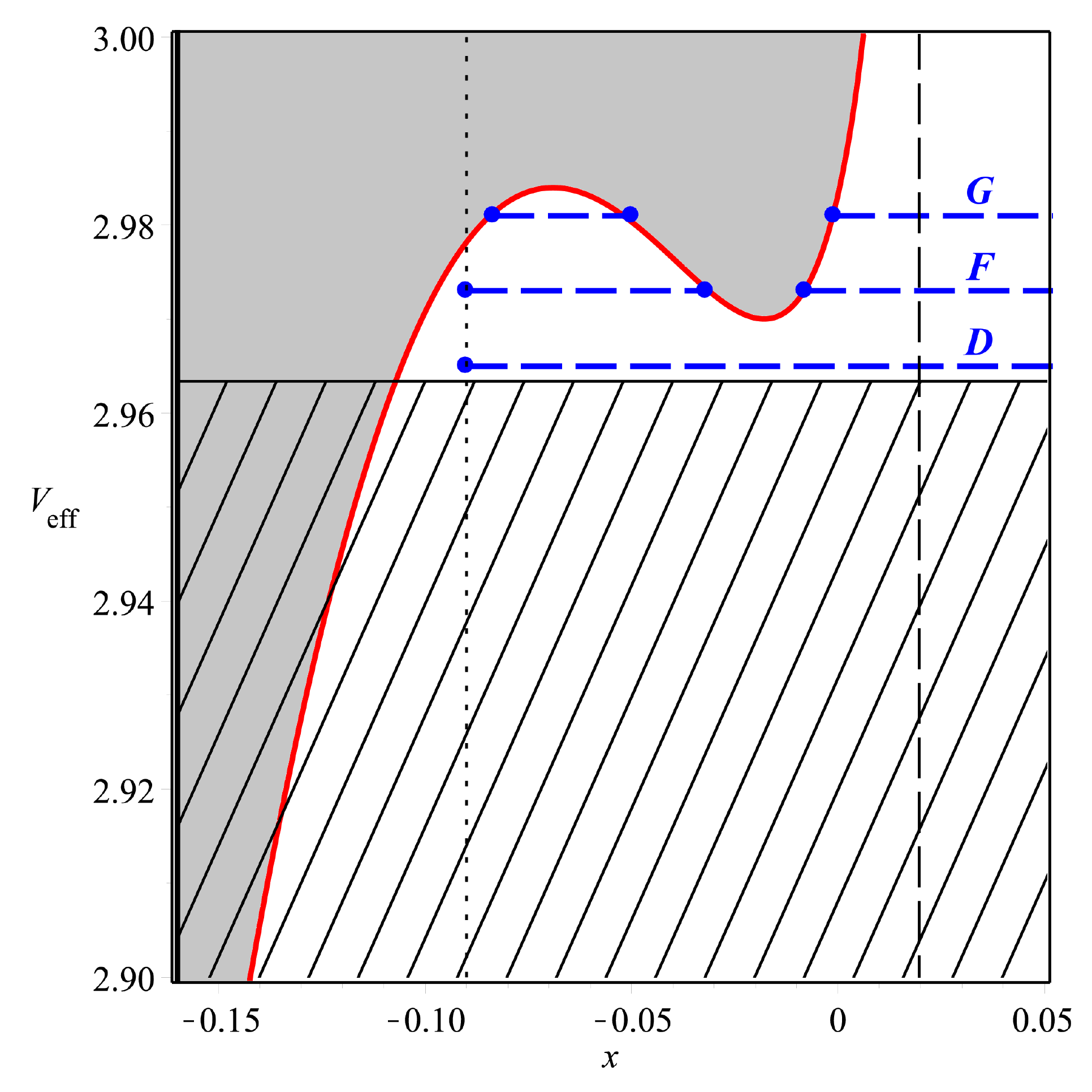}
 \subcaption{Close-up of blue dashed box in (c).}
    \label{fig:V2d}
   \end{minipage}
 \caption{Remaining orbit types $F, G$ for massive (a, b) and massless (c, d) test particle motion.}
   \label{fig:V2}
\end{figure*}

Finally, we will plot some effective potentials for various parameters and determine the possible orbit types. A typical effective potential for massive and massless test particles in the five-dimensional Myers-Perry spacetime is shown in Fig.~\ref{fig:V1a} and Fig.~\ref{fig:V1b}, respectively, illustrating orbit types $A$ to $E$. These orbit types are related to the regions (1), (2) and (3) of the $E$-$\Phi$-plots of $X$. Values of $x$ and $E$ where the polynomial $X$ becomes negative and thus no physical motion is possible are colored in gray.\\
There is a typical centrifugal barrier, which is well-known from the Kerr spacetime, that prevents a test particle from falling into the singularity. This barrier emerges due to a divergence of the effective potential at $\gamma_2 = 0$. However, we will see that there might be terminating orbits. Moreover, there is only a single maximum for positive test particle energy and a single minimum for negative energy, respectively, outside the event horizon, excluding the existence of stable bound orbits in this area. Never\-theless, these extrema indicate the presence of unstable circular bound orbits. For the case of circular null geodesics in the equatorial plane, the corresponding radius has been calculated explicitly in \cite{Cardoso:2008bp}.\\
As expected, both parts of the effective potential merge at the horizons and there may be orbits whose radial turning points coincide with the horizons as well.\\
The additional restrictions on the energy due to the $\theta$-equation, as discussed above, are shown in Fig.~\ref{fig:V1c},  \ref{fig:V1d}, \ref{fig:V2a} and \ref{fig:V2c} (hatched area). Because the zeros of $\Theta$ are $\theta_0 \notin [0, \frac{\pi}{2}]$ within the hatched area, there is no physical test particle motion within this energy threshold. In the case of $a=b$, we found out that the boundary conditions for $\Theta$ and the effective potential coincide with the singularity at $x =-a^2$ \cite{Kagramanova:2012hw}. This case is shown in Fig.~\ref{fig:V1c} for a massive test particle (orbit type $D$). The corresponding test particle's energy is given by \cite{Kagramanova:2012hw}
\begin{align}
E_{\rm crit} = V_{\rm eff}^\pm(-a^2) = \pm \frac{1}{a} \sqrt{a^2 \delta - K + (\Phi+\Psi)^2}.
\end{align}

Fig.~\ref{fig:V1d} exemplifies the general case of unequal rotation parameter for a massless test particle (orbit type $E$). The effective potential crosses the dotted line indicating $x=-b^2$, which allows the test particle to hit the singularity for values of the radial coordinate $x< -b^2$, depending on the $\theta$-value.\\
Finally, Fig.~\ref{fig:V2} indicates the remaining orbit types of the five-dimensional Myers-Perrry spacetime. Orbit type $D$ is shown for unequal rotation parameters. Orbit type $F$ belongs to terminating and two-world escape orbits, similar to orbit type $E$. Orbit type $G$ contains a bound orbit and a two-world escape orbit, which corresponds to parameter values of region (6). As we can see, these bound orbits are exclusively inside the Cauchy horizon.

For both massive and massless test particle motion, there are seven types of possible orbits in these effective potentials:

\begin{itemize}
\item[] {\bf Orbit type $\boldsymbol{A}$:} \tabto{2.57cm} A single two-world escape orbit.\\[-10pt]
\item[] {\bf Orbit type $\boldsymbol{B}$:} \tabto{2.57cm} A many-world bound orbit or an escape orbit, depending on the initial condition $x_{\rm in}$.\\[-10pt]
\item[] {\bf Orbit type $\boldsymbol{C}$:} \tabto{2.57cm} A single many-world bound orbit.\\[-10pt]
\item[] {\bf Orbit type $\boldsymbol{D}$:} \tabto{2.59cm} A single terminating orbit.\\[-10pt]
\item[] {\bf Orbit type $\boldsymbol{E}$:} \tabto{2.59cm} A terminating orbit or an escape orbit. \\[-10pt]
\item[] {\bf Orbit type $\boldsymbol{F}$:} \tabto{2.59cm} A terminating orbit or an two-world escape orbit.\\[-10pt]
\item[] {\bf Orbit type $\boldsymbol{G}$:} \tabto{2.59cm} A bound orbit or a two-world escape orbit.
\end{itemize}

Tab.~\ref{tab:orbittypes} provides a summary of all possible orbit types for massive and massless test particles in the five-dimensional Myers-Perry spacetime.

\begin{table*}[htbp]
\centering
\begin{tabular}{|c|l|c|c|c|}
\hline
Orbit type & \multicolumn{1}{c|}{Orbits} & Region & Physical zeros & Range of $x$\\ \hline \hline
A & TEO & (2) & 1 & \teo \\ 
A & TEO$_-$ & (2) & 1 & \teomin \\ \hline
B & MBO, EO & (5) & 3 & \mboeo \\ 
B & MBO$_+$, EO & (5) & 3 & \mboeomax \\ 
B & MBO$_-$, EO & (5) & 3 & \mboeomin \\ 
B & MBO$_\pm$, EO & (5) & 3 & \mboeomaxmin \\ \hline
C & MBO & (1) & 2 & \mbo \\
C & MBO$_+$ & (1) & 2 & \mbomax \\
C & MBO$_-$ & (1) & 2 & \mbomin \\
C & MBO$_\pm$ & (1) & 2 & \mbomaxmin \\ \hline
D & TO & (1), (2) & 0 & \tos \\ \hline
E & TO, EO & (5) & 2 & \toeo \\ \hline
F & TO, TEO & (6) & 2 & \toteo \\  \hline
G & BO, TEO & (6) & 3 & \boteo \\
G & BO, TEO$_-$ & (6) & 3 & \boteomin \\ \hline
\end{tabular}
\caption{Summary of the possible orbit types in the five-dimensional Myers-Perry spacetime. Thick horizontal lines represent possible orbits, whose radial turning points are illustrated by big dots. Horizons and the boundary of the singularity are shown as two thin vertical lines or one thick vertical line, respectively. Special cases, where a radial turning point coincides with a horizon, have an additional index ($x_{\rm \pm})$.}
\label{tab:orbittypes}
\end{table*}

\section{Analytical solutions}

In this section, we will solve the geodesic equations \eqref{eq:geo1} -  \eqref{eq:geo5}. The solutions will consist of the Weierstrass $\wp$-, $\zeta$- and $\sigma$- function. 

\subsection{$\theta$-equation}

In order to solve the geodesic equation \eqref{eq:geo2} for the $\theta$-motion, we will make use of the former substitution $\xi = \cos^2 \theta$, providing the differential equation
\begin{align}
\dot \xi^2 = a_3 \xi^3 + a_2 \xi^2 + a_1 \xi + a_0
\label{eq:thetasubs}
\end{align}
with coefficients given in Eq.~\eqref{eq:thetacoeffs}. Because the right-hand side is a polynomial of order three, the integration will yield an elliptic integral. These kind of integrals can be solved using the Weierstrass $\wp$-function. The first step is to obtain the usual Weierstrass-form. For a third order polynomial, this can be achieved by substituting
\begin{align}
\xi = \frac{1}{a_3} \left(4y - \frac{a_2}{3} \right).
\end{align}
This yields a differential equation in the needed form
\begin{align}
\dot y^2 = 4y^3 - g_2 y - g_3 =: Y
\label{eq:thetawei}
\end{align}
with Weierstrass invariants
\begin{align}
\begin{aligned}
g_2 &= \frac{a_2^2}{12} - \frac{a_1 a_3}{4},\\
g_3 &= \frac{a_1 a_2 a_3}{48} -\frac{a_0 a_3^2}{16} - \frac{a_2^3}{216}.
\end{aligned}
\end{align}
These invariants are associated with the roots $e_1^y$, $e_2^y$ and $e_3^y$ of $Y$ by
$g_2 = -4\left(e_1^y e_2^y + e_1^y e_3^y + e_2^y e_3^y\right)$ and $g_3 = 4 e_1^y e_2^y e_3^y$. Integration of the differential equation \eqref{eq:thetawei} leads to
\begin{align}
\begin{aligned}
\tau - \tau_{\rm in} &= \int_{y_{\rm in}}^y \frac{\mathrm du}{\sqrt{4u^3 - g_2 u - g_3}} \\ &= \int_{y_{\rm in}}^\infty \frac{\mathrm du}{\sqrt{4u^3 - g_2 u - g_3}} + \int_\infty^y \frac{\mathrm du}{\sqrt{4u^3 - g_2 u - g_3}},
\end{aligned}
\end{align}
where $\tau_{\rm in}$ and $y_{\rm in} = \frac{1}{4} \left(a_3 \xi_{\rm in} +\frac{a_2}{3} \right) =  \frac{1}{4} \left(a_3 \cos^2 \theta_{\rm in} +\frac{a_2}{3} \right)$ are the initial values.\\
The first integral yields a constant, whereas the second integral represents an elliptic integral of the first kind. By definition, the inversion of such an integral yields the Weierstrass $\wp$-function 
\begin{align}
y(\tau) = \wp\left(\tau - \tau_{\rm in}^\theta\right),
\end{align}
where we introduced
\begin{align}
\tau_{\rm in}^\theta :=  \tau_{\rm in} + \int_{y_{\rm in}}^\infty \frac{\mathrm du}{\sqrt{4u^3 - g_2 u - g_3}}.
\end{align}
Resubstitution of $y$ leads to
\begin{align}
\xi(\tau) = \frac{1}{a_3}  \left( 4 \wp \left(\tau - \tau_{\rm in}^\theta; g_2, g_3 \right) - \frac{a_2}{3} \right)
\label{eq:xisolution}
\end{align}
and another resubstitution of $\xi$ results in the final solution
\begin{align}
\theta(\tau) = \arccos \left( \pm \sqrt{\frac{1}{a_3}  \left( 4 \wp \left(\tau - \tau_{\rm in}^\theta; g_2, g_3 \right) - \frac{a_2}{3} \right)} \right).
\end{align}
Since $\theta \in  [0, \frac{\pi}{2}]$, we only consider the positive sign of the square root.

\subsection{$x$-equation}

The $x$-equation is a polynomial of third order, too. Thus, it can be solved by the same methods. The Weierstrass-form is again obtained by substituting
\begin{align}
x = \frac{1}{b_3} \left(4z - \frac{b_2}{3} \right),
\end{align}
leading to a differential equation
\begin{align}
\dot z^2 = 4z^3 - h_2 z - h_3 =: Z
\label{eq:xwei}
\end{align}
with Weierstrass invariants $h_2, h_3$ associated with the roots $e_1^z$, $e_2^z$ and $e_3^z$ of $Z$.\\
Splitting the integration of Eq.~\eqref{eq:xwei} and introducing the initial value 
\begin{align}
\tau_{\rm in}^x :=  \tau_{\rm in} + \int_{z_{\rm in}}^\infty \frac{\mathrm du}{\sqrt{4u^3 - h_2 u - h_3}}
\end{align}
 yields a solution of the following form
\begin{align}
z(\tau) = \wp(\tau -\tau_{\rm in}^x; h_2, h_3).
\end{align}
A simple resubstitution of $z$ leads to the final result
\begin{align}
x(\tau) = \frac{1}{b_3} \left( 4 \wp \left(\tau -  \tau_{\rm in}^x; h_2, h_3 \right) - \frac{b_2}{3} \right).
\label{eq:xsolution}
\end{align}
Furthermore, we can calculate the proper time as a function of the Mino time using Eq.~\eqref{eq:xisolution} and Eq.~\eqref{eq:xsolution}
\begin{align}
\begin{aligned}
\lambda(\tau) = &\,\int_{\tau_{\rm in}}^\tau \rho^2(\tau') d\tau' \\
=  &\,\int_{\tau_{\rm in}}^\tau \Big[x(\tau') + \left(a^2 -b^2 \right) \xi(\tau') + b^2 \Big] \mathrm d\tau'\\
= &- \bigg[ \frac{4}{b_3} \zeta \left(\tau' - \tau_{\rm in}^x \right)+\frac{4 \left(a^2 -b^2 \right)}{a_3} \zeta \left(\tau' - \tau_{\rm in}^\theta \right)\\ 
&+ \left(\frac{b_2}{3b_3} + \frac{a_2}{3a_3}- b^2 \right)\tau' \bigg]_{\tau_{\rm in}}^\tau,
\end{aligned}
\end{align}
where $\zeta \left(\tau' - \tau_{\rm in}^x \right)$ is, of course, related to the Weierstrass coefficients $h_1, h_2$ as $\zeta \left(\tau' - \tau_{\rm in}^\theta \right)$ is related to $g_1, g_2$.

\subsection{$\phi$-equation}

The $\phi$-equation comprises a $\theta$- and an $x$-dependent part. Therefore, the differential can be separated into two parts
\begin{align}
\mathrm d\phi = \mathrm d\phi_\theta + \mathrm d\phi_x
\end{align}
and reduced to their Weierstrass-forms
\begin{align}
\label{eq:phiparts}
\begin{aligned}
\mathrm d\phi_\theta &= \frac{\Phi}{\sin^2 \theta} \mathrm d\tau = \frac{\Phi}{1 - \xi} \frac{\mathrm d\xi}{\sqrt{\Xi}} = R^{\phi}(y) \frac{\mathrm dy}{\sqrt{Y}}, \\
\mathrm d\phi_x &= - \left(\frac{a \beta \mu}{\Delta} \mathcal{E} + \frac{a^2 - b^2}{\alpha} \Phi \right) \mathrm d\tau = \widetilde R^{\phi}(z) \frac{\mathrm dz}{\sqrt{Z}}.
\end{aligned}
\end{align}
Here, $R^{\phi}(y)$ and $\widetilde R^{\phi}(z)$ are rational functions that can be decomposed into partial fractions
\begin{align}
\begin{aligned}
R^{\phi}(y) &= \frac{G^\phi}{y-p_1},\\
 \widetilde R^{\phi}(z) &= \frac{H_1^\phi}{z-q_1} +  \frac{H_2^\phi}{z-q_2},
\end{aligned}
\end{align}
where $G^\phi, H_1^\phi,H_2^\phi, p_1, q_1$ and $q_2$ are constants which arise from the decomposition ansatz (see Appendix \ref{sec:anhang}). Therefore, we can rewrite these integrals in a compact form 
\begin{align}
\begin{aligned}
\phi_\theta - \phi_{\rm in}^\theta  &= \int_{y_{\rm in}}^y \frac{G^\phi}{y - p_1} \frac{\mathrm dy}{\sqrt{Y}},\\
\phi_x - \phi_{\rm in}^x &= \int_{z_{\rm in}}^z \sum_{j=1}^2 \frac{H_j^\phi}{y - q_j}\frac{\mathrm dz}{\sqrt{Z}}.
\end{aligned}
\end{align}
These are elliptic integrals of the third kind. Substituting $y(\tau) = \wp(v; g_2, g_3)$ with $v(\tau) = \tau - \tau_{\rm in}^\theta$ and $z(\tau) = \wp(w; h_2, h_3)$ with $w(\tau) = \tau - \tau_{\rm in}^x$ will eliminate the square root terms due to Eq.~\ref{eq:thetawei} and Eq.~\ref{eq:xwei}. This yields
\begin{align}
\phi_\theta  - \phi_{\rm in}^\theta &=  \int_{v_{\rm in}}^v \frac{G^\phi}{\wp(v) - \wp(v_1)} \mathrm dv,\\
\phi_x - \phi_{\rm in}^x &= \int_{w_{\rm in}}^w \sum_{j=1}^2 \frac{H_j^\phi}{\wp(w) - \wp(w_j)} \mathrm dw,
\end{align}
where $p_1 = \wp(v_1; g_2, g_3)$ and $v_{\rm in} = v(\tau_{\rm in})$ as well as $q_{1,2}= \wp(w_{1,2}; h_2, h_3)$ and $w_{\rm in} = w(\tau_{\rm in})$. Since the variable $v$ is always connected with the Weierstrass invariants $g_2, g_3$ as $w$ is connected with $h_2,h_3$, we may suppress these dependencies throughout. In order to finally solve these integrals, we need the following relation between the Weierstrass $\wp$- and $\zeta$-function \cite{Lawden:1989}
\begin{align}
\frac{\wp^\prime(y)}{\wp(z)-\wp(y)} = \zeta(z-y) - \zeta(z+y) + 2 \zeta(y).
\end{align}
Applying this relation will basically lead to integrals over Weierstrass $\zeta$-functions, which are in turn the logarithmic derivatives of the Weierstrass $\sigma$-functions. Thus, we can conclude that these integrals are solved by
\begin{align}
\begin{aligned}
\phi_\theta  = &\frac{G^\phi}{\wp^\prime(v_1)} \big[ \ln \sigma \left(v(\tau) - v_1 \right) - \ln \sigma(v_{\rm in} - v_1)\\ 
&- \ln \sigma \left(v(\tau) + v_1 \right) +  \ln \sigma(v_{\rm in} + v_1) \\
&+ 2 \zeta(v_1) \left(v(\tau) - v_1 \right) \big] + \phi_{\rm in}^\theta\\
=  &\frac{G^\phi}{\wp^\prime(v_1)} \bigg[ \ln \left(\frac{\sigma \left(v(\tau) - v_1 \right)}{\sigma \left(v(\tau) + v_1\right)} \right)\\
&- \ln \left(\frac{\sigma(v_{\rm in} - v_1)}{\sigma(v_{\rm in} + v_1)} \right) +  2 \zeta(v_1)\left(v(\tau) - v_1 \right) \bigg]\\
&+ \phi_{\rm in}^\theta.
\end{aligned}
\end{align}
and
\begin{align}
\begin{aligned}
\phi_x (\tau) = &\,\sum_{j=1}^2 \frac{H_j^\phi}{\wp^\prime(w_j)} \Bigg[ \ln \left(\frac{\sigma\big(w(\tau) - w_j\big)}{\sigma\big(w(\tau) + w_j\big)} \right) \\ &- \ln \left(\frac{\sigma\big(w_{\rm in} - w_j\big)}{\sigma\big(w_{\rm in} + w_j\big)} \right) +  2 \zeta(w_j)\big(w(\tau) - w_j \big) \Bigg]\\ &+ \phi_{\rm in}^x.
\end{aligned}
\end{align}
Taking these parts together results in the overall solution $\phi(\tau) = \phi_\theta(\tau) + \phi_x(\tau) + \phi_{\rm in}$, where $\phi_{\rm in} := \phi_{\rm in}^\theta + \phi_{\rm in}^x$.

\subsection{$\psi$-equation}

The $\psi$-equation can be obtained by replacing 
\begin{align}
\label{eq:exchange}
a \; \; \leftrightarrow \; \; b, \qquad \Phi \; \; \leftrightarrow \; \; \Psi \qquad \text{and} \qquad \theta(\tau) \; \; \leftrightarrow \; \; \theta(\tau) + \frac{\pi}{2}
\end{align}
in the $\phi$-equation. This means, the $\psi$-equation can also be rewritten in the following form
\begin{align}
\mathrm d\psi = \mathrm d \psi_\theta + \mathrm d\psi_x,
\end{align}
where
\begin{align}
\label{eq:psiparts}
\begin{aligned}
\mathrm d\psi_\theta &= \frac{\Psi}{\cos^2 \theta} \mathrm d\tau = \frac{\Psi}{\xi} \frac{\mathrm d\xi}{\sqrt{\Xi}} =R_{\psi}(y) \frac{\mathrm dy}{\sqrt{Y}},\\
\mathrm d\psi_x &= - \left(\frac{\alpha b \mu}{\Delta} \mathcal{E} + \frac{b^2 - a^2}{\alpha} \Psi \right) \mathrm d\tau = \widetilde R_{\psi}(z) \frac{\mathrm dz}{\sqrt{Z}}.
\end{aligned}
\end{align}
Again, a partial fraction decomposition of the rational functions $R^{\psi}(y)$ and $\widetilde R^{\psi}(z)$ leads to
\begin{align}
\begin{aligned}
R^{\psi}(y) &= \frac{G^\psi}{y-p_2},\\
\widetilde R^\psi(z) &= \frac{H_1^\psi}{z-q_1} +  \frac{H_2^\psi}{z-q_2},
 \end{aligned}
 \end{align}
where the constants $G^\psi, H_1^\psi,H_2^\psi$ and $p_2$ can be derived from the constants obtained in the former section by using Eq.~\eqref{eq:exchange} (see Appendix \ref{sec:anhang}). Consequently, the resulting elliptic integrals of the third kind are solved by
\begin{align}
\begin{aligned}
 \psi_\theta (\tau) = &\frac{G^\psi}{\wp^\prime(v_2)} \bigg[ \ln \left(\frac{\sigma\big(v(\tau) - v_2\big)}{\sigma\big(v(\tau) + v_2\big)} \right)\\ 
 &- \ln \left(\frac{\sigma\big(v_{\rm in} - v_2\big)}{\sigma\big(v_{\rm in} + v_2\big)} \right) +  2 \zeta(v_2)\big(v(\tau) - v_2 \big) \bigg]\\ &+ \psi_{\rm in}^\theta
 \end{aligned}
 \end{align}
and
\begin{align}
\begin{aligned}
\psi_x (\tau) = &\,\sum_{j=1}^2 \frac{H_j^\psi}{\wp^\prime(w_j)} \Bigg[ \ln \left(\frac{\sigma\big(w(\tau) - w_j\big)}{\sigma\big(w(\tau) + w_j\big)} \right)\\ 
&- \ln \left(\frac{\sigma\big(w_{\rm in} - w_j\big)}{\sigma\big(w_{\rm in} + w_j\big)} \right) +  2 \zeta(w_j)\big(w(\tau) - w_j \big) \Bigg]\\ &+ \psi_{\rm in}^x
\end{aligned}
\end{align}
yielding the total solution $\psi(\tau) = \psi_\theta(\tau) + \psi_x(\tau) + \psi_{\rm in}$, where $\psi_{\rm in} := \psi_{\rm in}^\theta + \psi_{\rm in}^x$.

\subsection{$t$-equation}

Basically, the $t$-equation can be solved with the same methods that were used for the $\phi$- and $\psi$-equation. We will split the integration in a $\theta$- and an $x$-dependent part again
\begin{align}
\mathrm dt = \mathrm dt_\theta + \mathrm dt_x
\end{align}
and substitute
\begin{align}
\label{eq:tparts}
\begin{aligned}
\mathrm dt_\theta &= E \left(a^2 \cos^2 \theta + b^2 \sin^2 \theta \right) \mathrm d\tau = F^t(y) \frac{\mathrm dy}{\sqrt{Y}} ,\\ 
\mathrm dt_x &= \left(Ex + \frac{\alpha \beta \mu}{\Delta} \mathcal{E} \right) \mathrm d\tau = R^t(z) \frac{\mathrm dz}{\sqrt{Z}}.
\end{aligned}
\end{align}
In contrast to the $\phi$- and $\psi$-equation, the $\theta$-dependent part only consists of a linear function $F^t(y)$, 
\begin{align}
F^t(y) = \frac{E}{a_3} \left(4y - \frac{a_2}{3} \right) \left(a^2 - b^2\right) + E b^2 := J^t_1\,y + J^t_0,
\end{align}
with coefficients $J^t_1, J^t_0$, while the rational function $R^t(z)$ of the $x$-dependent part can be decomposed into a linear function $\widetilde F^t(z)$ with coefficients $K^t_1, K^t_0$ and again two partial fraction terms (see Appendix \ref{sec:anhang})
\begin{align}
R^t(z) &= \widetilde F^t(z) +  \frac{H_1^t}{z-q_1} +  \frac{H_2^t}{z-q_2},\\
\widetilde F^t(z) &=  \frac{E}{b_3} \left(4 z - \frac{b_2}{3} + b_3 \mu \right) := K^t_1\,z + K^t_0.
\end{align}
After the usual substitution to the $v$ and $w$ coordinates, the linear terms will lead to integrals over the Weierstrass $\wp$-function, whose antiderivative is given by the negative Weierstrass $\zeta$-function. Altogether, we obtain
\begin{align}
\begin{aligned}
t_\theta &=  - J^t_1 \Big( \zeta \left(v(\tau) \right) - \zeta(v_{\rm in}) \Big) - J^t_0 \left(v(\tau) - v_{\rm in}\right) + t_{\rm in}^\theta
\end{aligned}
\end{align}
as well as
\begin{align}
\begin{aligned}
t_x (\tau) =&-K^t_1 \big(\zeta \left(w(\tau) \right) - \zeta(w_{\rm in})\big) - K^t_0 \left(w(\tau) - w_{\rm in}\right)\\
 &+ \sum_{j=1}^2 \frac{H_j^t}{\wp^\prime(w_j)} \Bigg[ \ln \left(\frac{\sigma\big(w(\tau) - w_j\big)}{\sigma\big(w(\tau) + w_j\big)} \right)\\ 
&- \ln \left(\frac{\sigma\big(w_{\rm in} - w_j\big)}{\sigma\big(w_{\rm in} + w_j\big)} \right) +  2 \zeta(w_j)\big(w(\tau) - w_j \big) \Bigg]\\ &+ t_{\rm in}^x.
\end{aligned}
\end{align}
Concludingly, the full solution is given by $t(\tau) = t_\theta(\tau) + t_x(\tau) + t_{\rm in}$, where again  $t_{\rm in} := t_{\rm in}^\theta+ t_{\rm in}^x$.

\section{Orbits}

In this section, we show examples of the types of orbits discussed in the previous sections. The orbits are plotted in the cartesian coordinates $(X,Y,Z,W)$ defined as
\begin{align}
\begin{aligned}
X &= \sqrt{|x+a^2|} \, \sin \theta \cos \phi, \;\;&Y &= \sqrt{|x+a^2|} \, \sin \theta \sin \phi,\\
Z &= \sqrt{|x+b^2|} \, \cos \theta \cos \psi,  &W &= \sqrt{|x+b^2|} \, \cos \theta \sin \psi.
\label{eq:trafos}
\end{aligned}
\end{align}
Since we cannot visualize a four-dimensional plot, we present two- and three-dimensional orbits, using special coordinate choices.

\subsection{Two-dimensional orbits}

\begin{figure*}[htbp]
\begin{minipage}[htbp]{0.5\textwidth}
\centering
   \includegraphics[width=0.73\textwidth]{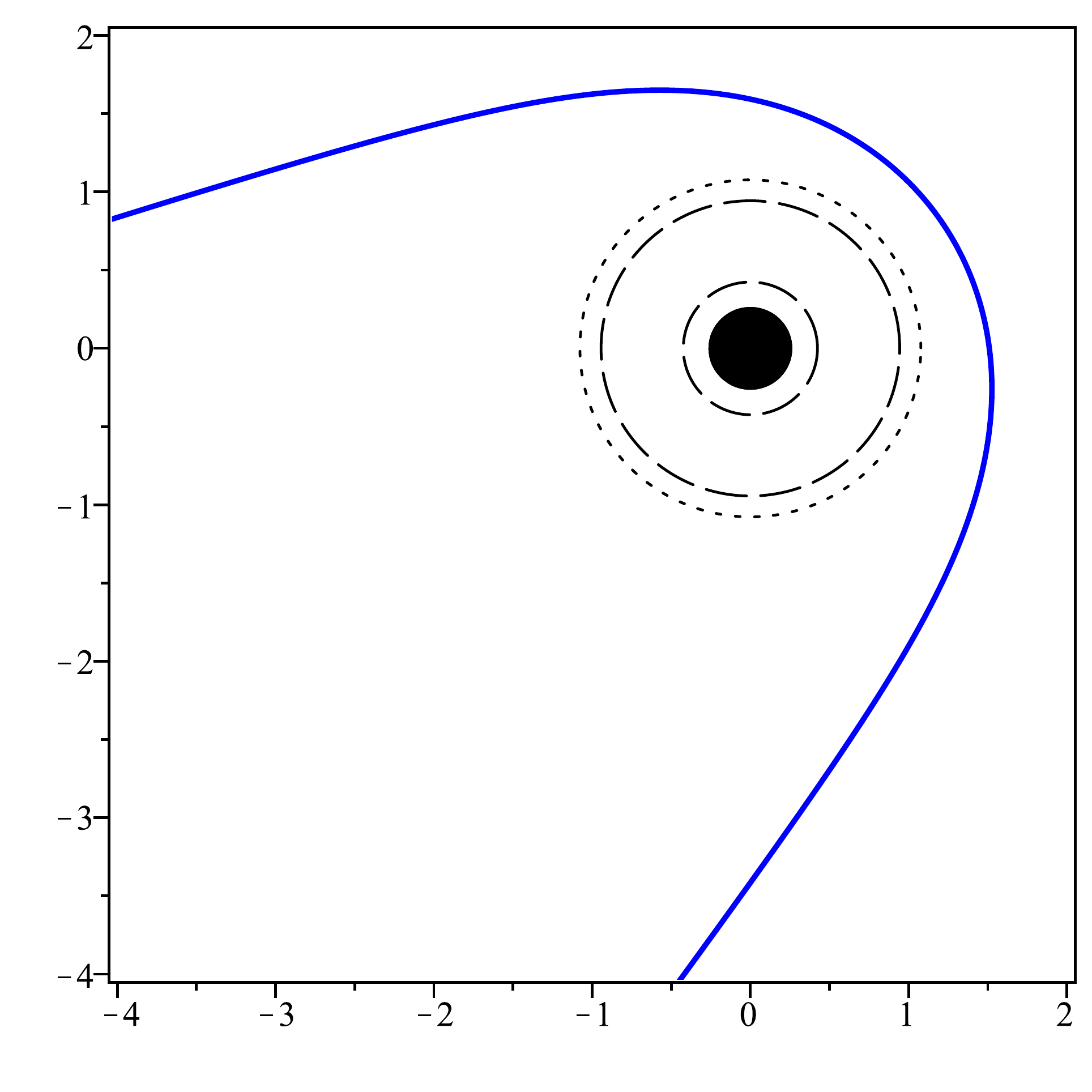}
 \subcaption{EO: $E=1.31$, $a=0.4$, $b=0.3$, $\Phi= -2$, $\mu=1$.}
    \label{fig:2Dorbitsa}
   \end{minipage}\hfill
   \begin{minipage}[htbp]{0.5\textwidth}
   \centering
   \includegraphics[width=0.73\textwidth]{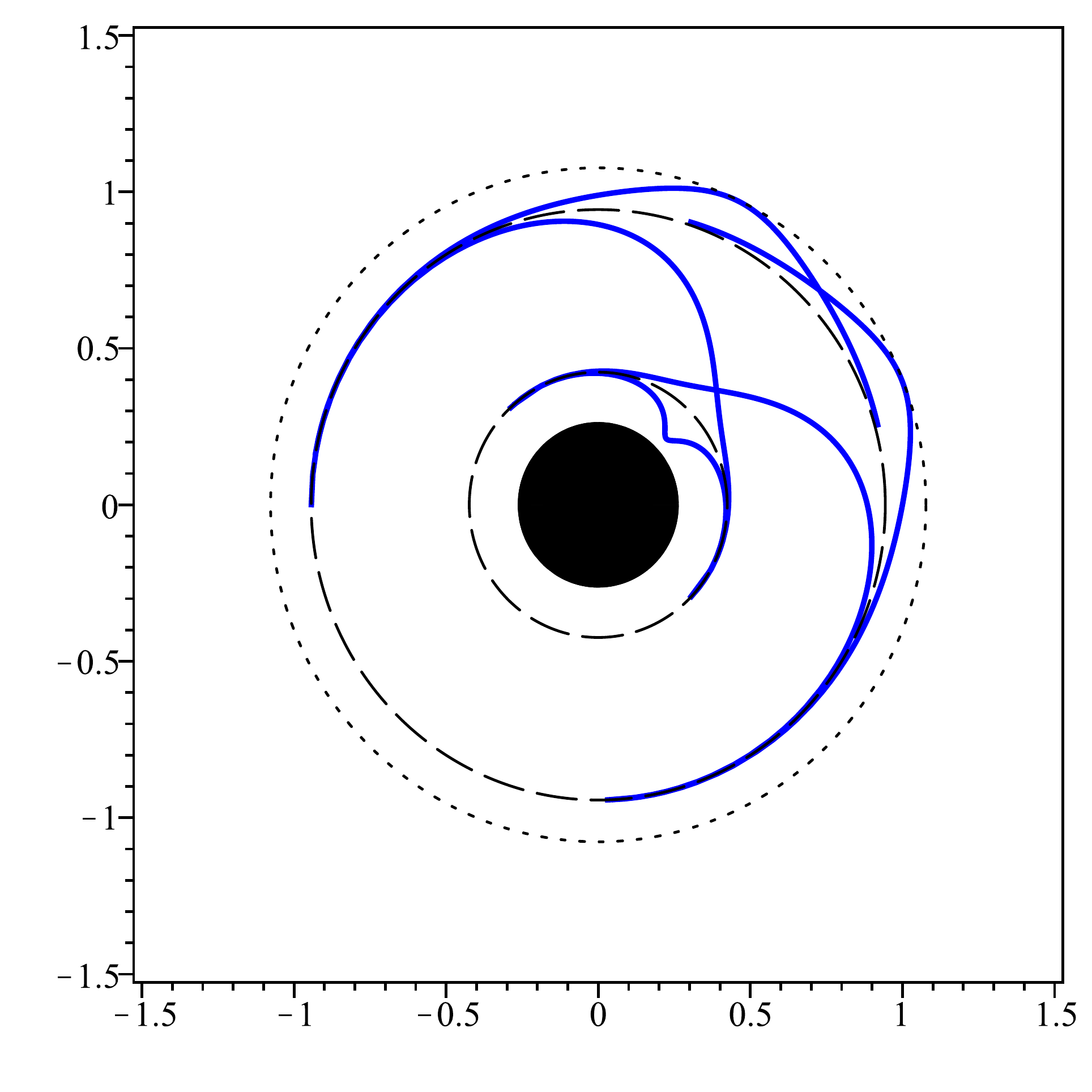} 
 \subcaption{MBO: $E=0.4$, $a=0.4$, $b=0.3$, $\Phi= 0.1$, $\mu=1$.}
    \label{fig:2Dorbitsb}
\end{minipage}\\[10pt]

\begin{minipage}[htbp]{0.5\textwidth}
\centering
   \includegraphics[width=0.73\textwidth]{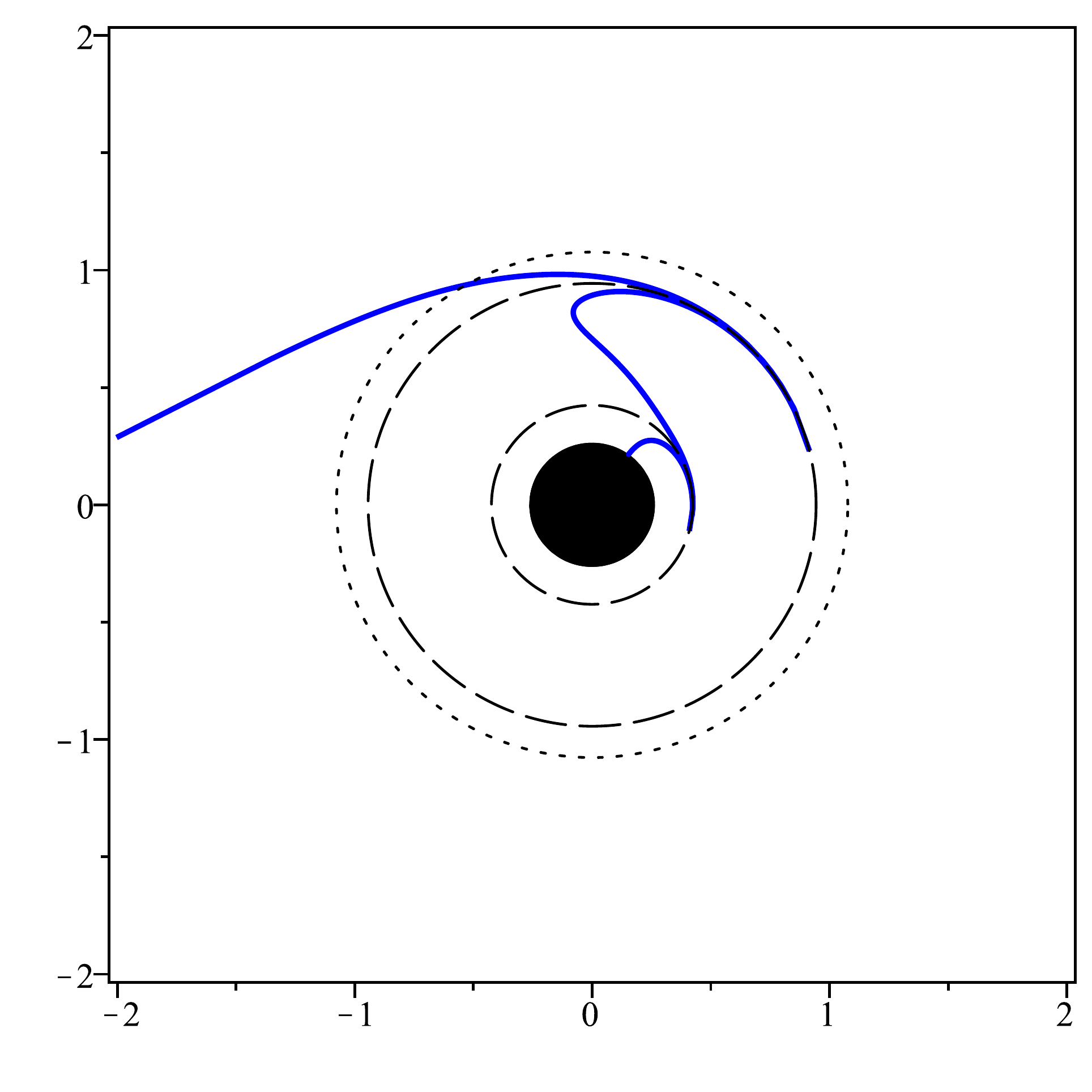}
 \subcaption{TO: $E=2$, $a=0.4$, $b=0.3$, $\Phi= -2$, $\mu=1$.}
    \label{fig:2Dorbitsc}
   \end{minipage}\hfill
   \begin{minipage}[htbp]{0.5\textwidth}
   \centering
   \includegraphics[width=0.73\textwidth]{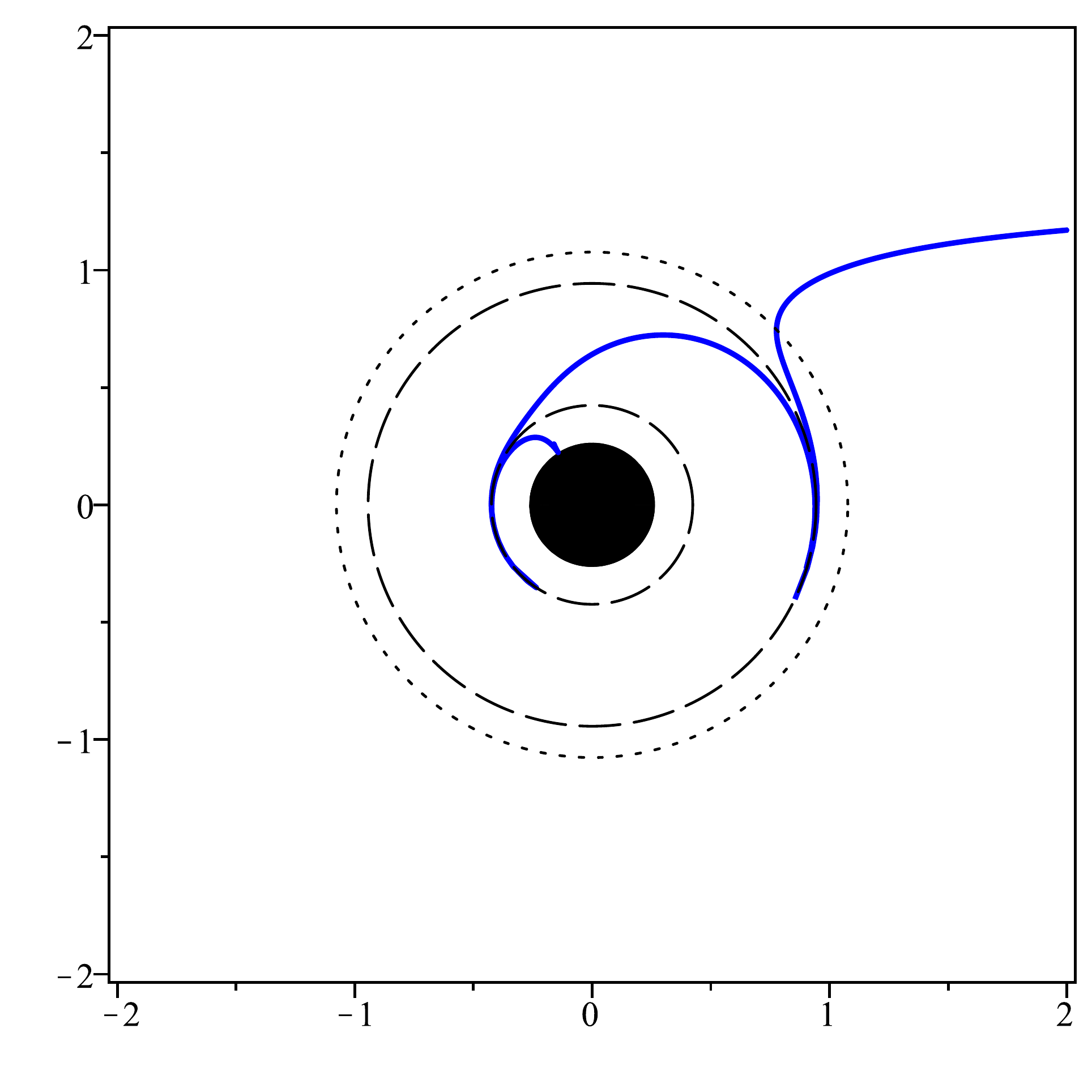} 
 \subcaption{TO: $E=2$, $a=0.4$, $b=0.3$, $\Phi= 2$, $\mu=1$.}
    \label{fig:2Dorbitsd}
\end{minipage}\\[10pt]

\begin{minipage}[htbp]{0.5\textwidth}
\centering
   \includegraphics[width=0.73\textwidth]{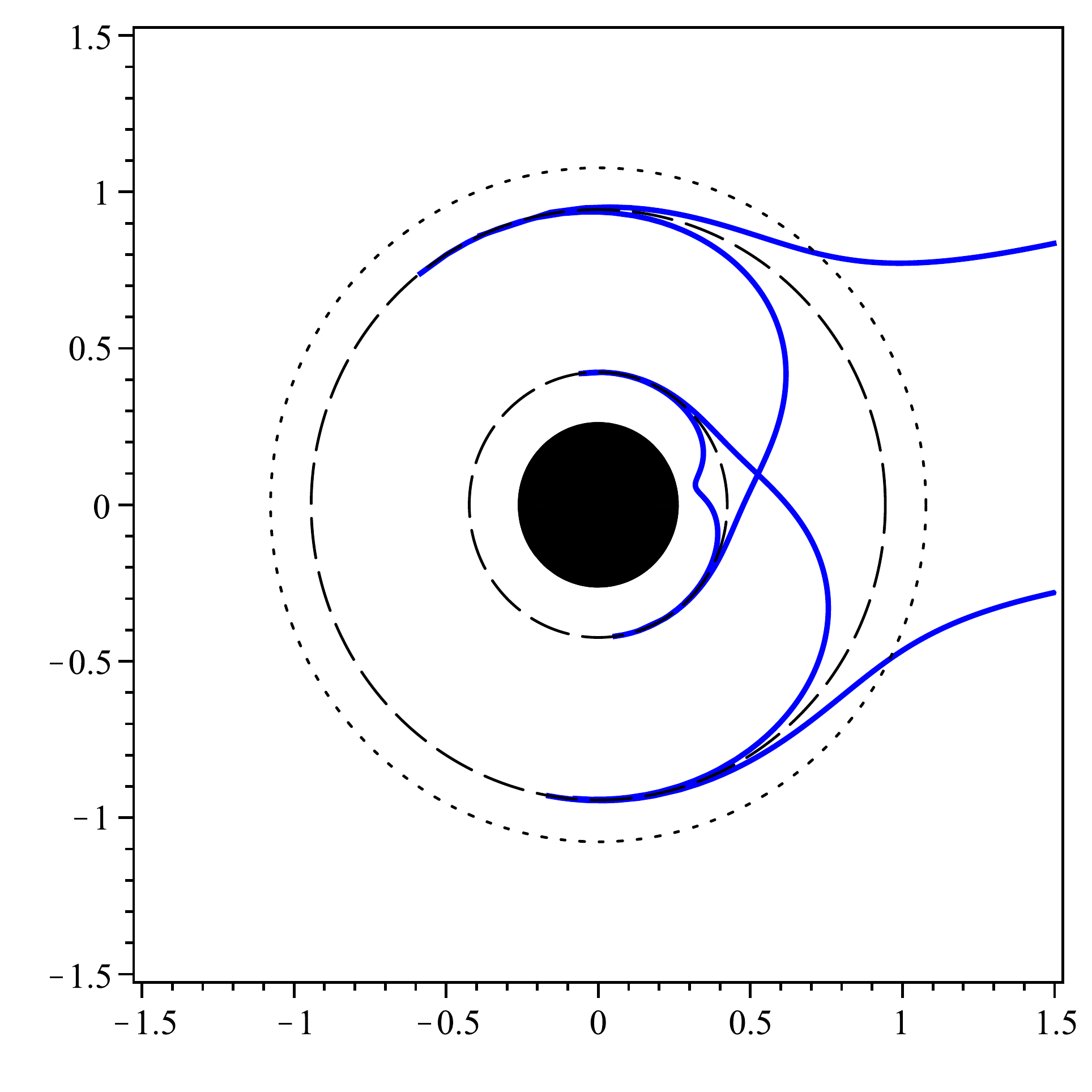}
 \subcaption{TEO: $E=4.4$, $a=0.4$, $b=0.3$, $\Phi= -1$, $\mu=1$.}
    \label{fig:2Dorbitse}
   \end{minipage}
   \caption{All possible types of two-dimensional, massive test particle orbits in the equatorial $\theta = \frac{\pi}{2}$-plane. The dotted line represents the static limit, the dashed lines indicate the horizons and the black circle represents the singularity.}
   \label{fig:2Dorbits}
\end{figure*}

Choosing $\theta = \frac{\pi}{2}$ will provide equatorial orbits including the restrictions $\psi=0$ and Eq.~\eqref{eq:Carterb}. Note that Eq.~\eqref{eq:Carterb} includes that the Carter constant $K$ is expressed by the test particle's energy $E$ and therefore the effective potential needs to be redefined. In Fig.~\ref{fig:2Dorbits} we present the possible orbits for massive test particles in this plane. The only difference to massless test particles is the existence of many-world bound orbits. As pointed out in \cite{Frolov:2004pr}, there are no bound orbits in both cases, though. Because of the bad choice of coordinates (Boyer-Lindquist coordinates), the geodesics diverge at the horizons. This is associated with the directional change of the orbits when crossing the horizons. This behavior is known from the four-dimensional Kerr spacetime \cite{Chandrasekhar:1983,Oneill:1995}. For reasons of clarity, we only indicated these divergences. As already mentioned, the geodesics enter another universe when leaving the event horizon again due to the principle of causality. In Fig.~\ref{fig:2Dorbitsd}, we can see the frame-dragging effect of the black hole, which leads to co-rotation for test particles that approach the black hole's ergosphere with appropriate angular momentum. The singularity in this plane is encountered at $x=-b^2$.

\subsection{Three-dimensional orbits}

\begin{figure}[ht]
\centering
   \includegraphics[width=0.48\textwidth]{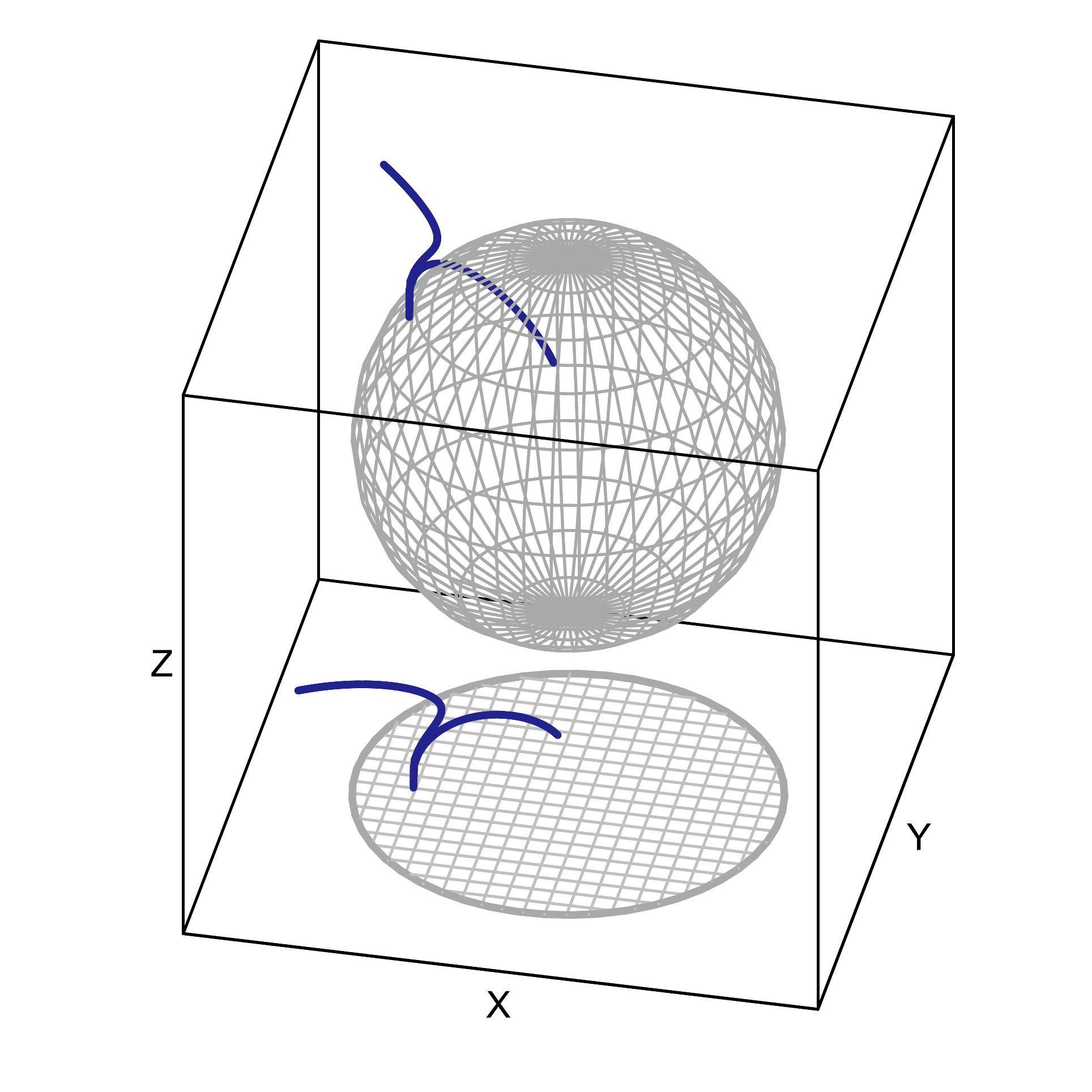}
\caption{Projection of a sphere $S^2 \subset \mathbb{R}^3$ and of an orbit diverging at the sphere into the $X$-$Y$-plane.}
\label{fig:projection}
\end{figure}

In order to obtain three-dimensional plots, we choose a projection of the four-dimensional motion into the three-dimensional space. This means, we won't obtain an actual three-dimensional orbit but a projected orbit, omitting one of the cartesian coordinates. In order to find a representation of the horizons and the singularity in this projection, we will calculate the boundary of this projection map to represent these quantities reasonably. One can easily see that the image of this map is completely contained in a spheroid of the form
\begin{align}
\frac{X^2}{|x+a^2|} + \frac{Y^2}{|x+a^2|} + \frac{Z^2}{|x+b^2|} \le 1,
\end{align}
where the equality is true for $\psi=0$. Therefore, the boundary of this map is obtained for $\psi=0$, which can be confirmed by calculating its Jacobian determinant. Consequently, we will represent the static limit, the horizons and the singularity by this special projection. Note that the interior of these projected boundaries will also represent the actual spacetime quantities in four dimensions. We may exemplify this by the coordinate-transformation from three-dimensional spherical coordinates $(r, \theta, \phi)$ to cartesian coordinates $(X,Y,Z)$ given by $X = r \sin \theta \cos \phi$, $Y = r \sin \theta \sin \phi$ and $Z = r \cos \theta$. The projection map into the $X$-$Y$-plane is bounded by the circle
\begin{align}
\frac{X^2}{r^2} + \frac{Y^2}{r^2} \le 1, 
\end{align}
where the equality is true for $\theta = \frac{\pi}{2}$. Of course, the map is also projecting into the inside of this circle. Thus, a three-dimensional orbit diverging at a sphere may now seem to cross the horizon smoothly and diverge inside the projected disc afterwards (see Fig.~\ref{fig:projection}).

\begin{figure*}[htbp]
\begin{minipage}[htbp]{0.5\textwidth}
\centering
   \includegraphics[width=0.73\textwidth]{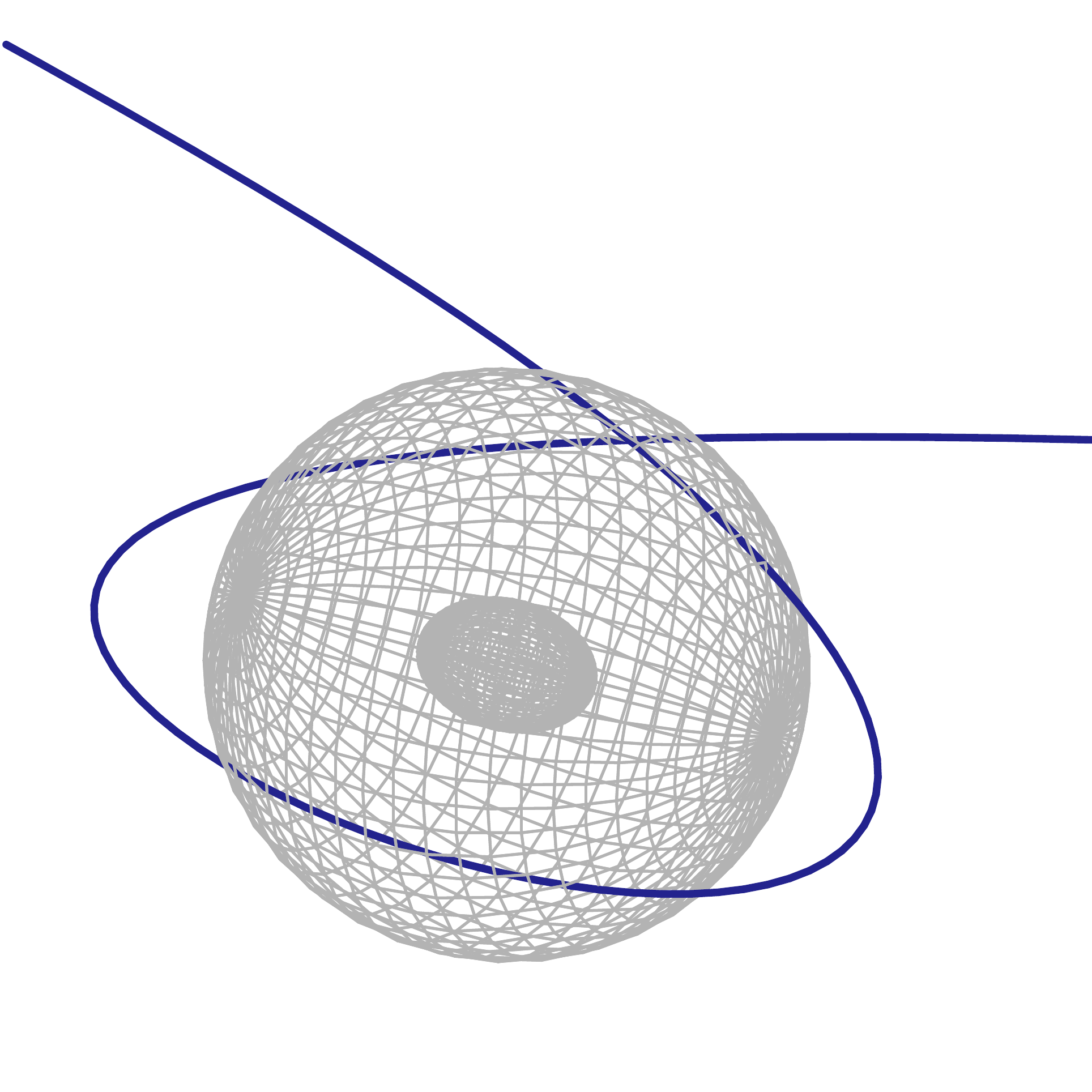}
 \subcaption{EO in $Y$-$Z$-$W$-space: $E=1.223$, $K=3$, $a=0.3$, $b=0.2$, $\Phi = -1.5$, $\Psi=0.1, \mu =1$.}
    \label{fig:3dorbitsa}
   \end{minipage}\hfill
   \begin{minipage}[htbp]{0.5\textwidth}
   \centering
   \includegraphics[width=0.73\textwidth]{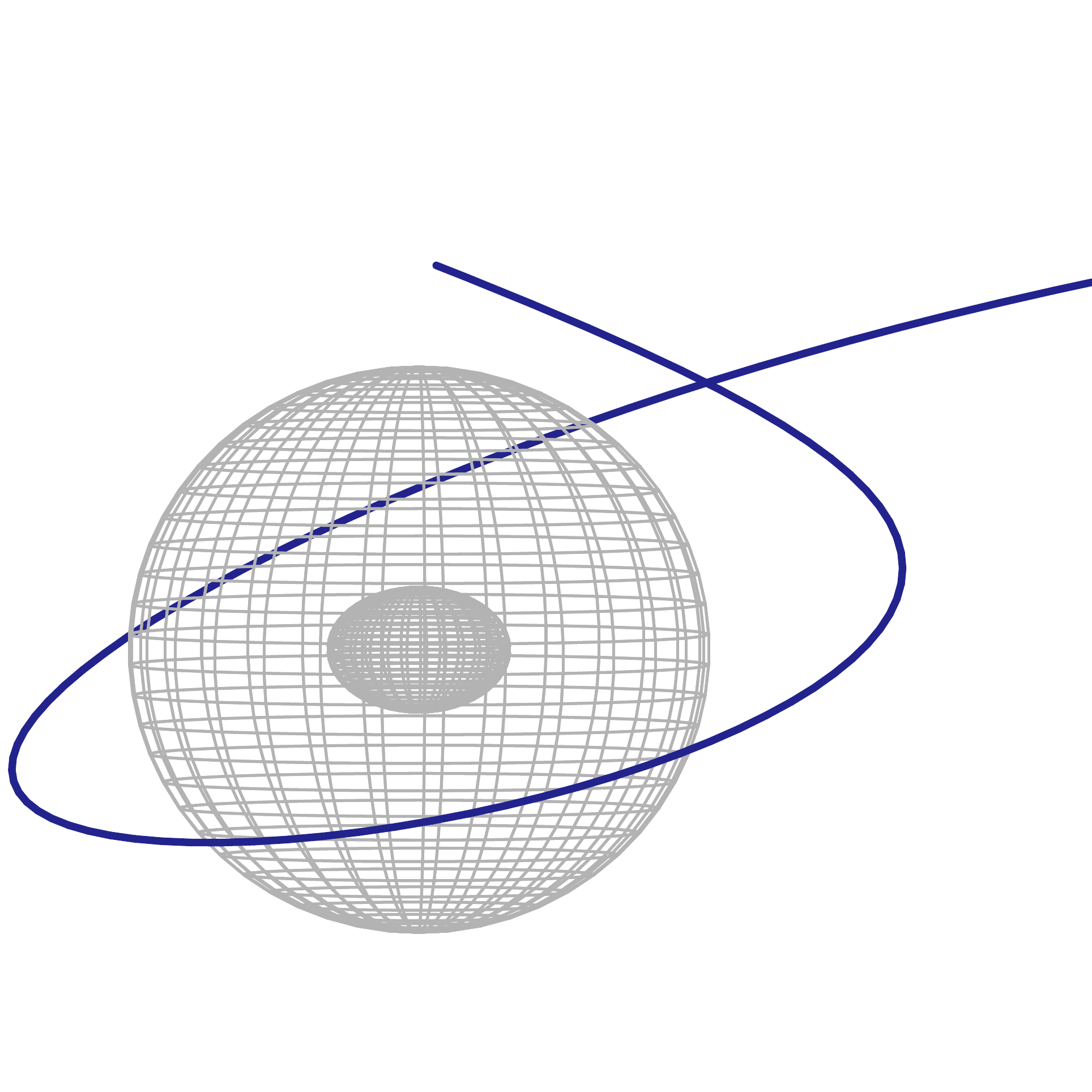} 
 \subcaption{Same orbit in $X$-$Y$-$Z$-space.}
    \label{fig:3dorbitsb}
\end{minipage}

\begin{minipage}[htbp]{0.5\textwidth}
\centering
   \includegraphics[width=0.73\textwidth]{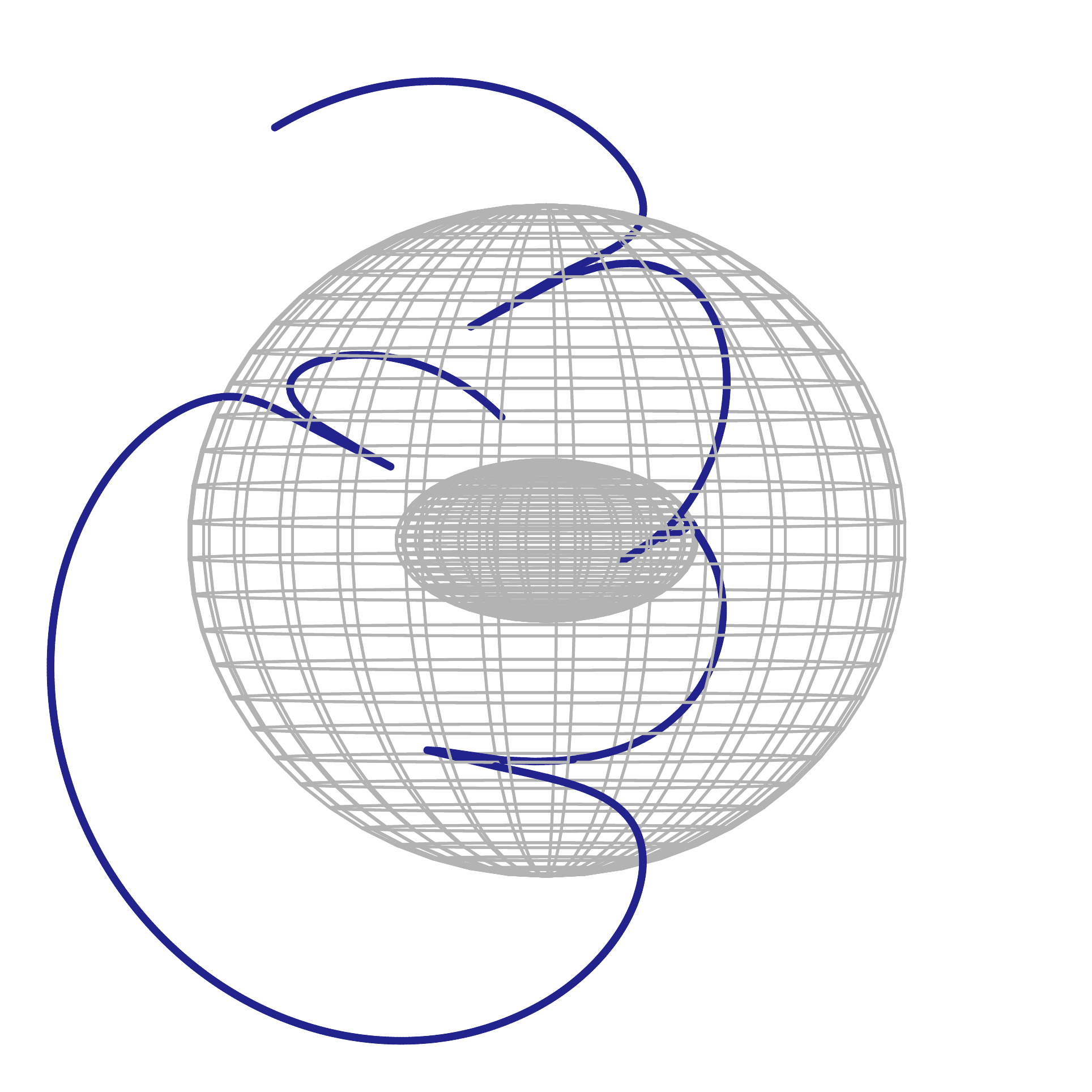}
 \subcaption{MBO in $X$-$Y$-$Z$-space: $E=1.1$, $K=3$, $a=0.4$, $b=0.2$, $\Phi= 0.5$, $\Psi=0.5$, $\mu=1$.}
    \label{fig:3dorbitsc}
   \end{minipage}\hfill
   \begin{minipage}[htbp]{0.5\textwidth}
   \centering
   \includegraphics[width=0.73\textwidth]{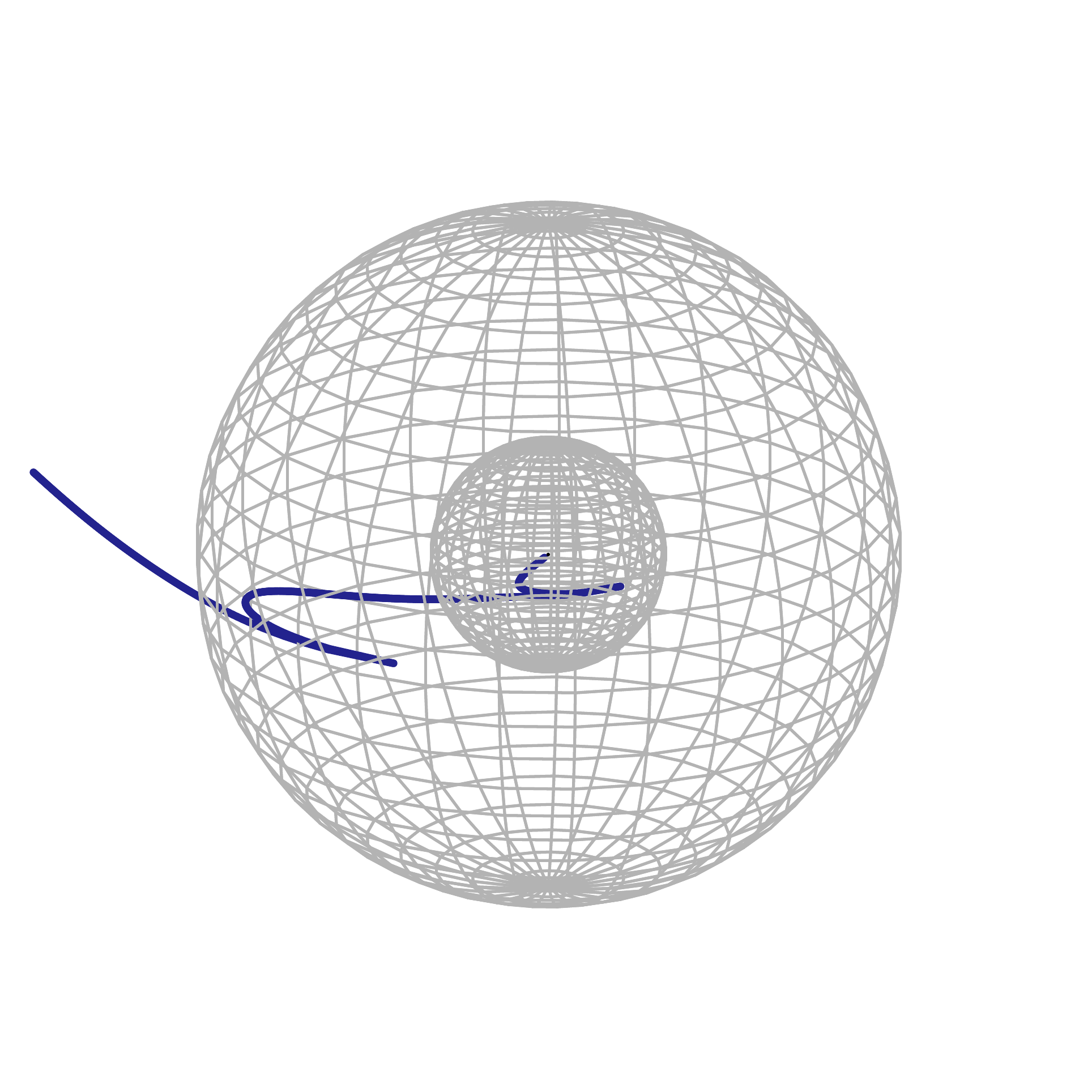} 
 \subcaption{TO in $X$-$Y$-$Z$-space: $E=1.666$, $K=1.8$, $a=b=0.3$, $\Phi= -1.5$, $\Psi=-0.1$, $\mu=1$.}
    \label{fig:3dorbitsd}
\end{minipage}\\[10pt]

\begin{minipage}[htbp]{0.5\textwidth}
\centering
   \includegraphics[width=0.73\textwidth]{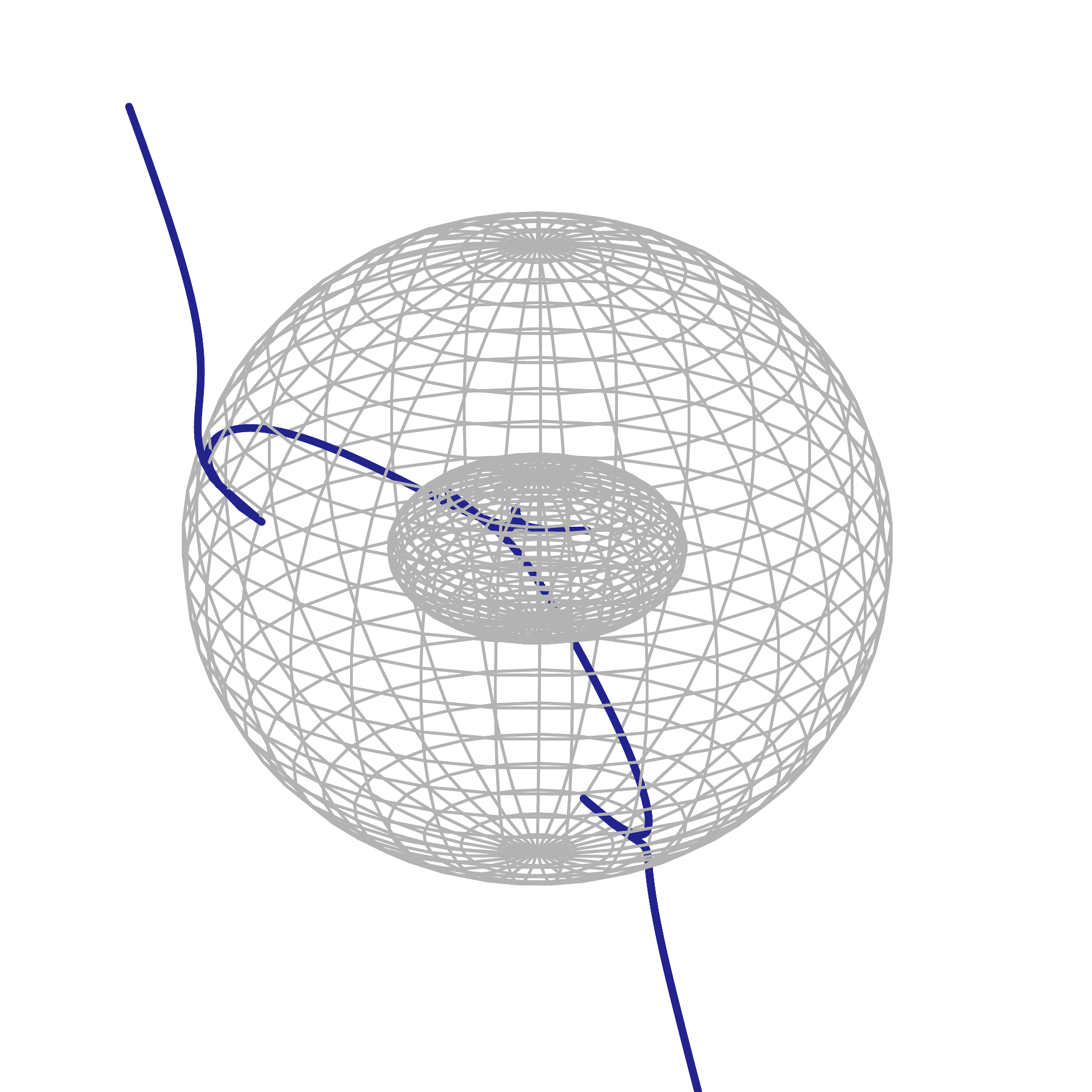}
 \subcaption{TEO in $X$-$Y$-$Z$-space: $E=1.8$, $K=3$, $a=0.4$, $b=0.2$, $\Phi=0.5$, $\Psi=0.5$, $\mu=1$.}
    \label{fig:3dorbitse}
   \end{minipage}\hfill
      \begin{minipage}[htbp]{0.5\textwidth}
      \centering
   \includegraphics[width=0.73\textwidth]{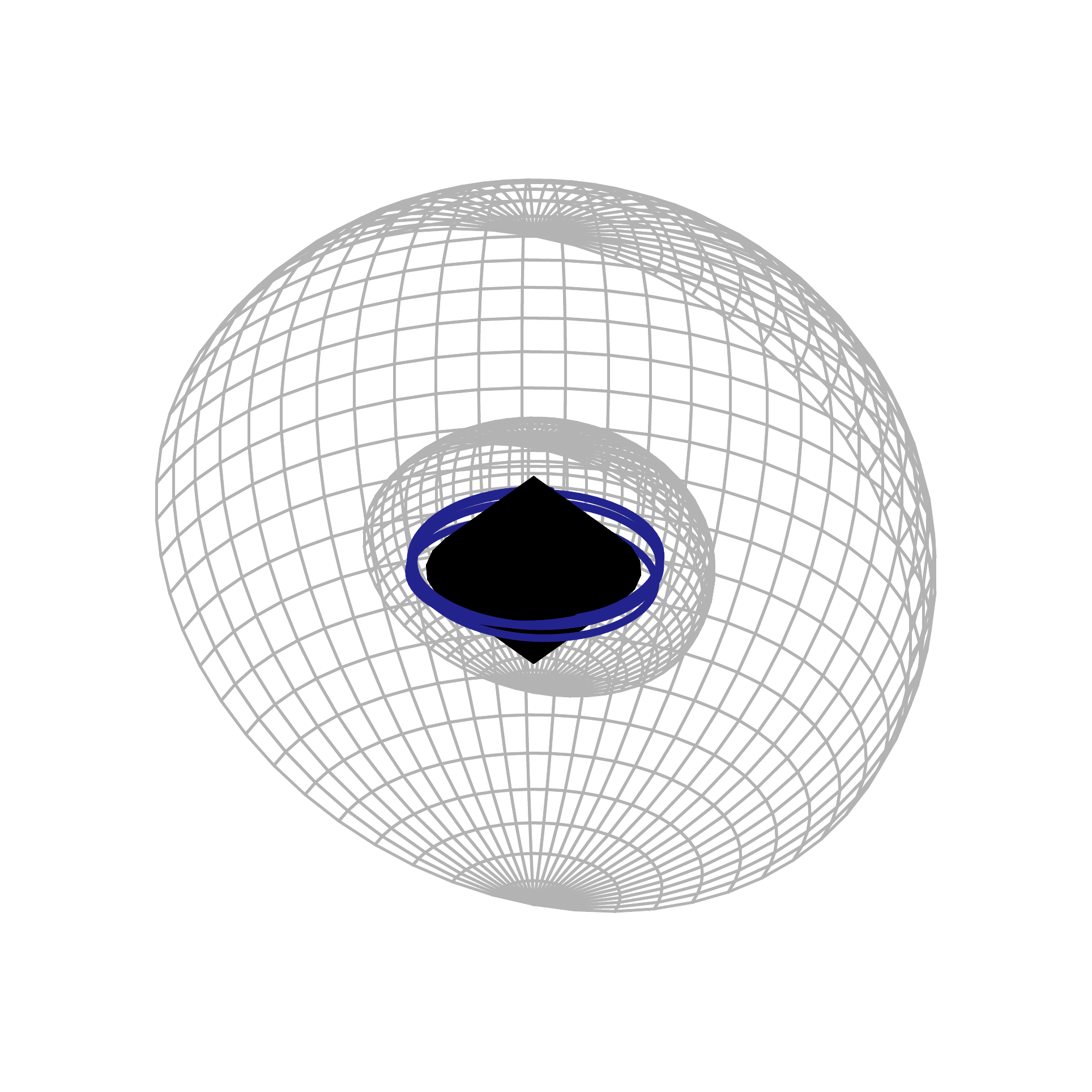} 
 \subcaption{BO in $X$-$Y$-$Z$-space: $E=6.01$, $K=5$, $a=0.4$, $b=0.3$, $\Phi= -2.6843$, $\Psi=-0.2$, $\mu=1$.}
    \label{fig:3dorbitsf}
\end{minipage}
   \caption{All possible types of massive test particle orbits in the five-dimensional Myers-Perry spacetime projected into the $X$-$Y$-$Z$-space (b-f) and an additional projection of the escape orbit into the $Y$-$Z$-$W$-space (a).}
   \label{fig:3dorbits}
\end{figure*}

Consequently, we will obtain these visual deficiencies for the projected three-dimensional orbits as well. In Fig.~\ref{fig:3dorbits} we present the possible orbits for massive test particles. In Fig.~\ref{fig:3dorbitsa} we show an escape orbit projected into the $Y$-$Z$-$W$-space and in Fig.~\ref{fig:3dorbitsb} the same orbit is projected into the $X$-$Y$-$W$-space. Apparently, the escape orbit is crossing the event horizon in Fig.~\ref{fig:3dorbitsa}, which is physically not the case, whereas the projection of Fig.~\ref{fig:3dorbitsb} yields a familiar representation of an escape orbit. Therefore, we also used this projection for the representation of the remaining orbit types in Fig.~\ref{fig:3dorbits}. As seen in Fig.~\ref{fig:3dorbitsf}, the singularity assumes the shape of a double cone in this case.

\section{Observables}

Concerning the geodesics of a spacetime, it is possible to calculate observable quantities. Observables are very useful in order to test the related theory. These quantities are e.g. the light deflection for escape orbits, the perihelion shift for bound orbits or the Lense-Thirring effect. We will follow along the lines of \cite{Drasco:2003ky,Fujita:2009bp,Hackmann:2011wp} to calculate these effects.

\subsection{Deflection angle}

\begin{figure}[ht]
\centering
   \includegraphics[width=0.45\textwidth]{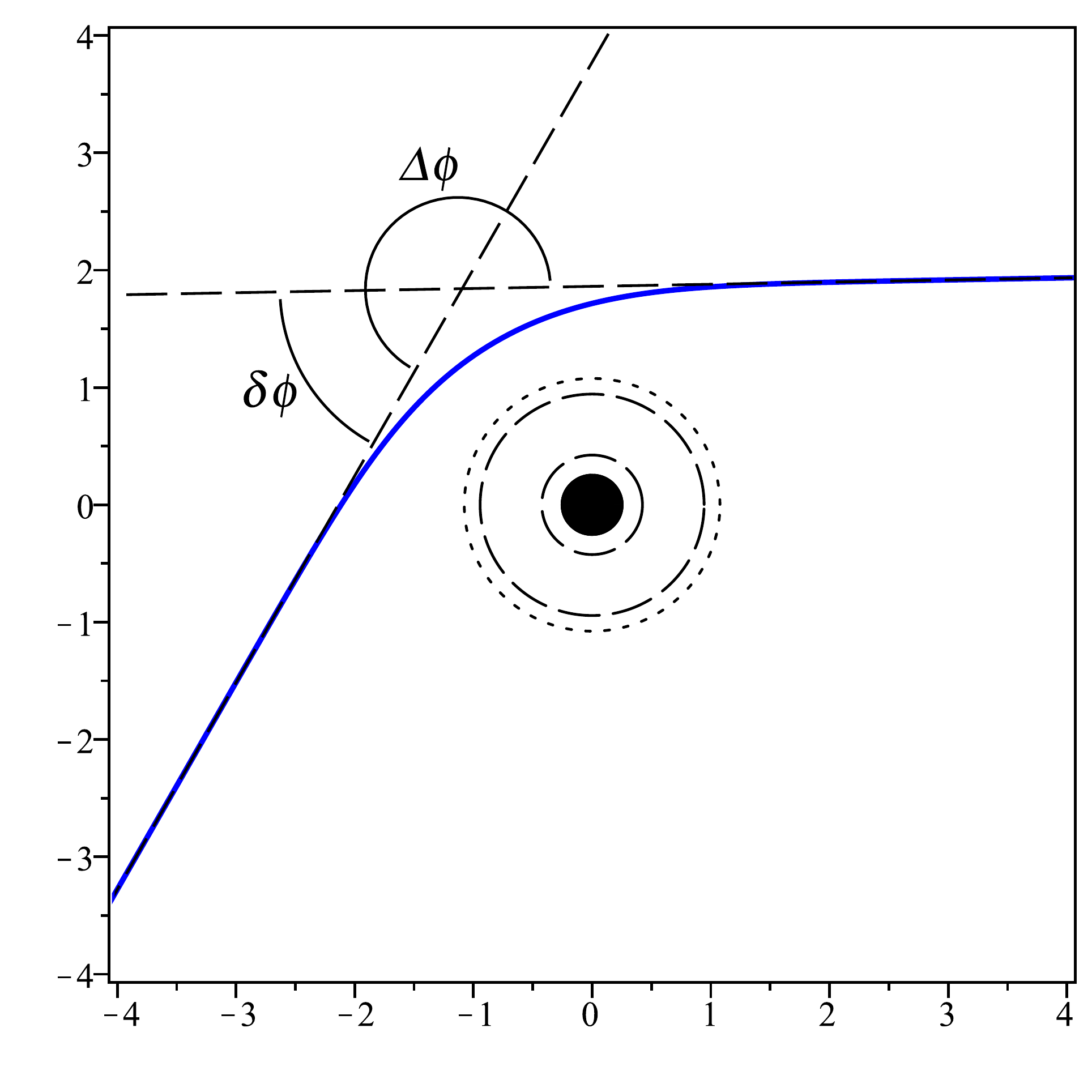}
   \caption{Total change of the $\phi$-coordinate $\Delta \phi$ and deflection angle $\delta \phi$ of a planar, massless escape orbit with parameter values: $E = 1.08, a = 0.4, b = 0.3, \Phi = -2, \mu = 1$.}
\label{fig:deflection}
\end{figure}

Let us first consider the deflection angle of an escape orbit with a radial turning point $x_0 = \frac{1}{b_3} \left(4 e_3^z - \frac{b_2}{3}\right)$.\\
The deflection angle can be determined by calculating the values $\tau_\pm^\infty$ of the Mino time for which the radial coordinate yields $x(\tau_\pm^\infty) = \infty$
\begin{align}
\tau_\pm^\infty = \int_{x_0}^\infty + \tau_{\rm in} = \int_{e_3^z}^\infty \frac{\mathrm dz}{\sqrt{Z}} + \tau_{\rm in}.
\label{eq:deflection}
\end{align}
The sign is related to both branches of $\sqrt{X}$ or $\sqrt{Z}$, respectively. The total change of the angular coordinates is given by
\begin{align}
\begin{aligned}
\Delta \theta &= \theta \left(\tau_+^\infty \right) - \theta \left(- \tau_-^\infty \right),\\
\Delta \phi &= \phi \left(\tau_+^\infty \right) - \phi \left(- \tau_-^\infty \right),\\
\Delta \psi &= \psi \left(\tau_+^\infty \right) - \psi \left(- \tau_-^\infty \right).
\end{aligned}
\end{align}
The related deflection angles $\delta \theta$, $\delta \phi$ and $\delta \psi$ are commonly defined as the total change of the angular coordinate minus $\pi$ \cite{Hartle:2003yu}. This is illustrated by Fig.~\ref{fig:deflection} for a two-dimensional light orbit of a massless particle:

Calculating the integral in Eq.~\eqref{eq:deflection} yields
\begin{align}
\tau_\pm^\infty = \pm 1.00957
\end{align}
and thus the deflection angle is approximately given by
\begin{align}
\delta \phi = \Delta \phi - \pi \approx 0.33 \pi.
\end{align}
This agrees with the deflection angle indicated in Fig.~\ref{fig:deflection}. Since in the two-dimensional case $\phi(\tau)$ only depends on $x$, we could alternatively make use of \cite{Hartle:2003yu}
\begin{align}
\frac{\mathrm d\phi}{\mathrm dz} = \frac{\mathrm d\phi}{\mathrm d\tau} \frac{\mathrm d\tau}{\mathrm dz} = \frac{\dot \phi}{\dot z} \quad \Rightarrow \quad \Delta \phi = \int_{e_3^z}^\infty \frac{R_\phi(z)}{\sqrt{Z}} \mathrm dz.
\end{align}

\subsection{Perihelion shift and Lense-Thirring effect}

Now we consider the perihelion shift and the Lense-Thirring effect for bound orbits or many-world bound orbits. The corresponding $x$- and the $\theta$-motions are periodic
\begin{align}
\begin{aligned}
\theta(\tau) &= \theta(\tau + \omega_\theta),\\
x(\tau) &= x(\tau + \omega_x),
\end{aligned}
\end{align}
whereby these periods are related to the first fundamental period $\omega_1^{y,z}$ of the $\wp$-function
\begin{align}
\begin{aligned}
\omega_\theta &= 2 \int_{\theta_{\rm min}}^{\theta_{\rm max}} \frac{\mathrm d\theta}{\sqrt{\Theta}} = 2 \int_{e_1^y}^{e_2^y} \frac{\mathrm dy}{\sqrt{Y}} = 2 \omega_1^y,\\
\omega_x &= 2 \int_{x_{\rm min}}^{x_{\rm max}} \frac{\mathrm dx}{\sqrt{X}} = 2 \int_{e_1^z}^{e_2^z} \frac{\mathrm dz}{\sqrt{Z}} = 2 \omega_1^z.
\label{eq:periods}
\end{aligned}
\end{align}
The corresponding orbital frequencies with respect to the Mino time $\tau$ are $\Upsilon_\theta = \frac{2\pi}{\omega_\theta}$ and $\Upsilon_x = \frac{2\pi}{\omega_x}$.\\
The orbital periods of the remaining coordinates $\phi$, $\psi$ and $t$ depend on both $\theta$ and $x$ and thus have to be treated differently. Therefore, the	solutions $\phi(\tau)$, $\psi(\tau)$ and $t(\tau)$ consist of two different parts, where one part represents the average rates $\Upsilon_\phi$, $\Upsilon_\psi$ and $\Gamma$ at which $\phi$, $\psi$ and $t$ accumulate with respect to $\tau$

\begin{align}
\begin{aligned}
\Upsilon_\phi &= \frac{2}{\omega_\theta} \int_{\theta_{\rm min}}^{\theta_{\rm max}} \mathrm d\phi_\theta + \frac{2}{\omega_x} \int_{x_{\rm min}}^{x_{\rm max}} \mathrm d\phi_x,\\
\Upsilon_\psi &= \frac{2}{\omega_\theta} \int_{\theta_{\rm min}}^{\theta_{\rm max}}\mathrm d\psi_\theta + \frac{2}{\omega_x} \int_{x_{\rm min}}^{x_{\rm max}} \mathrm d\psi_x,\\
\Gamma &= \frac{2}{\omega_\theta} \int_{\theta_{\rm min}}^{\theta_{\rm max}} \mathrm dt_\theta + \frac{2}{\omega_x} \int_{x_{\rm min}}^{x_{\rm max}} \mathrm dt_x,
\label{eq:frequencies}
\end{aligned}
\end{align}
and the other part represents oscillations around it with periods $\omega_\theta$ and $\omega_x$. The corresponding differentials are given by Eq.~\eqref{eq:phiparts}, Eq.~\eqref{eq:psiparts} and Eq.~\eqref{eq:tparts}. What we eventually need are the orbital frequencies with respect to the coordinate time $t$
\begin{align}
\begin{aligned}
\Omega_\theta =  \frac{\Upsilon_\theta}{\Gamma}, \quad \Omega_x = \frac{\Upsilon_x}{\Gamma}, \quad \Omega_\phi =\frac{\Upsilon_\phi}{\Gamma}, \quad \Omega_\psi = \frac{\Upsilon_\psi}{\Gamma}.
\end{aligned}
\end{align}
The perihelion shift and the Lense-Thirring effect are defined as differences between these orbital frequencies
\begin{align}
\begin{aligned}
\Delta_{\rm P}^\phi &= \Omega_\phi - \Omega_x = \frac{\Upsilon_\phi - \Upsilon_x}{\Gamma},\\
\Delta_{\rm P}^\psi &= \Omega_\psi - \Omega_x = \frac{\Upsilon_\psi - \Upsilon_x}{\Gamma},\\
\Delta_{\rm LT}^\phi &=  \Omega_\phi - \Omega_\theta =  \frac{\Upsilon_\phi -\Upsilon_\theta}{\Gamma},\\
\Delta_{\rm LT}^\psi &=  \Omega_\psi - \Omega_\theta =  \frac{\Upsilon_\psi -\Upsilon_\theta}{\Gamma}.
\end{aligned}
\end{align}

\section{Conclusions and outlook}

In this paper, we have discussed the motion of test particles 
and light in the general five-dimensional Myers-Perry spacetime. 
We have studied the general properties of the equations of motion 
and analyzed the structure of the resulting orbits. 
We have investigated the influence of the test particle's energy 
and angular momenta on its orbital motion by means of $E$-$\Phi$-plots.
This allows us to classify the possible types of orbits 
in this spacetime.
In parallel, we have introduced the effective potential,
associated with the radial coordinate, 
yielding the same classification.\\
To obtain the orbits of test particles and light
we have integrated the equations of motions explicitly
in terms of the Weierstrass elliptic $\wp$-, $\zeta$- and $\sigma$-functions.
The existence of such analytical solutions allows for
systematic applications 
and offers a frame for testing the accuracy 
and reliability of numerical integrations
in other contexts.\\
We have used these solutions in order to illustrate
some examples of typical orbits for massive and massless test particles.
Here we have chosen either two-dimensional plots for orbits lying
in an equatorial plane or three-dimensional plots for the general case,
projecting the general four-dimensional motion into three dimensions.
Furthermore, we have presented expressions for spacetime observables,
such as the light deflection angle, the perihelion shift
and the Lense-Thirring effect.\\
As in four dimensions, also in the five-dimensional Myers-Perry spacetime there is a 
Killing tensor, which is related to the Carter constant allowing to separate the equations of motion.\\
We could point out the typical presence of a centrifugal barrier, preventing most particles from reaching the singularity.\\
However, the additional spacetime dimension of the Myers-Perry solution
causes also differences concerning the properties of the black
hole spacetimes and the respective orbits of test particles and light. 
In five dimensions a general black hole solution possesses
two independent rotation parameters,
because four spatial dimensions imply the existence of two independent planes of rotation. 
Moreover, the structure of the singularity is much more complex. \\
We have confirmed the absence of stable bound orbits
outside the black hole's event horizon, which represents
a generic feature of higher-dimensional spacetimes
with event horizons of spherical symmetry
(see also \cite{Hackmann:2008tu}). Note, though, that there are unstable circular orbits inside and outside the black hole's event horizon. As an interesting fact, we have shown the existence of stable bound orbits 
between the singularity and the Cauchy horizon. 
This has also been seen for
the Reissner-Nordstr\"{o}m \cite{Grunau:2010gd} 
and the Kerr-Newman spacetime \cite{Hackmann:2013pva}. 
However, such orbits would be hidden from an external observer. \\
As a future step, it would be interesting to 
derive the analytical solutions of the geodesics equations 
in the six-dimensional Myers-Perry spacetime. 
The related geodesic equations are solvable in terms of hyperelliptic functions 
\cite{Enolski:2010if,Enolski:2011id}.
Since the six-dimensional Myers-Perry spacetime is even-dimensional, 
it may share a different set of similarities with the Kerr spacetime 
than the five-dimensional case. 
But there are also new features that arise only in higher dimensions. 
For example, if the Myers-Perry spacetime has only one non-vanishing 
angular momentum, then there is no extremal black hole 
in more than five dimensions \cite{Myers:1986un}.\\
Furthermore, the discussion of the five-dimensional de Sitter 
or anti-de Sitter Myers-Perry spacetime \cite{Gibbons:2004uw} 
would be instructive. 
The corresponding analytical solutions may still be expressed 
in terms of hyperelliptic functions. 
The analysis of other charged rotating spacetimes
in higher dimensions might be interesting, as well.
This includes black hole
solutions with horizons of spherical topology, such as
the supersymmetric spacetimes \cite{Chong:2005hr}
or the charged rotating black holes in
Einstein-Maxwell-dilaton theory \cite{Kunz:2006jd},
but also spacetimes with a different horizon topology.\\
In black ring spacetimes, for instance,
the equations of motion could be separated only in special cases
corresponding to
geodesics on the two rotational axes, or zero energy null geodesics
\cite{Hoskisson:2007zk,Durkee:2008an}.
Consequently, the geodesic motion was studied analytically as well as
numerically
  \cite{Hoskisson:2007zk,Elvang:2006dd,Igata:2010ye,Armas:2010pw,Grunau:2012ai,Durkee:2008an,Grunau:2012ri}.
Interestingly, in contrast to Myers-Perry black hole spacetimes,
there are stable bound orbits in black ring spacetimes
\cite{Igata:2010ye,Grunau:2012ai,Grunau:2012ri,Igata:2013be}. Moreover, recent analysis \cite{Igata:2010cd}
revealed, that such ring space-times may possess chaotic bound orbits.
It should be interesting to see whether chaotic motion will also appear in
further higher-dimensional spacetimes. 

\section*{Acknowledgement}

We gratefully acknowledge support by the Deutsche Forschungsgemeinschaft (DFG), in particular, within the framework of the DFG Research Training group 1620 {\it Models of gravity}. We also acknowledge fruitful discussions with Daniel Grieser.

\newpage

\begin{appendix}
\parbox{\textwidth}{
\section{CONSTANTS OF THE PARTIAL FRACTION DECOMPOSITION}
\label{sec:anhang}
The constants of the partial fraction decomposition, used in the analytical solutions for $\phi(\tau)$, $\psi(\tau)$ and $t(\tau)$, are given by
\begin{align*}
p_1 &= \frac{a_3}{4} + \frac{a_2}{12},\\[10pt]
p_2 &= \frac{a_2}{12},\\[10pt]
q_1 &= \frac{b_3}{8} \left(\mu + \frac{2}{3} \frac{b_2}{b_3} - a^2 - b^2 + \sqrt{\left(\mu - a^2 - b^2 \right)^2 - 4 a^2 b^2} \right) = \frac{b_3}{4} x_+ + \frac{b_2}{12},\\[10pt]
q_2 &= \frac{b_3}{8} \left(\mu + \frac{2}{3} \frac{b_2}{b_3} - a^2 - b^2 - \sqrt{\left(\mu - a^2 - b^2 \right)^2 - 4 a^2 b^2} \right) =  \frac{b_3}{4} x_- + \frac{b_2}{12}
\end{align*}
and
\begin{align*}
G^\phi &= - \frac{a_3}{4} \Phi,\\[10pt]
G^\psi &=  \frac{a_3}{4} \Psi,\\[10pt]
H_1^\phi &= -\,b_3 \frac{\Phi \big(3 a^2 b^2 b_3 + 3 b^2 b_3 \mu - 3 b^4 b_3 - a^2 b_2 + 12 a^2 q_1 - 12 b^2 q_1 + b^2 b_2 \big)+ E \big( 3 a b^2 b_3 \mu - a b_2 \mu + 12 a \mu q_1 \big)}{48 q_1 - 48 q_2}\\ &\hspace{11.5pt} - b_3 \frac{\Psi \big(3 a b b_3 \mu \big)}{48q_1 - 48q_2},\\[10pt]
H_2^\phi &= b_3 \frac{\Phi \big(3 a^2 b^2 b_3 + 3 b^2 b_3 \mu - 3 b^4 b_3 - a^2 b_2 + 12 a^2 q_2 - 12 b^2 q_2 + b^2 b_2 \big) + E \big( 3 a b^2 b_3 \mu - a b_2 \mu + 12 a \mu q_2 \big)}{48 q_1 - 48 q_2}\\ &\hspace{11.5pt} + b_3 \frac{\Psi \big(3 a b b_3 \mu \big)}{48q_1-48q_2},\\[10pt]
H_1^\psi &= -\,b_3 \frac{\Psi \big(3 a^2 b^2 b_3 + 3 a^2 b_3 \mu - 3 a^4 b_3 - b^2 b_2 + 12 b^2 q_1 - 12 a^2 q_1 + a^2 b_2 \big) + E \big( 3 a^2 b b_3 \mu - b b_2 \mu + 12 b \mu q_1 \big)}{48 q_1 - 48 q_2}\\ &\hspace{11.5pt} -b_3 \frac{\Phi \big(3 a b b_3 \mu \big)}{48q_1-48q_2},   \\[10pt]
H_2^\psi &= b_3 \frac{\Psi \big(3 a^2 b^2 b_3 + 3 a^2 b_3 \mu - 3 a^4 b_3 - b^2 b_2 + 12 b^2 q_2 - 12 a^2 q_2 + a^2 b_2 \big) + E \big( 3 a^2 b b_3 \mu - b b_2 \mu + 12 b \mu q_2 \big)}{48 q_1 - 48 q_2}\\ &\hspace{11.5pt} +b_3 \frac{\Phi \big(3 a b b_3 \mu \big)}{48q_1-48q_2},\\[10pt]
H_1^t &= b_3 \mu \frac{\Phi \big(12 a q_1 - a b_2 + 3 a b^2 b_3 \big) + \Psi \big(12 b q_1 - b b_2 + 3 a^2 b b_3 \big) + E \big(12 \mu q_1 - b_2 \mu \big)}{48 q_1 - 48 q_2},\\[10pt]
H_2^t &= -\,b_3 \mu \frac{\Phi \big(12 a q_2 - a b_2 + 3 a b^2 b_3 \big) + \Psi \big(12 b q_2 - b b_2 + 3 a^2 b b_3 \big) + E \big(12 \mu q_2 - b_2 \mu \big)}{48 q_1 - 48 q_2}.
\end{align*}}
\end{appendix}

\newpage
\quad

\end{document}